\documentclass[aps,pre,twocolumn,amssymb,amsmath]{revtex4-1}
\usepackage{graphicx,latexsym,pifont,dsfont,amsmath}
\usepackage{times}
\usepackage{tikz}
\usetikzlibrary{snakes}
\definecolor{labelkey}{cmyk}{.4,.2,0,0}
\usepackage{color}
\definecolor{Blue}{rgb}{0.00, 0.00, 1.00}
\definecolor{Red}{rgb}{1.00, 0.00, 0.00}

\newcommand{\rme}{{\mathrm{e}}}
\newcommand{\rmd}{{\mathrm{d}}}

\newcommand{\half}{\frac12}

\newcommand{\nn}{\nonumber}

\newcommand{\fig}[2]{\includegraphics[width=#1]{./#2}}
\newcommand{\Fig}[1]{\includegraphics[width=8.7cm]{./#1}}

\newlength{\bilderlength}

\newcommand{\mtilde}{\tilde}

\renewcommand{\epsilon}{\varepsilon}
\newcommand{\asinh}{\mbox{asinh}}

\def\be{\begin{equation}}
\def\ee{\end{equation}}

\def\bal{\begin{align}}
\def\eal{\end{align}}

\def\bea{\begin{eqnarray}}
\def\eea{\end{eqnarray}}
\renewcommand{\log}{\ln }

\usepackage{color}

\begin{document}

\title{Perturbative Expansion for the Maximum of  Fractional Brownian Motion}

\author{Mathieu Delorme and Kay J\"org Wiese}
\address{CNRS-Laboratoire de Physique Th\'eorique de l'Ecole
Normale Sup\'erieure, 24 rue Lhomond, 75005 Paris, France
}

\begin{abstract}
Brownian motion is the only random process which is {\em Gaussian}, {\em stationary} and {\em Markovian}.  Dropping the Markovian property, i.e.\ allowing for memory, one obtains a class of processes called 
{\em Fractional Brownian motion},   indexed by the  {\em Hurst} exponent $H$. For $H=1/2$,   Brownian motion is recovered. We develop a perturbative approach to treat the non-locality  in time  in an expansion in  $\varepsilon = H-1/2$. This allows us to derive analytic results beyond  scaling exponents for various observables related to extreme value statistics: The maximum $m$ of the process and the time $t_{\rm max}$ at which this maximum is reached, as well as their joint distribution. We test  our analytical predictions with extensive numerical simulations for different values of $H$. They    show excellent agreement, even for  $H$ far from $1/2$.
\end{abstract}

\maketitle

\section{Introduction}

Random processes are ubiquitous in nature. 
Though many processes can successfully be mExtensive numerical simulations for different values of $H$  test these analytical predictions and show excellent agreement, even for large $\varepsilon$.odeled by Markov chains and are well analyzed by   tools of statistical mechanics, there are also interesting and realistic systems which do not evolve with independent increments, and thus are non-Markovian, \textit{i.e.}\ history dependent. Dropping the Markov property, but demanding that a continuous process be scale-invariant and Gaussian with stationary increments  defines an enlarged class of random processes, known as fractional Brownian motion (fBm). Such processes appear in a broad range of contexts: Anomalous diffusion \cite{BouchaudGeorges1990}, diffusion of a marked monomer inside a polymer 
\cite{WalterFerrantiniCarlonVanderzande2012,AmitaiKantorKardar2010}, 
polymer translocation through a pore \cite{AmitaiKantorKardar2010,ZoiaRossoMajumdar2009,DubbeldamRostiashvili2011,PalyulinAlaNissilaMetzler2014}, single-file diffusion in ion channels \cite{KuklaKornatowskiDemuthGirnusal1996,WeiBechingerLeiderer2000}, the dynamics of a tagged monomer \cite{GuptaRossoTexier2013,Panja2011}, finance (fractional Black-Scholes, fractional stochastic volatility models, and their limitations) \cite{CutlandKoppWillinger1995,Rogersothers1997,RostekSchobel2013}, hydrology  \cite{MandelbrotWallis1968,MolzLiuSzulga1997},  and many more.

\begin{figure}[t]
        \Fig{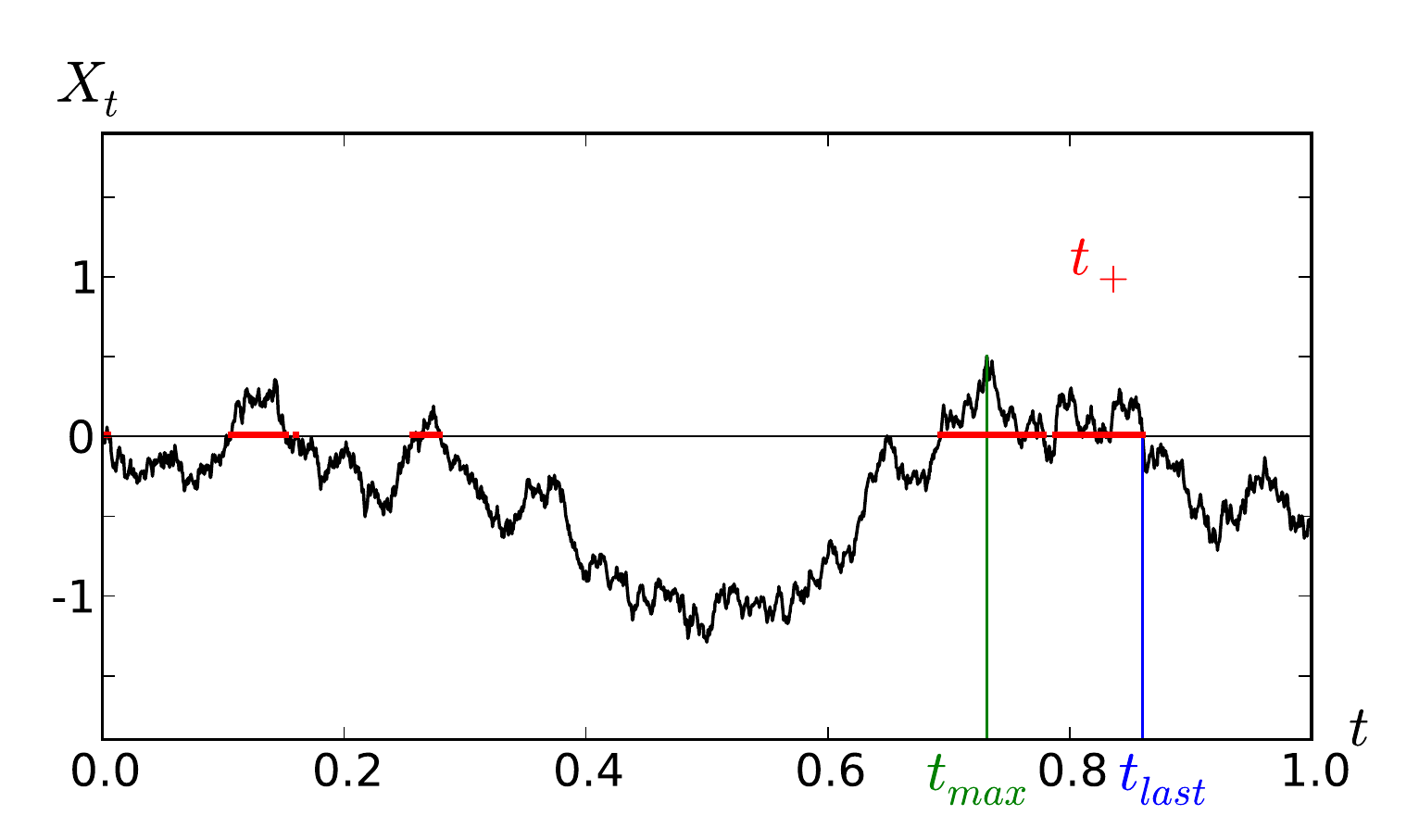}
        \caption{The three arcsine laws discussed in the main text. $t_{\rm max}$, in green, is the time where the process achieves its maximum. $t_{\rm last}$, in blue, is the last time the process is at its starting value $X_0 =0$. Finally,  $t_+$, in red, is the  time spend in the positive half space, which is the sum of the red intervals .}
        \label{f:3-arcsine-laws}
\end{figure}

While  averaged quantities have been studied extensively and are well characterized, it is often more important to understand  the extremal behavior of these processes, or the time the process satisfies a given criterion \cite{GumbelBook}. These quantities are associated with failure in fracture or earthquakes, a crash in the stock market, the breakage of dams, the time one has to heat, etc. 
For Brownian motion the three {\em arcsine laws} are well studied examples. They state that for a Brownian process $X_t$, with  $0<t<1$, and $X_0=0$,   three observables $  Y$ have the same probability distribution, namely 
\bea \label{ArcsinDistrib}
{\cal P}(Y<y) &=& \frac2\pi \,\mbox{arcsin}(\sqrt{y}) \\
\Leftrightarrow~~ {\cal P}(y) &=& \frac1{\pi \sqrt{y(1-y)}}\ .
\eea
The observables in question are (see Fig.\ \ref{f:3-arcsine-laws})
\begin{enumerate}
\item First (L\'evy's) arcsine law: The time the   process $X_t$ is positive, (red in Fig.\ \ref{f:3-arcsine-laws}),  
\be t_+:= \int_0^1 \Theta (X_t)\,\rmd t\ .
\ee
\item Second arcsine law: The last time the   process is at  its initial position, (blue in Fig.\ \ref{f:3-arcsine-laws}), 
\be
t_{\rm last}:= \sup \,\{ t\in [0,1], X_t=0 \}\ .
\ee
\item Third arcsine law:
The time at which the process $X_t$ achieves its maximum (which is almost surely unique), (green in Fig.\ \ref{f:3-arcsine-laws})
\be  
t_{\rm max}:= t, \ {\rm s.t.\ } X_t =  \sup \,\{ X_s, s \in [0,1] \}\ .
\ee
\end{enumerate}
While these laws are well-studied for Brownian motion, little is known about their generalization to other random processes. In this article, we will generalize the third arcsine law to fractional Brownian motion, and obtain the distribution of the achieved maximum. 

Fractional Brownian motion (fBm) is a random process $X_t$ characterized by the Hurst exponent $H$ which quantifies the growth of the 2-point function in time, \begin{equation}
\left< (X_t - X_s)^2 \right>= 2 |t-s|^{2H}\ .
\end{equation}
Up to now, analytical tools to  study its extreme value statistics were  available only for Brownian motion, \textit{i.e.}\ $H=1/2$. In this article, we aim     to extend this to $H\neq1/ 2$. This is achieved by constructing a path integral, and evaluating it  perturbatively around a Brownian,  setting $H=1/2
+ \varepsilon$. This technique has been introduced in Ref.~\cite{WieseMajumdarRosso2010}. We will calculate  the probability
distribution of the maximum $m$  of the process and the time
$t_{\rm max}$ at which the maximum is reached, as well as their
joint distribution.  A short account of this work was published in Ref.~\cite{DelormeWiese2015}.

The article is structured as follows:\ Section \ref{s:fBm}   defines the fBm,   discusses its relation to anomalous diffusion, and defines the observables related to extremal value statistics we wish to study. 

Section \ref{s:pert.approach} introduces the path integral we need to calculate, followed by its perturbative expansion in $\epsilon = H-1/2$. This defines the main integrals to be calculated, for which we also give a diagrammatic representation. As the calculations are rather tedious, they are relegated to appendix \ref{AppendixZgamma}. 

Section \ref{SectionResluts} presents our  results: We start by recalling scaling relations in section \ref{s:scaling}, before introducing  our most general formula, the  probability to start at $m_1>0$, to reach the minimum $x_0\approx 0$ at time time $t$, and to finish at time $T>t$ in $m_2$. This allows us to derive several simpler results: First the distribution of times when the maximum is achieved, for a Brownian known as the {\em third arcsine law} (section \ref{s:3rd-arcsine-law}). Second, the distribution of the value of this maximum. And third, the joint distribution of maximum, and the time when this maximum is taken.

 Extensive numerical simulations for different
values of $H$   test these analytical predictions   in section \ref{s:Numerics}.

Conclusions are given in Section \ref{s:Conclusions}, followed by several appendices: Appendix \ref{AppendixDetails} gives details on the perturbation expansion.  Appendix \ref{AppendixOldResult} reviews results from \cite{WieseMajumdarRosso2010}, including a new derivation of the latter. Appendix \ref{AppendixZgamma} calculates the  main  new, and most difficult, contribution.   
Appendix \ref{AppendixArcsine_Law} gives details on the corrections to the third Arcsine Law, while      for the attained maximum and its cumulative distribution this is done in appendices \ref{AppendixMaxDistrib} and \ref{AppendixSurvival}. 
Appendix \ref{AppendixLaplace} gives a list of used inverse Laplace transforms. Finally, in  appendix \ref{AppendixCheck} is verified that the second cumulant is correctly reproduced. 

\section{Fractional Brownian motion and  Observables}
\label{s:fBm}
\subsection{Definition of the fBm}

FBm is a generalization of  standard Brownian motion to other fractal dimensions, introduced in its final form by Mandelbrot and Van Ness \cite{MandelbrotVanNess1968}. It is a  Gaussian process $(X_t)_{t\in \mathbb{R}}$, starting at zero, $X_{0}=0$, with  mean $\left<X_{t}\right>=0$  and  covariance function (variance)
\begin{equation}
\langle X_t X_s \rangle = s^{2H} + t^{2H} - |t-s|^{2H} \ .
\label{covariance}
\end{equation}
A fBm $Y_t$ starting at a non-zero value $y=Y_0$ is   defined as $Y_t = X_t +y$, with $X_t$ as above. The parameter $H\in(0,1)$ appearing in \eqref{covariance} is the Hurst exponent. Standard Brownian motion corresponds to $H=1/2$; there the covariance function \eqref{covariance}  reduces to $\langle X_t X_s \rangle = 2 \min(s,t) $. 
Unless $H=1/2$, the process is non-Markovian , \textit{i.e.}\ its increments are not independent: For $H>1/2$ they are correlated, whereas for $H<1/2$ they are anti-correlated:
\be
\begin{split}
\langle \partial_{t} X_t \,\partial_{s} X_s \rangle =& 2H (2H-1) |t-s|^{2(H-1)} .\\
\end{split}
\ee
It is   important to note that the process is stationary, as the second moment (and thus the whole distribution) of the increments is a function  of the time difference $|t-s|$ only, 
\begin{equation}
\langle(X_t - X_s)^2 \rangle= 2 |t-s|^{2H}\ .
\end{equation}
The fact that a fBm process is non-Markovian makes its study difficult, as most of the standard stochastic-process tools (decomposing transition probabilities into products of propagators, or writing the evolution of a density using a Fokker-Plank equation) rely on the Markov property.

\subsection{Anomalous diffusion}

Anomalous diffusion is another interesting property of the fBm. It is caracterized by the non-linear growth (for $H \neq 0.5$) of the second moment of the process,
\begin{equation}
\langle X_t^2 \rangle = 2 t^{2H}\ .
\end{equation}
For $H<1/2$, a fBm is a sub-diffusive process, while for $H>1/2$, it is   super-diffusive.

Anomalous diffusion is usually implied by a stronger property (but equivalent in the case of a Gaussian process):  self-similarity of exponent $H$. It means that rescaling time by $\lambda >0$ and space by $\lambda^{-H}$ leaves every averaged observable $\langle\mathcal{O}[X_t]\rangle$ defined on the process invariant, 
\begin{equation}
\langle\mathcal{O}[\lambda^{-H} X_{\lambda t}]\rangle=\langle\mathcal{O}[X_t]\rangle\ .
\end{equation}
This property is stronger in the sense that the growth of every moment,  and not only the second one, is governed by the same exponent $H$: $\langle X_t^n \rangle \sim t^{n H}$.

It is well known that standard Brownian motion is the only {\em continuous} process with {\em stationary}, independent (Markovian) and {\em Gaussian} increments. 
As a consequence,  every process in this class is $\half$-self-similar, \textit{i.e.} exhibits normal diffusion.
To obtain an anomalous diffusive process, one of these three hypotheses has to be removed. This gives three main classes of anomalous diffusion:
\begin{itemize}
\item heavy tails of the increments (Levy-flight process) or heavy tails in the waiting time between increments (CTRW processes); these  processes are {\em non-Gaussian}.
\item time dependence of the diffusive constant: the distribution of the increments is time dependent, \textit{i.e.} the process is {\em non-stationary}.
\item correlations between increments: the process is {\em non-Markovian}
\end{itemize}
FBm is the only process which is {\em Gaussian, stationary, and statistically self-similar}. As the  first two hypotheses are natural in a large class of  processes appearing in nature, and self-similarity with exponent $H \neq 1/2$ is equivalent to anomalous diffusion for a Gaussian process,  fBm appears as an important representative for anomalous diffusion.

Interestingly, several processes commonly used in physics, mathematics, and computer science belong to the fBm class. For expample, it was recently proven that the dynamics of a tagged particle in single-file diffussion  (\textit{cf.} \cite{WeiBechingerLeiderer2000,KrapivkyMallickSadhu2014,KrapivkyMallickSadhu2015a,KrapivkyMallickSadhu2015}) has at large times the fBm covariance function \eqref{covariance} with Hurst exponent $H=1/4$.

\subsection{Extreme-value statistics (EVS)} 
\begin{figure}[t]
        \Fig{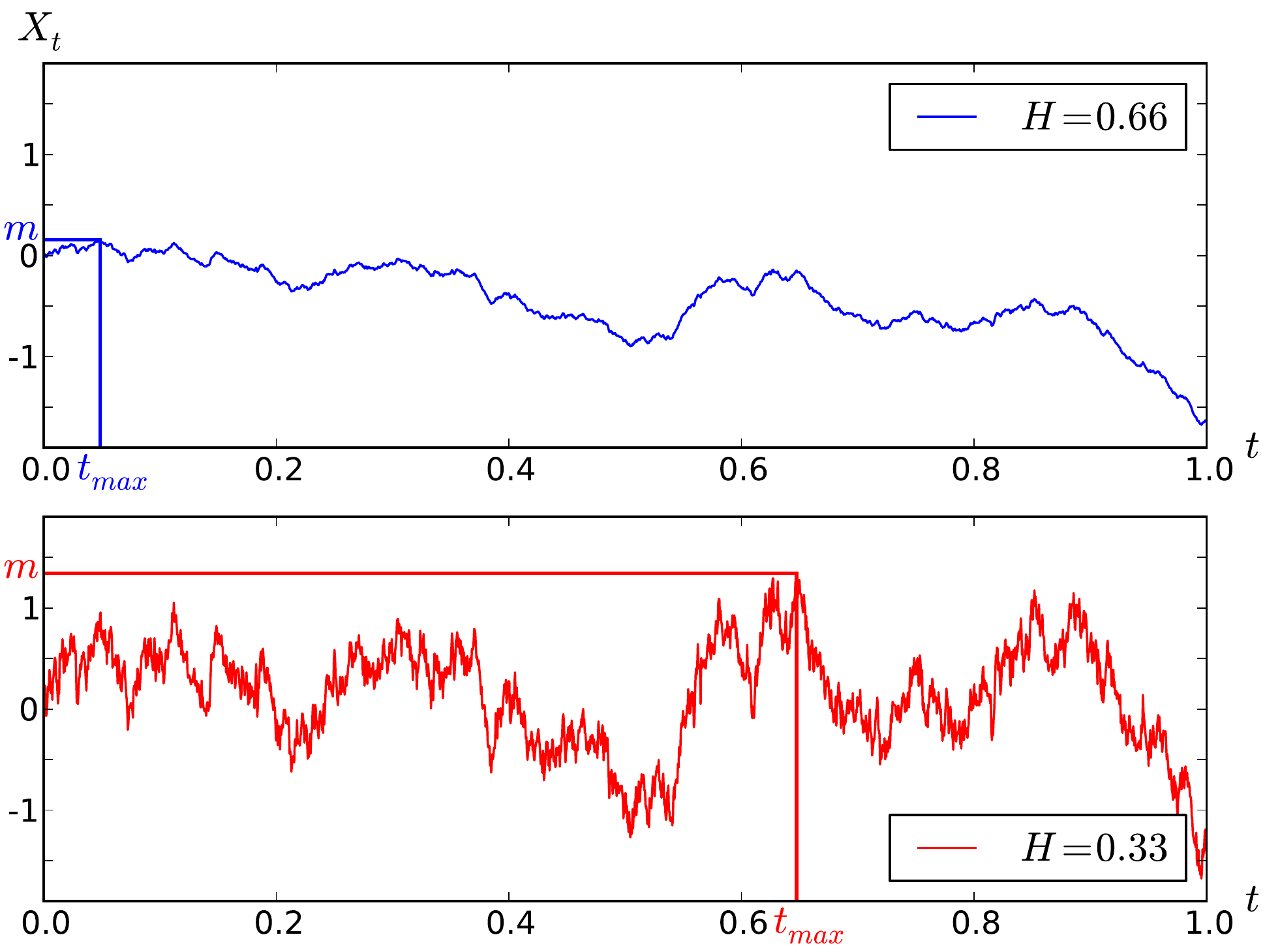}
        \caption{Two realisations of fBm paths for different values of $H$, generated using the same random numbers for the Fourier modes in the    Davis and Harte  procedure  \cite{Dieker}. The observables $m$ and $t_{\rm max}$ are given.}
        \label{Illustration}
\end{figure}
The objective of this article is to study  fBm in the context of what is now called {\em extreme-value statistics}. While the knowledge of averages or of the typical behavior is an important step in understanding and comparing stochastic models to experiments or data, there are  situations were the interest lies in the extremes or rare events. For example, the physics of disordered systems at low temperatures is governed by the states with a (close to) minimal energy in the random energy landscape. Extreme weather conditions are of   importance in the dimensioning  of infrastructures such as dams and  bridges. More generally, extreme value questions appear naturally in many optimization problem.
         
The simplest and first case studied for these extreme-value statistics was the distribution of the maximum of a large number $N$ of independent and identically distributed random variables, which is now well understood in the large-$N$ limit thanks to the classification of the Fisher-Tippett-Gnedenko theorem: Depending on the initial distribution of the variables, the rescaled maximum follows either a Weibull, Gumbel or Fr\'echet distribution \cite{GumbelBook,BouchaudMezard1997}. This is the equivalent of the central-limit theorem, which classifies the sums, or equivalently averages, of a large number of independent identically distributed (i.i.d.) variables.

The case of strongly correlated variables was a natural extension to this problem, as many physically relevant situations present significant deviations from the i.i.d. case. Many results were derived for random walks and Brownian motion \cite{SchehrLeDoussal2010,DeanMajumdar2001}. The distribution of the largest eigenvalue is also a central question in random matrix theory \cite{MajumdarSchehr2014}. Finally, some previous study in the context of non-Markovian processes can be found in Ref.~\cite{DerridaHakimZeitak1996,Majumdar1999,MajumdarRossoZoia2010}.
 
In this article we   study the extremal properties of a fractionnal Brownian motion $X_t$. The main observables are the {\em maximum} $m= \max_{t\in[0,T]}X_t$ and the time $t_{\rm max}$ when this maximum is reached. Figure \ref{Illustration} shows an illustration for different values of $H$, using the same random numbers for the Fourier modes. We will denote $P^T_H(m)$ and $P^T_H(t)$ their respective probability distributions. Previous studies on these distributions, focusing on the small-scale behaviour, can be found in Refs.\ \cite{Sinai1997,Molchan1999}.

These observables are closely linked to other quantities of interest, such as the first-return time, the survival probability, the persistence exponent, and the statistics of records.

\section{The perturbative approach}
\label{s:pert.approach}
\subsection{Path integral formulation and the Action}
Following the ideas of \cite{MajumdarSire1996,OerdingCornellBray1997,WieseMajumdarRosso2010}, we start with the  path-integral, 
\begin{equation}\label{PathIntegral}
\begin{split}
 Z^+&(m_1,t_1;x_0;m_2,t_2) = \\
&\int_{X_0 =m_1}^{X_{t_1+t_2}=m_2} \mathcal{D}[X] \,\Theta[X] \, \delta(X_{t_1} - x_0)\, e^{-S[X]}\ .
\end{split}
\end{equation}
It   sums over all paths $X_t$, weighted by their probability $e^{-S[X]}$,  starting at $X_0=m_1>0$, passing through $x_0$ (close to $0$) at time $t_1$, and ending in $X_{t_1+t_2}=m_2>0$, while staying {\em positive} for all $t\in \left[0,T=t_1+t_2 \right]$.  The latter is enforced by the product of  Heaviside functions $\Theta[X]:= \prod_{s=0}^{t_1+t_2}\Theta(X_s)$. This path integral depends on the Hurst exponent $H$ through the action.
Since
$X_t$ is a Gaussian process, the action $S$ can  (at least formally) be constructed from the covariance function of $X_t$,
\begin{equation}
S[X]= \frac12\int_{t_1,t_2} X_{t_1} G(t_1,t_2) X_{t_2}\ .
\end{equation}
Here $\langle X_{t_1} X_{t_2}\rangle = G^{-1}(t_1,t_2)$. This, however, is not enough to evaluate the path integral \eqref{PathIntegral}, since it is not evident how to implement the product of $\Theta$-functions. Following the formalism of Ref.~\cite{WieseMajumdarRosso2010}, we use standard Brownian motion as a starting point for a perturbative expansion, setting  $H = \frac{1}{2} + \varepsilon$ with $\varepsilon$ a small parameter; then  the action at first order in $\varepsilon$ is (we refer to the appendix of Ref.~\cite{WieseMajumdarRosso2010} for the derivation)
\begin{align}
 S\left[ X \right] =&\; \frac{1}{4 D_{\varepsilon,\tau}} \int_0^T \dot{X}_{\tau_1}^2 \rmd\tau_1 \nn\\
& -\frac{\varepsilon}{2} \int_0^{T-\tau} \rmd\tau_1 \int_{\tau_1+\tau}^T \rmd\tau_2 \,\frac{ \dot{X}_{\tau_1} \dot{X}_{\tau_2}}{|\tau_2-\tau_1|}  +\mathcal{O}(\varepsilon^2)\ .
\label{ActionExpansion}
\end{align}
The time $\tau$ is a regularization cutoff for coinciding times (a UV cutoff). We will see that it has no impact on the distribution of observables which  can be extracted from the path integral. (One can also introduce discrete times  spaced by $\tau$ \cite{WieseMajumdarRosso2010}).

The first line of Eq.~\eqref{ActionExpansion}, which we   denote $S_0[X]$, is the action for standard Brownian motion, with a rescaled diffusion constant
\begin{equation}\label{Diffusive_constant}
D_{\varepsilon,\tau} = 1 + 2 \varepsilon [1 + \ln(\tau)] + \mathcal{O}(\varepsilon^2) \simeq (\rme\tau)^{2 \varepsilon}.
\end{equation}
It is a dimensionfull constant, as fBm and standard Brownian motion  do not  have the same {\em time dimension}. The second line, which we denote $S_1[X]$, is the first  correction to the action. It is non-local in time, which implies that the process is   non-Markovian (even if we neglect $\mathcal{O}(\varepsilon^2)$ terms).
We check this expansion of the action in appendix \ref{AppendixCheck}, where we compute the covariance of the process from a path integral, and recover Eq.~\eqref{covariance} at first order in $\varepsilon$.

As we will see in section \ref{SectionResluts}, this path integral $ Z^+(m_1,t_1;x_0;m_2,t_2)$, in the limit of $x_0 \rightarrow 0$, encodes a plethora of information about the maximum of the process: both distributions $P^T_H(m)$ and $P^T_H(t)$ can be extracted from it, as well as the joint distribution. Further,  the same distributions in the case of a fBm bridge.

 It is important to note that the limit of $x_0 \rightarrow 0$ is non-trivial, as it forces the process to go close to an absorbing boundary which   leads to non-trivial scaling involving the persistence exponent $\theta$ defined below in section \ref{s:scaling}.

\subsection{The order-$0$ term}

Having expressed the perturbative expansion of the action, the main task is to compute the path integral \eqref{PathIntegral}, at first order in $\varepsilon$, and in the limit of small $x_0$. Expanding the exponential of the action in \eqref{PathIntegral}, 
\begin{equation}
e^{-S[X]}=e^{-S_0[X]}\left(1-S_1[X]+... \right)\ ,
\end{equation}
allows us to compute the path integral perturbatively in the non-local interaction $S_1[X]$,  defined as the second line of Eq.\ \eqref{ActionExpansion},
\begin{equation}
S_1[X]=-\frac{\varepsilon}{2} \int_0^{T-\tau} \rmd\tau_1 \int_{\tau_1+\tau}^T \rmd\tau_2 \,\frac{ \dot{X}_{\tau_1} \dot{X}_{\tau_2}}{|\tau_2-\tau_1|} \ .
\end{equation}
This gives
\begin{align}\label{PathIntegralExpansion} \nn
Z^+(m_1,t_1;x_0;m_2,t_2) =&\,Z^+_0(m_1,t_1;x_0;m_2,t_2)\\ \nn
&+\varepsilon Z^+_1(m_1,t_1;x_0;m_2,t_2)\\
&+\mathcal{O}(\varepsilon^2)\ .
\end{align}
$Z_0^+$ is the term with no non-local interaction, while $\varepsilon Z_1^+$ is the term with one interaction (it is proportional to $\varepsilon$ because the non-local interaction itself has an amplitude of order $\varepsilon$). Formally, the order-$0$ term is  \begin{equation}
\begin{split}
Z_0^+&(m_1,t_1;x_0;m_2,t_2) = \\
&\int_{X_0 =m_1}^{X_{t_1+t_2}=m_2} \mathcal{D}[X] \,\Theta[X] \, \delta(X_{t_1} - x_0)\, e^{-S_0[X]}\ ,
\end{split}
\end{equation}
where $S_0$ is the action of a standard Brownian motion,
\begin{equation}
S_0\left[ X \right] =\; \frac{1}{4 D_{\varepsilon,\tau}} \int_0^t \!\dot{X}_{\tau_1}^2 \rmd\tau_1 \ .
\label{Action0}
\end{equation}
Since Brownian motion is a Markov process, this action is local in time. It allows us to write the path integral as a product,
\begin{align}
&Z_0^+(m_1,t_1;x_0;m_2,t_2)\nn \\
&=\int\limits_{X_0 =m_1}^{X_{t_1}=x_0}\!\!\! \mathcal{D}[X]\Theta[X]  e^{-S_0[X]} \int\limits_{X_{t_1} =x_0}^{X_{T}=m_2} \!\!\!\mathcal{D}[X] \Theta[X]   e^{-S_0[X]} \nn\\
&=P_0^+(m_1,x_0,t_1) P_0^+(x_0,m_2,t_2)\ . 
\end{align}
In the second line, the constraint $\delta(X_{t_1}-x_0)$ is  enforced by the boundary conditions of the path integral. In the last line, we expressed each path integral in terms of the propagator $P_0^+(x_1,x_2,t)$ of  standard Brownian motion,  constraint to the positive half space. It is  obtained via the method of images, 
\begin{equation}\label{ConstrainedPropagator}
\begin{split}
P_0^+(x_1,x_2,t) &= \frac{1}{\sqrt{4 \pi D t}}\left(e^{-\frac{(x_1-x_2)^2}{4 D t}}-e^{-\frac{(x_1+x_2)^2}{4 D t}}\right)\\
& \underset{x_1 \rightarrow 0}{\simeq} x_1 x_2 \frac{e^{-\frac{x_2^2}{4 D t}}}{\sqrt{4 \pi  D^3 t^3}}\ ,
\end{split}
\end{equation}
for an arbitrary diffusive constant $D$. We   now use that the diffusive constant is   $D_{\varepsilon,\tau}=1+\mathcal{O}(\varepsilon)$.
This allows us to express the path integral \eqref{PathIntegral} at leading order in $\varepsilon$, and in the limit of small $x_0$, as 
\begin{equation}\label{Zorder0}
Z_0^+(m_1,t_1;x_0;m_2,t_2)\underset{x_0 \rightarrow 0}{\simeq}  x_0^2 \frac{m_1 m_2 e^{-\frac{m_1^2}{4 t_1}-\frac{m_2^2}{4 t_2}}}{4 \pi t_1^{3/2}t_2^{3/2}}+\mathcal{O}(\varepsilon)\ .
\end{equation}
To include the order-$\varepsilon$ term in the diffusive constant to get the full result for $Z_+$ at  order $\varepsilon$, we   use Eq.~\eqref{Diffusive_constant}  expanded in $\varepsilon$,
\begin{equation}\label{Zorder0Bis}
\begin{split}
Z_0^+&\underset{x_0 \rightarrow 0}{\simeq} x_0^2 \frac{m_1 m_2 e^{-\frac{m_1^2}{4 t_1}-\frac{m_2^2}{4 t_2}}}{4 \pi t_1^{3/2}t_2^{3/2}} \times \\
&\left\{ 1 +  \varepsilon\left[1+\ln(\tau)\right] \left(\frac{m_1^2}{2t_1}+\frac{m_2^2}{2t_2}-6\right)\right\} +\mathcal{O}(\varepsilon^2)\ .
\end{split}
\end{equation}
It is interesting to note that the order-$\varepsilon$ term appearing here can also be computed from the result \eqref{Zorder0} as
\begin{equation}
2(1 + \log(\tau)) (t_1 \partial_{t_1}+t_2 \partial_{t_2})Z_0^+ .
\end{equation}
\subsection{The first-order terms}

To go beyond   Brownian motion and include non-Markovian effects, i.e. interactions non-local in time, we need to compute the first-order correction in the expansion \eqref{PathIntegralExpansion}, which is called $Z_1^+$ and reads
\begin{widetext}        
\begin{equation}\label{PathIntegralOrdre1}
\begin{split}
Z^+_1(m_1,t_1;x_0;m_2,t_2)=&\frac{1}{2}\int_0^{T-\tau} \rmd\tau_1 \int_{\tau_1+\tau}^T \rmd\tau_2 \int_{X_0 =m_1}^{X_{T}=m_2} \mathcal{D}[X] \frac{ \dot{X}_{\tau_1} \dot{X}_{\tau_2}}{|\tau_2-\tau_1|} \, \delta(X_{t_1} - x_0)\,\Theta[X] \, e^{-S_0[X]} .
\end{split}
\end{equation}  
\end{widetext}
As before, we denote $T=t_1+t_2$. To compute $Z_1^+$,  we  decompose it into three terms, distinguished by their  time ordering. Denote $Z^+_{\alpha}$ the part where $\tau_1<\tau_2 <t_1$, $Z^+_{\beta}$ the part where $t_1<\tau_1 <\tau_2 $, and  $Z^+_{\gamma}$ the term where $\tau_1<t_1<\tau_2$. Then
\begin{equation}\label{Z1split}
\begin{split}
Z_1^+ (m_1,t_1;x_0;m_2,t_2)=&\,Z_\alpha^+(m_1,t_1;x_0;m_2,t_2) \\
& + Z_\beta^+ (m_1,t_1;x_0;m_2,t_2) \\
&+ Z_\gamma^+(m_1,t_1;x_0;m_2,t_2)\ .\\
\end{split}
\end{equation}
In the first term, the \textit{interaction} affects only the process in the time interval $[0,t_1]$, and there is no coupling with the process on the time interval $[t_1,t_1+t_2]$. This leads, as shown in appendix \ref{AppendixDetails}, to a factorized expression for $Z^+_{\alpha}$,
\begin{equation}\label{Zalpha}
\begin{split}
Z^+_{\alpha}(m_1,t_1;&x_0;m_2,t_2)= P_1^+(m_1,x_0,t_1) P_0^+(x_0,m_2,t_2) .
\end{split}
\end{equation}
Here $P_1^+(m,x_0,t)$ is the order-$\varepsilon$  correction to the propagator of fBm in the half space (i.e. constrained to remain positive). This object, which we need   in the limit of $x_0 \rightarrow 0$, was studied and computed in Ref.\ \cite{WieseMajumdarRosso2010}. The result is recalled in appendix \ref{AppendixOldResult}, and recalculated  using more efficient technology developed here. The second term is similar to the first,  swapping the two time intervals,
\begin{equation}\label{Zbeta}
Z^+_{\beta}(m_1,t_1;x_0;m_2,t_2)=P_0^+(m_1,x_0,t_1) P_1^+(x_0,m_2,t_2) .
\end{equation}
The third term, $Z_{\gamma}^+$, is more complicated as the interaction couples the two time intervals $[0,t_1]$ and $[t_1,T=t_1+t_2]$. We can still take advantage of  locality in time of the action $S_0$ to write the path integral  \eqref{PathIntegralOrdre1}, with time integrals restricted to $0<\tau_1<t_1<\tau_2<T$, as a product of simpler path integrals:
\begin{widetext}        
\begin{equation}
\begin{split}
Z^+_{\gamma}(m_1,t_1;x_0;m_2,t_2)=\frac{1}{2}\int_0^{t_1} \!\!\rmd\tau_1 \int_{t_1}^T \!  \frac{\rmd\tau_2}{\tau_2-\tau_1} \int_{x_1,x_2>0} & \int_{X_0 =m_1}^{X_{\tau_1}=x_1} \mathcal{D}[X] \Theta[X]  e^{-S_0[X]} \int_{X_{\tau_1} =x_1}^{X_{t_1}=x_0} \mathcal{D}[X] \Theta[X] \dot{X}_{\tau_1}  e^{-S_0[X]}\\
&\times \int_{X_{t_1}=x_0}^{X_{\tau_2}=x_2}\mathcal{D}[X] \Theta[X]  e^{-S_0[X]} \int_{X_{\tau_2}=x_2}^{X_{T}=m_2} \mathcal{D}[X] \Theta[X] \dot{X}_{\tau_2} e^{-S_0[X]}\ .
\label{17}
\end{split}
\end{equation}
In this expression, all path integrals can be expressed in terms of the bare propagator $P_0^+$; we refer to appendix \ref{AppendixDetails} for how to deal with the terms containing $\dot{X}$. We have not written the  cut-off $\tau$   as there are no short-time divergences that need to be regularized (contrary to the terms $Z^+_{\alpha}$ and $Z^+_{\beta}$). The structure of the time integrals, which are products of convolutions, suggests to use Laplace transforms (with respect to the time variables:  $t_1 \rightarrow s_1$, $t_2 \rightarrow s_2$). This, and the identity 
\begin{equation}
\frac{1}{\tau_2-\tau_1} = \int_{y>0} e^{-y (\tau_2-\tau_1)}
\end{equation}
give us a simple form for the double Laplace transform of $Z^+_{\gamma}$, which we will denote with a tilde (for details see appendix \ref{AppendixDetails}), 
\begin{equation}\label{ZgammaLaplace}
\tilde Z^+_{\gamma}(m_1,s_1;x_0;m_2,s_2)=2\int_{x_1,x_2,y>0}  \tilde P_0^+(m_1,x_1;s_1)\,\partial_{x_1} \tilde P_0^+(x_1,x_0;s_1+y)\tilde P_0^+(x_0,x_2;s_2+y)\,\partial_{x_2}\tilde P_0^+(x_2,m_2;s_2)\ .
\end{equation}
\end{widetext}
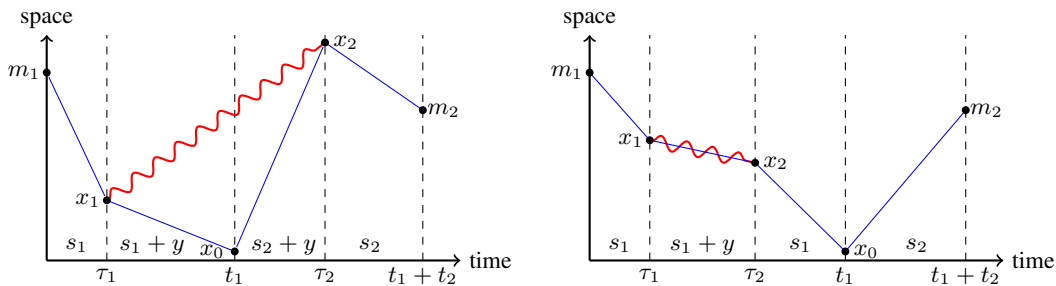
\begin{figure*}[t]
	{{\begin{tikzpicture}
			\draw [->,thick] (0,0) -- (5.5,0);
			\draw [->,thick] (0,0) -- (0,3);
			\draw [dashed] (2.5,0) -- (2.5,3);
			\draw [dashed] (5,0) -- (5,3);
			\draw [dashed] (0.8,0) -- (0.8,3);
			\draw [dashed] (3.7,0) -- (3.7,3);
			\node (m1) at  (0,2.5) {$\hspace{-6mm}m_1$};
			\node (x1) at (0.8,0.8) {$\hspace{-5mm}x_1$};
			\node (x2) at (3.7,2.9) {$\hspace{-5mm}\parbox{0mm}{$~~~~~~x_2$}$};
			\node (x0) at (2.5,0.12) {$\hspace{-6mm}x_0$};
			\node (m2) at  (5,2) {$\hspace{-6mm}\parbox{0mm}{$~~~~~~m_2$}$};
			\draw [snake=snake,red,thick] (x1) -- (x2);
			\draw [blue] (m1) -- (x1);
			\draw [blue] (x1) -- (x0);
			\draw [blue] (x0) -- (x2);
			\draw [blue] (x2) -- (m2);
			\fill (m1) circle (1.5pt);
			\fill (x0) circle (1.5pt);
			\fill (x1) circle (1.5pt);
			\fill (x2) circle (1.5pt);
			\fill (m2) circle (1.5pt);
			\node (s1) at (0.4,0.2) {$s_1$};
			\node (tau1) at (0.8,-0.2) {$\tau_1$};
			\node (s1+y) at (1.4,0.2) {$s_1+y$};
			\node (t1) at (2.5,-0.2) {$t_1$};
			\node (s2+y) at (3.15,0.2) {$s_2+y$};
			\node (t2) at (5,-0.2) {$t_1+t_2$};
			\node (s2) at (4.3,0.2) {$s_2$};
			\node (tau2) at (3.7,-0.2) {$\tau_2$};
			\node (time) at (5.9,0) {time};
			\node (space) at (0,3.2) {space};
			\end{tikzpicture}}}
	~~~~
	{{\begin{tikzpicture}
			\draw [->,thick] (0,0) -- (5.5,0);
			\draw [->,thick] (0,0) -- (0,3);
			\draw [dashed] (2.2,0) -- (2.2,3);
			\draw [dashed] (5,0) -- (5,3);
			\draw [dashed] (0.8,0) -- (0.8,3);
			\draw [dashed] (3.4,0) -- (3.4,3);
			\node (m1) at  (0,2.5) {$\hspace{-6mm}m_1$};
			\node (x1) at (0.8,1.6) {$\hspace{-5mm}x_1$};
			\node (x2) at (2.2,1.3) {$\hspace{-5mm}\parbox{0mm}{$~~~~~~x_2$}$};
			\node (x0) at (3.4,0.12) {$\hspace{-5mm}\parbox{0mm}{$~~~~~~x_0$}$};
			\node (m2) at  (5,2) {$\hspace{-6mm}\parbox{0mm}{$~~~~~~m_2$}$};
			\draw [snake=snake,red,thick] (x1) -- (x2);
			\draw [blue] (m1) -- (x1);
			\draw [blue] (x1) -- (x2);
			\draw [blue] (x2) -- (x0);
			\draw [blue] (x0) -- (m2);
			\fill (m1) circle (1.5pt);
			\fill (x0) circle (1.5pt);
			\fill (x1) circle (1.5pt);
			\fill (x2) circle (1.5pt);
			\fill (m2) circle (1.5pt);
			\node (s1) at (0.4,0.2) {$s_1$};
			\node (tau1) at (0.8,-0.2) {$\tau_1$};
			\node (s1+y) at (1.5,0.2) {$s_1+y$};
			\node (tau2) at (2.2,-0.2) {$\tau_2$};
			\node (s1) at (2.8,0.2) {$s_1$};
			\node (t2) at (5,-0.2) {$t_1+t_2$};
			\node (s2) at (4.35,0.2) {$s_2$};
			\node (tau2) at (3.4,-0.2) {$t_1$};
			\node (time) at (5.9,0) {time};
			\node (space) at (0,3.2) {space};
			\end{tikzpicture}}}
	\caption{Left: Graphical representation of the contribution $Z^+_{\gamma}$ to the path-integral $Z^+\! (m_1,t_1;x_0;m_2,t_2)$ given in Eq.~\eqref{PathIntegral}. The red curve represents the non-local interaction in the action, second line of Eq.~\eqref{ActionExpansion}, while blue lines are bare propagators. We also indicate the Laplace variable which appears in each time slice in Eq.\ \eqref{ZgammaLaplace}. 
		Right: Graphical representation of $Z^+_{\alpha}$}\label{PathIntegralFigure}
	\label{Diagrams}
\end{figure*}
The Laplace-transformed constrained propagator   appearing in this expression is
\begin{align}\label{propagator}
\tilde P_0^+(x_1,x_2;s)&=\int_0^{\infty} \rmd t\, e^{-s t} P^+_0(x_1,x_2,t)\\
&=\frac{e^{-\sqrt{s } \left| x_1-x_2\right| }-e^{-\sqrt{s } (x_1+x_2)}}{2 \sqrt{s }}\!\underset{x_1 \rightarrow 0}{\simeq}\! x_1 e^{-\sqrt{s}x_2}\nn\ .
\end{align}
The Laplace transformation gives another simplification: the space dependence is now exponential, as compared to the Gaussian form of $P_0^+(x_1,x_2,t)$, which renders the space integrations elementary. (Without the  Laplace transform, already the first space integration  gives an error-function, and the remaining integrations are highly non-trivial). Nevertheless, the final result for $Z_{\gamma}^+(m_1,t_1;x_0;m_2,t_2)$ is complicated, and requires to compute the three integrals in Eq.~\eqref{ZgammaLaplace}, and   two inverse Laplace transformations. These steps are  performed in appendix \ref{AppendixZgamma}.

\subsection{Graphical representation}
It is useful to give a diagrammatic representation for the terms of the perturbative expansion. We denote bare propagators \eqref{propagator} with a solid blue lines. The interaction between two points $ (\tau_1,x_1)$ and $  (\tau_2,x_2)$ is represented in red. As can be seen from Eq.~(\ref{ZgammaLaplace}),
it acts as $2\partial_{x_1}$ on the propagator starting at $x_1$, $2\partial_{x_2}$ on the propagator starting at $x_2$; it also translates the Laplace variable of each time slice   between these two points  by $+y$. The space variables $x_1$, $x_2$ and the \textit{interaction} variable $y$ (which has the inverse dimension of  time) have to be integrated  from $0$ to $\infty$.  In case of divergences, the integration has to be cut off with a large-$y$ cutoff (\textit{c.f.} appendix \ref{AppendixLaplace} for the link between the short time cutoff $\tau$ and the large $y$ cutoff).

The contribution of $Z_{\gamma}^+$,   is computed in detail in Appendix \ref{AppendixZgamma}, and represented in Figure \ref{PathIntegralFigure} (left), together with the contribution   $Z^+_{\alpha}$(right).

\section{Analytical Results}
\label{SectionResluts}
We present here some known scaling results about extremal properties of the fBm. We then  show how our perturbative expansion, and the computation of $Z^+(m_1,t_1;x_0;m_2,t_2)$, allows us to obtain analytical results on the distributions beyond these scaling arguments.  Some of our results were already presented in a Letter \cite{DelormeWiese2015}.
\subsection{Scaling results}
\label{s:scaling}
Let us start with the the survival probability $S(T,x)$, and the  persistence exponent $\theta$, defined for any random process $X_t$ with $X_0=x>0$ as
\begin{equation}
\begin{split}
S(T,x)& := \mbox{prob}\left(X_t\geq 0 \mbox{ for all }t \in [0,T]\right)\\
& \underset{T \rightarrow \infty}{\sim} T^{-\theta_x+o(1)}\ .
\end{split}
\end{equation}
For a review of these concepts in the context of theoretical physics, we refer to \cite{BrayMajumdarSchehr2013}. In a large class of processes the exponent $\theta$ is independent of $x$, and characterizes the power-law decay for the probability of  long positive excursions. For fractional Brownian motion with Hurst exponent $H$ it was shown that $\theta_x=\theta = 1- H$ \cite{Molchan1999,Aurzada2011}. To understand the link of $S(T,x)$ with the maximum distribution for fBm, we use  self affinity of the process $X_t$ to write $P_H^T(m)$ as 
\begin{equation}\label{14}
P^T_H(m)=\frac{1}{\sqrt{2} T^H}f_H \!\left(y=\frac{m}{\sqrt{2} T^H}\right)\ .
\end{equation}
Here $f$ is a scaling function depending on $H$. The survival probability is related to the maximum distribution by
\begin{equation}
S(T,x)=\int_0^xP^T(m)\,\rmd m = \int_0^{\frac{x}{ \sqrt{2} T^H}}f_H(y)\,\rmd y\ .
\end{equation}
This   states that due to  translational invariance a realisation of a fBm starting at $x$ and remaining positive is the same as a realisation starting at $0$ and having   a minimum larger than $-x$. Finally, the symmetry $x \rightarrow -x$ (for a fBm starting at $X_0=0$) gives the correspondence between minima and maxima.
These considerations allow  us to predict  the scaling behavior of $P^T_H(m)$ at small $m$ from the large-$T$ behaviour of $S(T,x)$ \cite{Molchan1999}, 
\begin{equation}
f(y) \underset{y \rightarrow 0}{\sim} y^{\alpha}  ~\Leftrightarrow~ S(T) \sim T^{-(\alpha+1)H}
\ ,\end{equation}
and finally
\begin{equation}
P^T_H(m) \underset{m \rightarrow 0}{\sim} m^{\frac{\theta}{H}-1}= m^{ \frac{1}{H}-2}\ .
\label{MaxScaling}
\end{equation}
For the distribution of the time at which the  maximum is achieved we can estimate the behavior close to the origin by assuming that small values of the maximum are reached close to the origin. Starting with
\begin{equation}
P^T_H(m) \rmd m = P^T_H(t) \rmd t\ ,
\end{equation}
and using  scaling, $m \sim t^{H}$, we obtain
\begin{equation}
P^T_H(t) \sim P^T_H(m) \frac{\rmd m}{\rmd t} \sim \left(t^{H}\right)^{ \frac{1}{H}-2} t^{H-1} \sim t^{-H}
\label{scalingTmax}
\ .\end{equation}
This should be valid when $t \rightarrow 0$ (or $m \rightarrow 0$). By time reversal symmetry $t \rightarrow T-t$, we  also have
\begin{equation}
P^T_H(t) \underset{t\rightarrow T}{\sim} (T-t)^{-H}\ .
\label{28}
\end{equation}

\subsection{The complete result for $Z^+(m_1,t_1;x_0;m_2,t_2)$}
\label{s:P+}

We present here the final result for $Z^+$, defined in Eq.~\eqref{PathIntegral}, at order $\varepsilon$. This path integral was first expanded, \textit{c.f.}~Eq.~\eqref{PathIntegralExpansion}, by treating the non-local term in the action (\ref{ActionExpansion}) perturbatively. The first term $Z_0^+$ of this expansion is given in Eq.~\eqref{Zorder0Bis}, while the second term $Z_1^+$ was split into three contributions $Z_{\alpha}^+$, $Z_{\beta}^+$ and $Z_{\gamma}^+$, see Eq.~\eqref{Z1split}. The first two terms can be obtained explicitly from \eqref{PropagatorOrdre1}, while the third one is computed in appendix \ref{AppendixZgamma}, the result being split between \eqref{ResultZA}, \eqref{ResultZB} and \eqref{ResultZC}.

In order to display a compact form, we choose $T\equiv t_1+t_2 = 1$ (which is equivalent to rescaling $m_1$ and $m_2$ by $T^{-H}$ and $t_1$ and $t_2$ by $T^{-1}$) and introduce new rescaled (dimensionless) variables,
\bea
y_1 &=& \frac{m_1}{\sqrt{2} t_1^H}\ , ~~~~~
y_2 = \frac{m_2}{\sqrt{2} t_2^H} \\
t_1 &=& \vartheta\ , ~~~~~~~~~~t_2 = 1- \vartheta\ . 
\eea In these new variables, the final result is 
\begin{widetext}
\noindent 
\begin{align}\label{mainresult}
Z^+\!(m_1,&t_1;x_0;m_2,t_2)\underset{x_0 \rightarrow 0}{\simeq}x_0^{  2-4 \varepsilon} \frac{y_1 y_2 \exp\!\left(-\half y_1^2-\half y_2^2\right)}{2 \pi \left[\vartheta(1-\vartheta)\right]^{2H}}\times\nn\\
\Bigg\{ 1+\varepsilon &\Bigg[
{\cal I}(y_1)\left(1+ \sqrt{\frac{1-\vartheta}{\vartheta}} \frac{y_2}{y_1}\right)
+{\cal I}(y_2)\left(1+ \sqrt{\frac{\vartheta}{1-\vartheta}} \frac{y_1}{y_2}\right)
+\frac{\left(1-y_2^2\right){\cal I}\big(\sqrt{1-\vartheta} y_1\big)} {\sqrt{\vartheta(1-\vartheta)} y_1 y_2}
+\frac{\left(1-y_1^2\right){\cal I}\big(\sqrt{\vartheta} y_2\big)} {\sqrt{   \vartheta (1-\vartheta)} y_1 y_2}\\
&-\frac{{\cal I}\!\left(\sqrt{1- \vartheta } y_1+\sqrt{ \vartheta} y_2\right)}{\sqrt{\vartheta (1- \vartheta ) } y_1 y_2}
+2\frac{ (1-\vartheta ) y_1^2+ \vartheta  y_2^2-1} {\sqrt{\vartheta(1-\vartheta)} y_1 y_2}+(y_1^2-2) \big(\ln(2y_1^2)+\gamma_{\rm E}\big)+(y_2^2-2) \big(\ln(2y_2^2)+\gamma_{\rm E}\big)\nn\\
&-4-2 \gamma_{\rm E}\Bigg]\Bigg\}+\mathcal{O}(\varepsilon^2)\nn\ .
\end{align}
\end{widetext}
The special function $\mathcal{I}$ appearing in this expression is
\begin{eqnarray}\label{defI_maintext}
\mathcal{I}(z) &=& \frac{z^4}{6}  \,_2F_2\! \left(1,1; \frac{5}{2},3; \frac{z^2}{2} \right) + \pi (1-z^2)\, \mathrm{erfi}\!\left( \frac{z}{\sqrt{2}} \right) \nn\\&&- 3z^2 + \sqrt{2 \pi} e^{\frac{z^2}{2}}z +2\ .
\end{eqnarray}

\subsection{The third Arcsine Law: Distribution of the time when the maximum is reached}
\label{s:3rd-arcsine-law}
To simplify   the result \eqref{mainresult}, we can  extract from it the distribution of a single observable. We start with the probability distribution $P_H(t)$ of $t_{\rm max}$, the time when the fBm achieves its maximum. For Brownian motion ($H=1/2$), this distribution is well known as the third arcsine law, because the cumulative distribution involves the arcsin function \textit{c.f.} Eq.~\eqref{ArcsinDistrib}, 
\begin{equation}
P^T_{\half}(t) = \frac{1}{\sqrt{\pi t(T-t)}} \ , \mbox{~for~} t\in [0,T] \ .
\label{ArcSinLaw1}
\end{equation}
Until now, only scaling properties were known for this distribution in the general case \cite{MajumdarRossoZoia2010b}, as recalled in Eq.~\eqref{scalingTmax}.

The path integral \eqref{PathIntegral}, in the limit of $x_0 \rightarrow 0$, selects paths which go through $x_0 \approx 0^+$ at time $t_1$ while staying positive. This means that we sum  over paths reaching their minimum (in the interval $[0,t_1+t_2]$, and which  is almost surely unique) at $t_1$, starting at $m_1$ and ending at $m_2$. This is equivalent to summing over paths starting at $0$, reaching their minimum with value $-m_1$ at time $t_1$, and ending in $m_2-m_1$. Integrating over $m_1$ and $m_2$ finally gives the sum over all paths reaching their minimum in $t_1$, independent of the value of this minimum, and the end point. Up to a normalization, this is the probability distribution of $t_{\rm min}$. By symmetry, this is   the same as the distribution of $t_{\rm max}$.  Formally, it reads
\begin{equation}
P^T_H(t) = \lim\limits_{x_0\rightarrow 0}\frac{1}{Z^N} \int_{m_1,m_2>0} Z^{+}(m_1,t;x_0;m_2,T-t)\ .
\label{TimeDistribDef}
\end{equation}
The normalization $Z^N$ depends  on $x_0$ and $T$. It ensures that $P^T_H(t)$ is normalized; it can be expressed in terms of $Z^+$ as 
\begin{equation}
Z^N(x_0,T)=\int_0^T \rmd t \int_{m_1,m_2>0}\,Z^{+}(m_1,t;x_0;m_2,T-t)\ .
\end{equation}
At order $0$, starting from Eq.\ \eqref{Zorder0} and  integrating over $m_1$ and $m_2$ allows us to recover Eq.~\eqref{ArcSinLaw1} with  normalisation $Z_N = x_0^2$.
\begin{figure}[t]
      \Fig{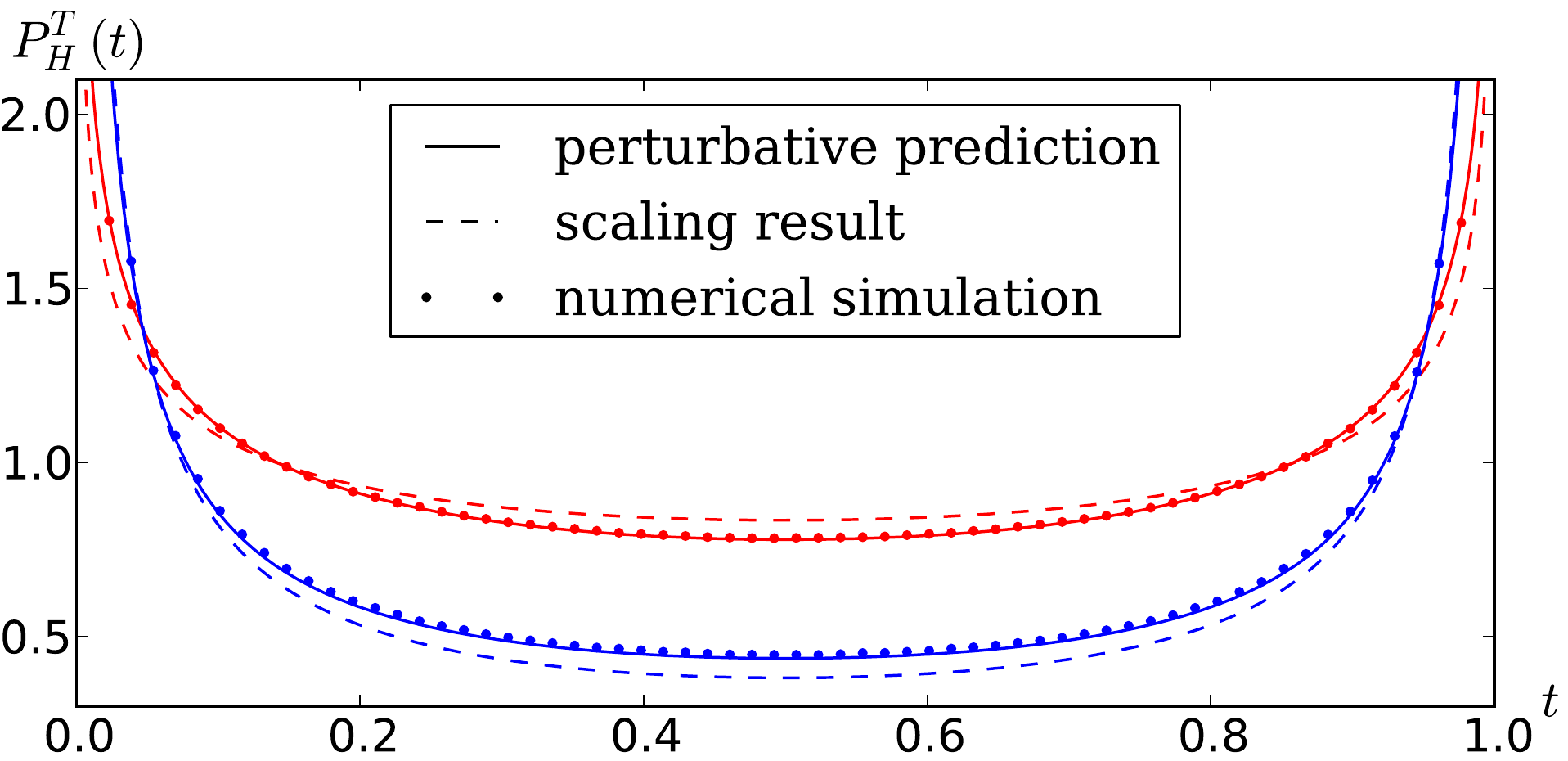}
        \caption{Distribution of $t_{\rm max}$ for $T=1$ and $H=0.25$ (red) or $H=0.75$ (blue) given in Eq.~\eqref{MaxPosDistrib} (plain lines) compared to the scaling ansatz, \textit{i.e.}\ $\mathcal{F}= \mbox{cst.}$ (dashed lines) and  numerical simulations (dots). For $H<0.5$ realisations with $t_{\rm max}\approx T/2$  are less probable (by about $10\%$) than expected from  scaling. For $H>0.5$ the correction has the opposite sign.}\label{ArcSinLawfBm}
\end{figure}

For the order-$\varepsilon$ correction, the integrations over $m_1$ and $m_2$ are  lengthy. This is done in appendix \ref{AppendixArcsine_Law}. It allows us to write an $\varepsilon$-expansion for the distribution of $t_{\rm max}$ in the form \begin{equation}
P^T_H(t)=P^T_{\half}(t)+\varepsilon\, \delta P^T(t) + \mathcal{O}(\varepsilon^2)
\ .\end{equation}
The result  \eqref{ArcineLawExpension2} reads
\begin{equation}
\begin{split}
\delta P^T\!(t)= \frac{1}{\pi \sqrt{t_1 t_2}} \Bigg\{& \sqrt{\frac{t_1}{t_2}}  \left[\pi -2 \arctan\!\left(\sqrt{\frac{t_1}{t_2}}\right)\right]\\&+\sqrt{\frac{t_2}{t_1}} \left[\pi -2 \arctan\!\left(\sqrt{\frac{t_2}{t_1}}\right)\right]\\
 &\qquad-\log(t_1 t_2)+ {\rm cst} \Bigg\}\ ,
\end{split}
\end{equation}
where $t_1 = t$ and $t_2 =T-t$. It takes a simple form if we exponentiate this order-$\varepsilon$ correction,
\begin{equation}
P^T_H(t)= \frac{1}{\pi[t (T-t)]^H} \, \exp\! \left(\varepsilon\, \mathcal{F} \!\left(\frac{t}{T-t}\right)\right) +\mathcal{O}\!\left(\varepsilon^2\right).
\label{MaxPosDistrib}
\end{equation}
The term $\log(t_1 t_2)=\log\!\big(t(T-t)\big)$   in $\delta P^T\!(t)$ gives the expected change, from Eq.~\eqref{scalingTmax} and \eqref{28}, in the scaling form of the Arcsine law, $\sqrt{t(T-t)} \rightarrow [t(T-t)]^H$. The regular part induces a non-trivial change in the shape,
\begin{equation}
\begin{split}
\mathcal{F} (u)= &   \sqrt{u} \left[\pi-2 \arctan\left(\sqrt{u}\right)\right]\\
&+ \frac{1}{\sqrt{u}}\left[\pi-2 \arctan\left(  \frac{1}{\sqrt{u}}\right)\right]+ \mbox{cst}\ .
\label{Fprediction}
\end{split}
\end{equation}
The time reversal symmetry $t\to T-t$  (corresponding to $u \rightarrow u^{-1}$) is explicit and the constant ensures normalization. 
The contribution of $\mathcal{F}(u)$ to the probability that the maximum is attained at time $t$ is quite noticeable, as shown in Fig.~\ref{ArcSinLawfBm}.

\begin{figure}[t]
        \Fig{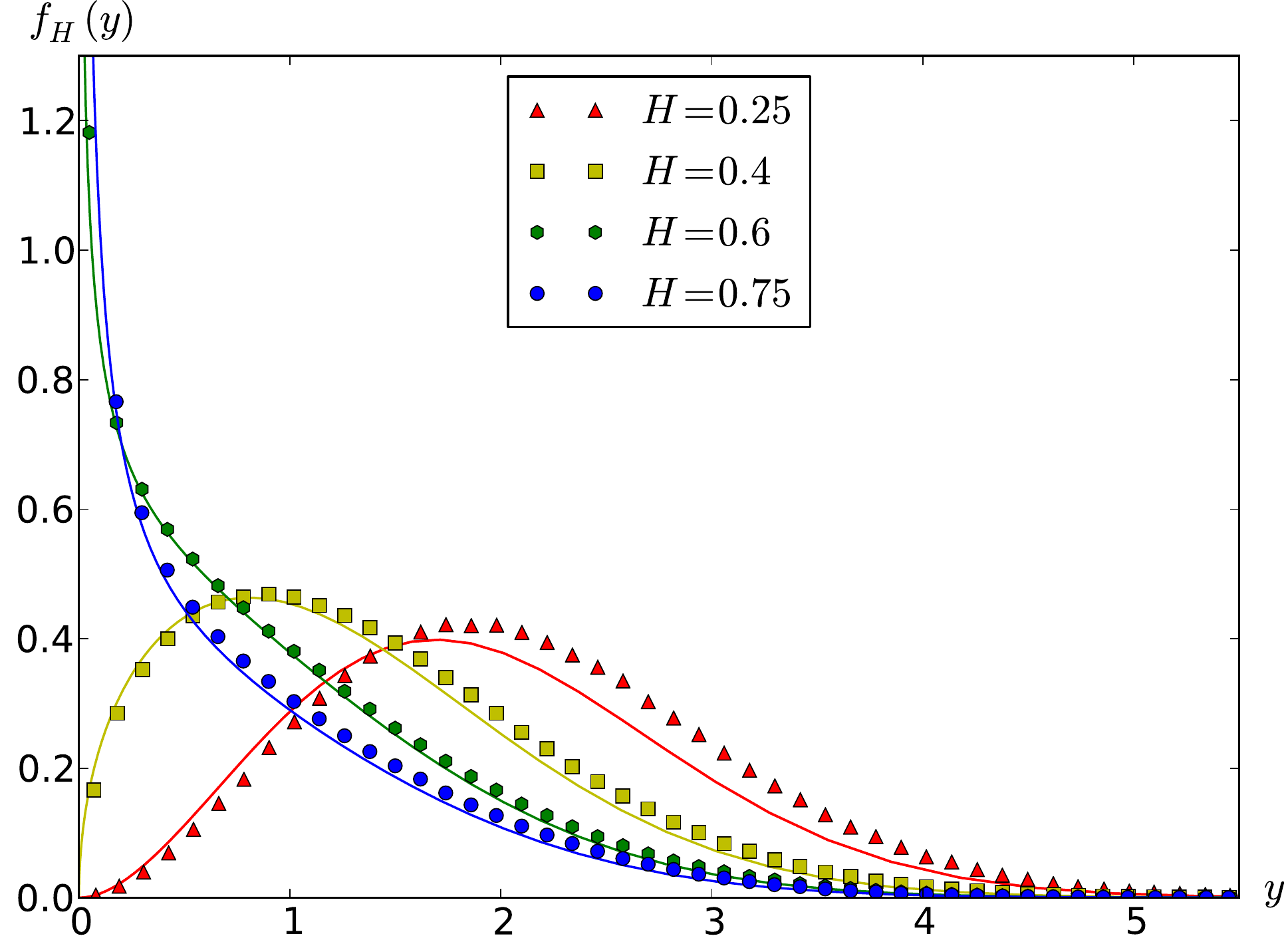}
        \caption{Scaling function $f_H(y)$ for the distribution of the maximum, as defined in Eq.~\eqref{14}, for different values of $H$: $H=0.25$ in red, $H=0.4$ in yellow, $H=0.6$ in green, and $H=0.75$ in blue. The plain lines represent the analytic prediction from our perturbative theory (at first order in $\varepsilon$) given in Eq.~\eqref{MaxDistribPrediction};  the symbols are results from numerical simulations, \textit{c.f.} section \ref{s:Numerics}.}\label{MaxDistribPlot}
\end{figure}

\begin{figure*}[t]
\Fig{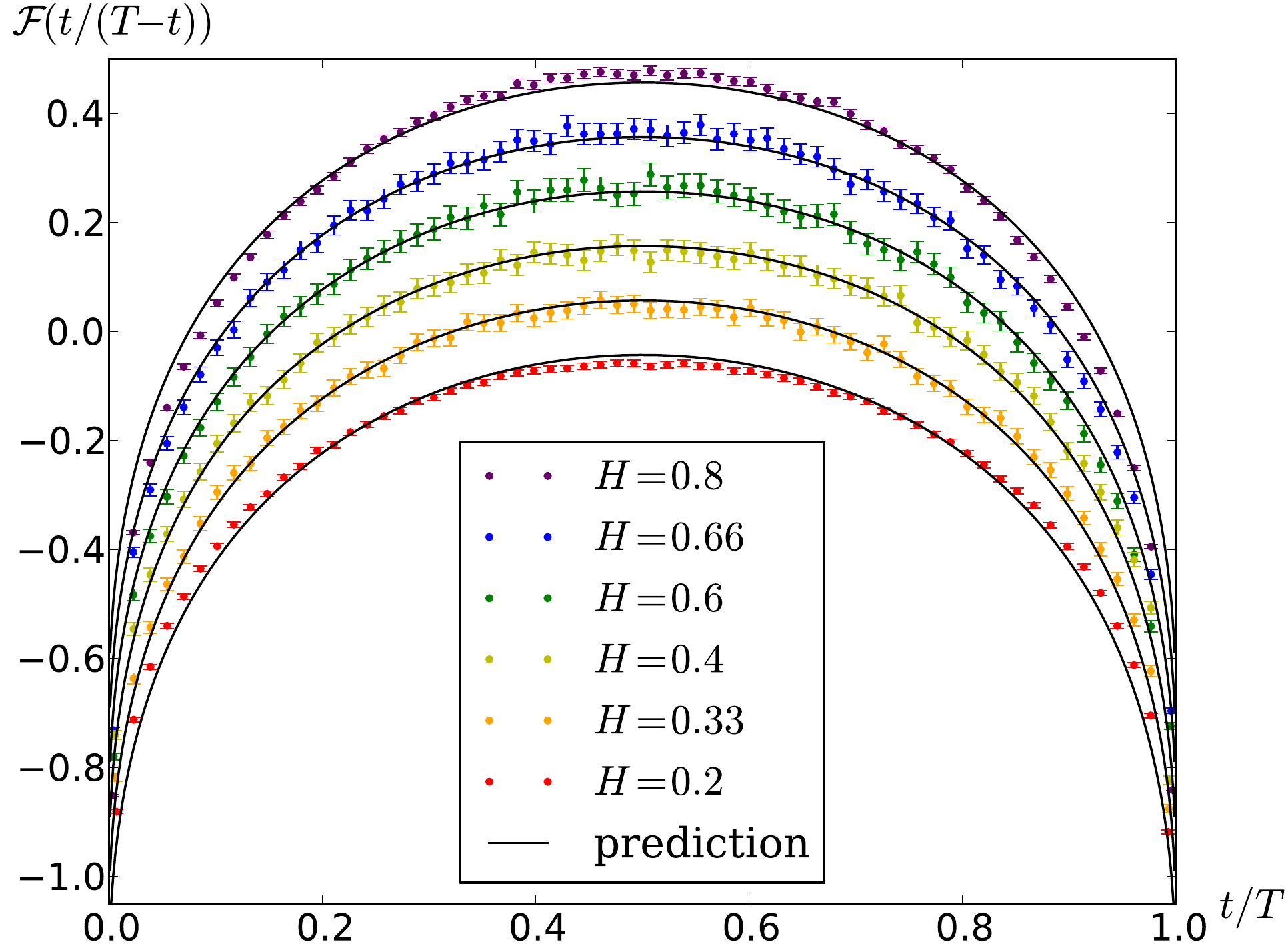}
\Fig{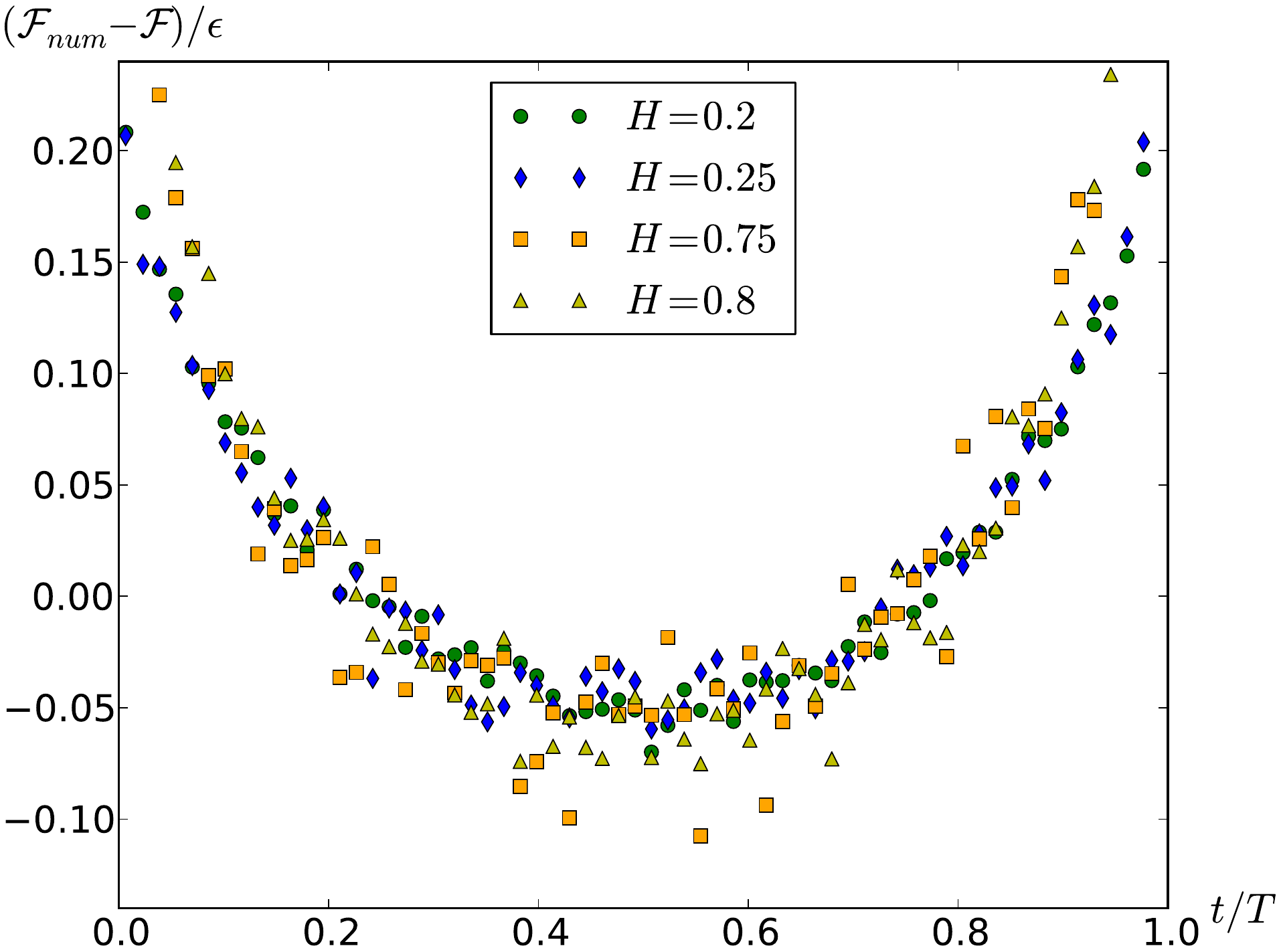}
\caption{Left: Numerical estimation of $\mathcal{F}$ for different values of $H$ on a discrete system of size $N=2^{12}$, using $10^8$ realizations. Plain curves  represent  the  theoretical prediction \eqref{Fprediction},   vertically translated for  better visualization. Error bars are $2\sigma$ estimates. Note that for $H=0.6$, $H=0.66$  and $H=0.8$ the expansion parameter $\epsilon$ is positive, while for $H=0.4 $, $H=0.33$  and $H=0.2$ it is negative.
\\
Right: Deviation  for large $|\varepsilon|$ between the theoretical prediction \eqref{Fprediction} and the numerical estimations \eqref{Fnum}, rescaled by $\varepsilon$, \textit{c.f.} Eq.~\eqref{ArcsinLawOrdre2}. These curves collapse for different  values of $H$, allowing for an estimate of the $\mathcal{O}(\varepsilon^2)$ correction  to $P_H^T(t)$, as written in Eq.~\eqref{ArcsinLaw2}.}
\label{ArcSinLawfBmnum}
\end{figure*}

\subsection{The  distribution of the maximum}
\label{s:dist-max}

We now present results for the distribution of the maximum $P^T_H(m)$. For standard Brownian motion
\begin{equation}
P^{T}_{\half}(m) = \frac{e^{-\frac{m^2}{4T}}}{\sqrt{\pi T}}  \ , \qquad  m>0 \ . \label{MaxDistribSBM}
\end{equation}
On the other hand, the scaling results presented in \ref{s:scaling} predict that for any $H$, $P^T_H(m)$ behaves at small scale as  $m^{1/H-2}$, as given in Eq.~\eqref{MaxScaling}.

Using our path integral, we can go further. Similarly to the distribution of $t_{\rm max}$, the  distribution of the maximum $m$ itself can be extracted from $Z^+$, defined in Eq.~\eqref{PathIntegral}, 
\begin{equation}
P^T_{H}(m) = \lim\limits_{x_0\rightarrow 0}\frac{1}{Z^N} \int_{0}^{T} {\rm d}t\int_{m_2>0} Z^{+}(m,t;x_0;m_2,T-t)\ .
\end{equation}
The details of these computations (integrations over $t$ and $m_2$) are given in appendix \ref{AppendixMaxDistrib}. Its $\varepsilon$-expansion, recast in exponential form, leads to the scaling form of Eq.~(\ref{14}), with
\begin{eqnarray}\label{MaxDistribPrediction}
f_{H}(y)= \sqrt{\frac{2}{\pi}}  e^{-\frac{y^2}{2}}  \,e^{\varepsilon \left[\mathcal{G}(y) + {\rm  cst}\right]} + O(\varepsilon^2)\ .
\end{eqnarray}
The constant term ensures  normalization.   Figure \ref{MaxDistribPlot} shows the form of this scaling function for different values of $H$, as well as a {\em first comparison} to numerical simulations. The function $\mathcal{G}$ involves a combination of special functions denoted  $\mathcal{I}$ in Eq.\ \eqref{defI_maintext} , and logarithmic terms, 
\begin{equation}
\mathcal{G}(y) = \mathcal{I}(y) + (y^2-2)[\gamma_{\rm E}+\log(2y^2)]
\ .\end{equation}
It has a different   asymptotics for small and large $y$,
\begin{equation}\label{G_asymptotics}
        \mathcal{G}(y) \sim
        \begin{cases} 
                \hfill -2 \ln(y)  \hfill & \text{ for } y \rightarrow \infty \\
                \hfill -4\ln(y) \hfill & \text{ for } y \rightarrow 0\\
        \end{cases}.
\end{equation}
The second line implies that
$P^T_H(m)  \sim m^{-4\varepsilon}$ when $m\to 0$, 
which is consistent (at order $\varepsilon$) with the scaling result  \eqref{MaxScaling},  $\frac{1}{H}-2= -4 \varepsilon + \mathcal{O}(\varepsilon^2)$.
Formulas \eqref{MaxDistribPrediction}-\eqref{G_asymptotics} also predict the  distribution at large $m$.  It is   known that the leading behavior of $P^T_H(m)$ is Gaussian, which can be formalized as
\begin{equation} \label{GaussianAsyptotic}
\lim\limits_{y \rightarrow \infty}  \frac{\ln\!\big(f_H(y)\big)}{y^2}=-\frac{1}{2} \ .
\end{equation}
This is a direct consequence of an important theorem in the theory of Gaussian processes,   the {\em Borrel inequality}. It states that for any Gaussian process $X_t$ the cumulative distribution of its maximum value over the interval $[0,T]$, $m=\sup_{t\in[0,T]}X_t$, verifies
\begin{equation}
{\rm Prob}(m>u) \leq \exp\!\left({-\frac{(u-\langle m\rangle)^2}{2 \sigma^2}} \right)
\ ,\end{equation}
where $\langle m \rangle$ and $\sigma^2 =\sup_{t\in[0,T]} \langle X_t^2 \rangle$ are assumed to be finite. Specifying this to   fBm   with $T=1$ allows us to derive Eq.~\eqref{GaussianAsyptotic}. A proof of this theorem and a derivation of its implications for fBm can be found in Ref.~\cite{NourdinBook}.

Our result \eqref{MaxDistribPrediction} goes further, and gives the subleading term in the large-$m$ (and equivalently large-$y$) regime,   a power law with exponent $-2\varepsilon + \mathcal{O}(\varepsilon^2)$. It can be written   as
\begin{equation}
\lim\limits_{y \rightarrow \infty} \frac{\log\!\left(f_H(y) \exp(\frac{y^2}{2}{ )} \right)}{\log(y)} = -2 \varepsilon + \mathcal{O}(\varepsilon^2)  \ .
\label{asymp}
\end{equation} 
Comparison of our full prediction (\textit{i.e.}\ not only the asymptotics) with numerical simulations of the fBm are presented in the next section~\ref{s:Numerics}.

\subsection{Survival probability}

The survival probability $S(x,T)$ is defined as the probability for a process $X_t$ to stay positive up to time $t$, while starting at $X_0=x$,
\begin{equation}
S(x,t):=\text{prob}\left(X_t\!>\!0,\,\forall t\in[0,T]\,|\, X_0=x\right)\ .
\end{equation}
As before, scaling properties of the fBm allow us to write this as a function of $y = \frac{x}{\sqrt{2}T^H}$. As mentioned, the survival probability is   the cumulative distribution of the maximum value, and reads
\begin{equation}
S(y)= \int_0^y \rmd u f_H(u) 
\end{equation}
with $f_H$ defined in Eq.~\eqref{14}. Similarly to the other distributions, we can compute its $\varepsilon$-expension and recast it into an exponential form to get
\begin{equation}\label{survivalscaling}
S(y) = {\rm erf}\!\left(\frac{y}{\sqrt{2}}\right)\exp\!\left(\varepsilon \frac{\mathcal{M}(y)}{{\rm erf}\!\left(\frac{y}{\sqrt{2}}\right)}\right)+\mathcal{O}(\varepsilon^2)\ .
\end{equation}

The function $\mathcal{M}(y)$ is 
\begin{eqnarray}\label{Mexpression}
\mathcal{M}(y)&=& \sqrt{\frac{8}{\pi}}y\,_2F_2\!\left(\frac{1}{2},\frac{1}{2};\frac{3}{2},\frac{3}{2};-\frac{y^2}{2}\right)  \\ \nn
&& -\sqrt{\frac{2}{\pi}}e^{-\frac{y^2}{2}}y^3\,_2F_2\!\left(1,1;\frac{3}{2},2;\frac{y^2}{2}\right)\\ \nn
&&+\sqrt{2\pi}e^{-\frac{y^2}{2}}y\,\textrm{erfi}\!\left(\frac{y}{\sqrt{2}}\right) \\
&& -\left[\textrm{erf}\!\left(\frac{y}{\sqrt{2}}\right)+\sqrt{\frac{2}{\pi}}e^{-\frac{y^2}{2}}y\right]\left[\ln\!\left(2y^2\right)+\gamma_{\rm E}\right] \nn
\end{eqnarray}  
Some details of its derivation are given in appendix \ref{AppendixSurvival}.
\begin{figure}
\fig{8.7cm}{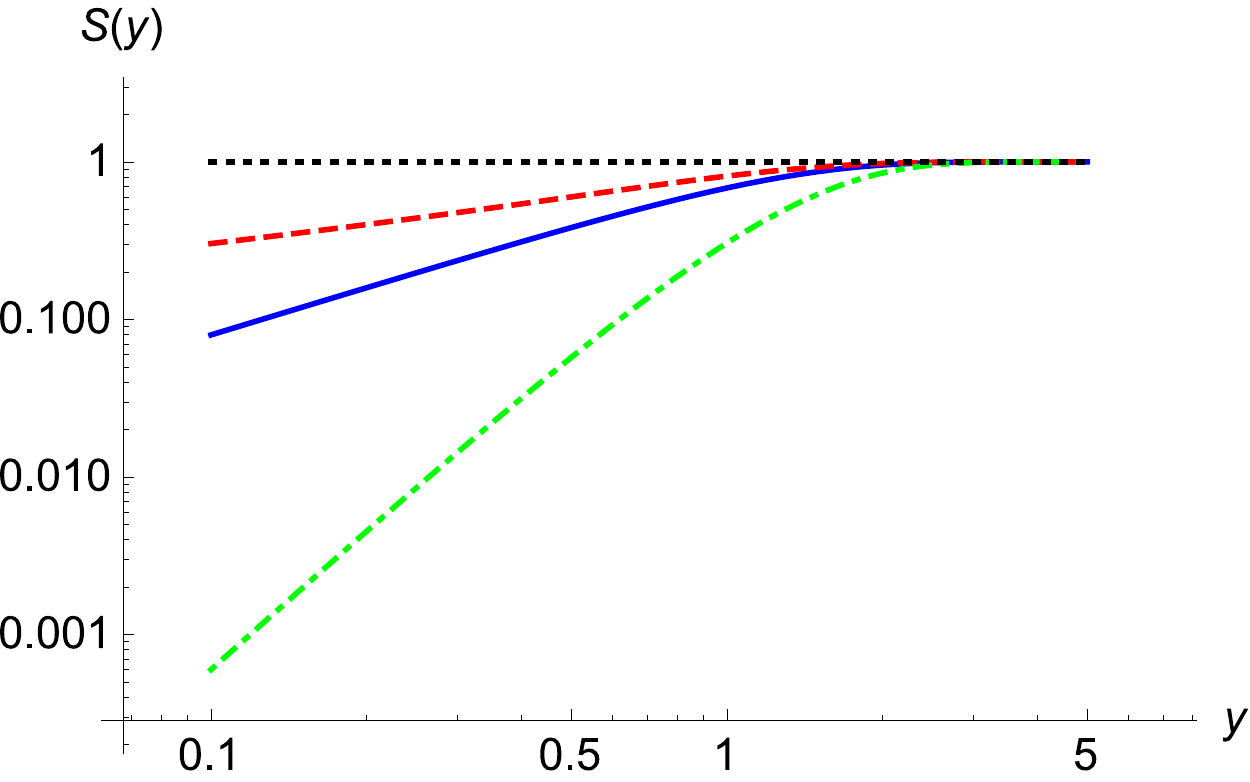}
\caption{The survival probability $S(y)$ for $H=1/2$ (blue solid line), $H=0.75$ (red, dashed), $H=0.25$ (green, dot-dashed), and asymptotics $S(y)=1$ (black, dotted), in a log-log plot.}
\label{f:Sofy}
\end{figure}

\subsection{The joint distribution for $t_{\rm max}$ and $m$}

The result \eqref{mainresult} was obtained by considering paths starting at $X_0=m_1>0$   with an absorbing boundary at $x=0$ constraining the process to stay positive, as can be seen from the path-integral definition \eqref{PathIntegral}. Using translational invariance, and the symmetry $x \leftrightarrow -x$ of the fBm, we can reinterpret this as the sum over paths starting at $X_0=0$, reaching their maximum (over the interval $[0,T=t_1+t_2]$) of value $m_1$ at time $t_1$, and ending in $X_T=m_1-m_2<m_1$.

The integral over $m_2$ is then, in the limit $x_0 \to 0$ and up to a normalisation factor $Z^N$, the joint  probability density   for a fBm to have a maximum value $m=m_1$ at a time $t=t_ {\rm max}=t_1$ over the interval $[0,T]$; this we can write as

\begin{equation}
P^T_H(m,t)=\lim\limits_{x_0 \to 0}\frac{1}{Z^N}\! \int_0^{\infty}\!\!\!\rmd m_2 Z^+(m,t;x_0;m_2,T-t)\ .
\end{equation}

We recall the result for Brownian motion that we recover    for $\varepsilon=0$, 
\begin{equation}\label{jointBrowniany}
P^T_{\frac12}(m,t)=\frac{m e^{-\frac{m^2}{4 t}}}{2 \pi  t^{3/2} \sqrt{T-t}}\ .
\end{equation}
To simplify the ensuing discussion, we now consider the conditional probability 
\begin{equation}
P^T_H(m|t):=\frac{P^T_H(m,t)}{\int_{m>0} P^T_H(t,m)}=\frac{P_H^T(m,t)}{P_H^T(t)}\ .
\end{equation}
Interestingly, in the case of the Brownian motion, we can make a change of variables $m \to y:= m/\sqrt{2 t}$ such that this conditional distribution function becomes independent of $t$ (or equivalently, independent of $\vartheta = t/T$)
\begin{equation}
P^T_{\frac12}(m|t)=m \frac{e^{-\frac{m^2}{4 t}}}{2t}=\frac{1}{\sqrt{2t}} y e^{-\frac{y^2}{2}}=\frac{\rmd y}{\rmd m}P_{\frac12}(y|\vartheta)\,
\end{equation}
with
\begin{equation}\label{jointBrownian}
P_{\frac12}(y| \vartheta )= {y e^{-\frac{y^2}{2}}} \ .
\end{equation}
For $H\neq \frac12$, this independence is broken, and the result at  order $\varepsilon$ can be written as
\begin{equation}\label{70}
P_H(y| \vartheta )  = {y e^{-\frac{y^2}{2}}}  \rme^{ \varepsilon {\cal G}(y|\vartheta)} +\mathcal{O}(\varepsilon^2) \ ,
\end{equation}
where now $y=\frac{m}{\sqrt{2}t^H}$ (to keep $y$ a dimensionless variable). It is important to note that the variable $y$ here is not the same as in Eq.~\eqref{MaxDistribPrediction}, as the maximum $m$ is rescaled by $t$ (the time at which the maximum is reached), and not  by $T$ (the total time of the process).

The non-trivial correction ${\cal G}(y| \vartheta ) $ is obtained from the   result \eqref{mainresult} as
\begin{equation}\label{71}
  {\cal G}(y_1|\vartheta)=\int_{y_2>0} y_2 e^{-\frac{y_2^2}{2}} \left[...\right]  \ ,
\end{equation}
where $[...]$ are the   terms  in  the  square brackets of Eq.~\eqref{mainresult}.

While we can integrate Eq.~(\ref{mainresult}) over $y_1$ and $y_2$ to obtain the probability that the maximum is attained at time $t$,  we were {\em in general} not able to   analytically integrate it solely over $y_2$, due to the presence of the term ${{\cal I} (\sqrt{1- \vartheta } y_1+\sqrt{ \vartheta} y_2)}$. An exception are the  two limiting cases
$\vartheta=0$ and $\vartheta=1$, for which 
\begin{figure*}
\fig{6.0cm}{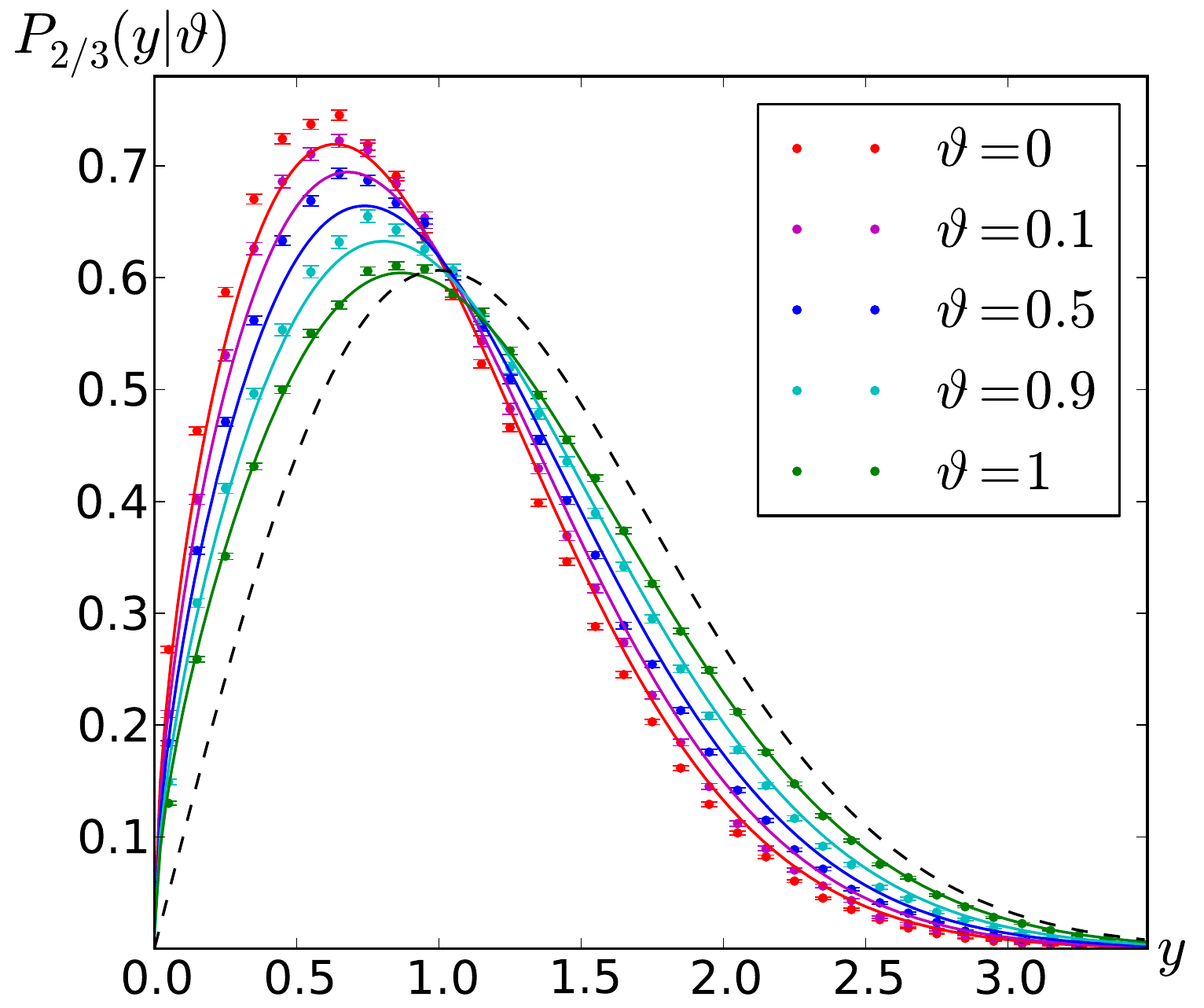}\fig{6cm}{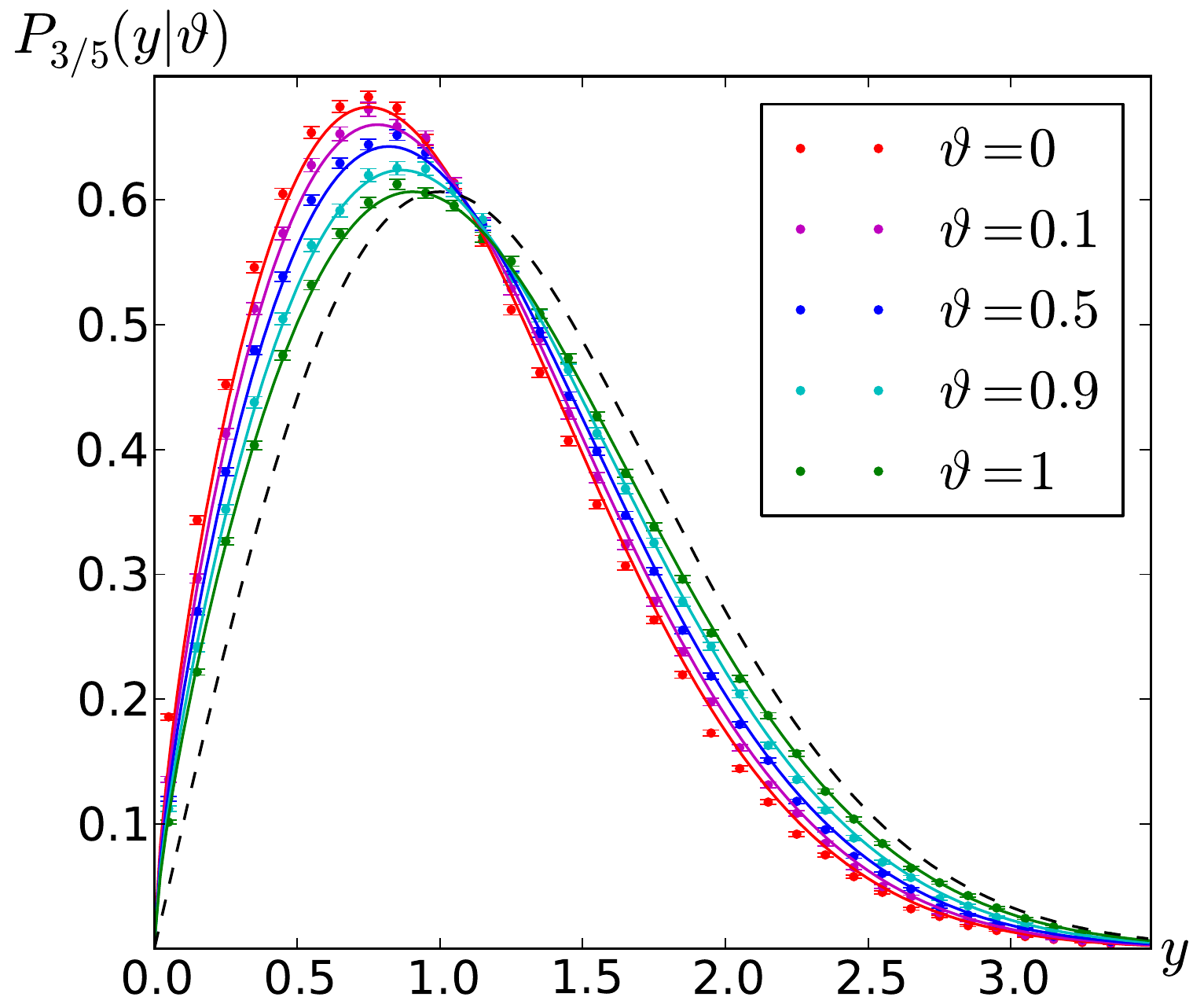}\fig{6cm}{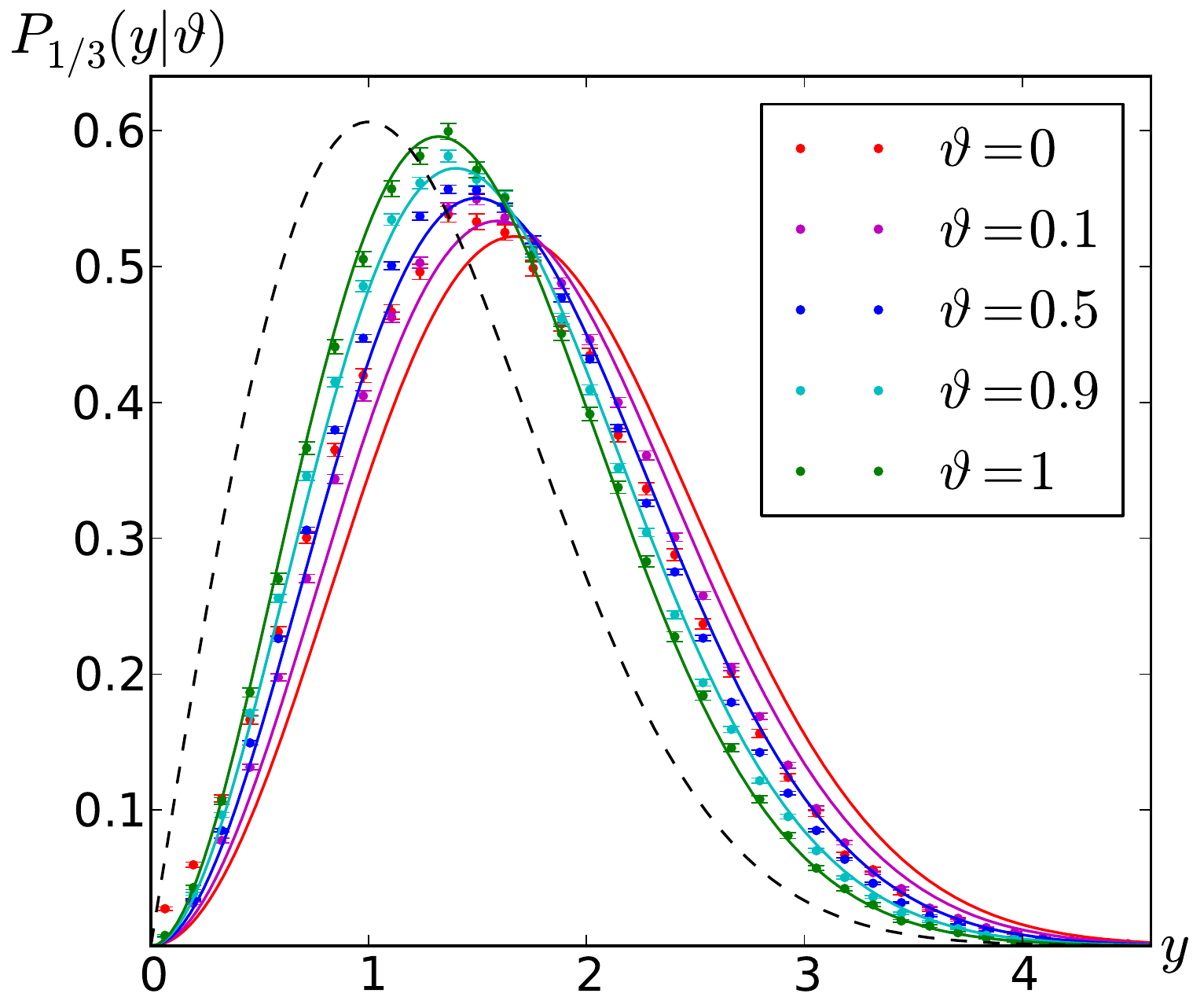}
\caption{Left : The conditional probability $ P_H(y|\vartheta)$  for $ H=\frac23$ and various values of $\vartheta$. Middle and right: ibid with  $ H=\frac35$ and $ H=\frac13$. The plain curves are the analytical prediction \eqref{70}, where the scaling functions are given analytically for the two extremal cases, $\vartheta=0$ and $\vartheta=1$ \textit{c.f.} Eqs.~(\ref{cond-G-0})-(\ref{cond-G-1}); for $0<\vartheta<1$ the curves are obtained via numerical integration.  The predicted spread of the curves (which collapse for $H=\frac12$ to  Eq.~\eqref{jointBrowniany}, plotted  in black dots) is well reproduced in the numerics, both for $\varepsilon>0$ and $\varepsilon <0$. For  $\vartheta\to 1$ the agreement with numerics is remarkable, while for $\vartheta$ close to zero, we see significant deviations. These deviations may be due to both discretisation effects and $\varepsilon^2$ corrections (they have the same sign for both $\varepsilon>0$ and $\varepsilon <0$).}
\label{f:cond-proba}
\end{figure*}
\begin{align}\label{cond-G-0}
&{\cal G}(y|0) =  (y^2-2)[\gamma_{\rm E}+\log(2y^2)]+  { \frac{(3-y^2)[\mathcal{I}(y)-2]}{1-y^2}} \nn\\
&\hphantom{{\cal G}(y|0) =} +\frac{2 \sqrt{2 \pi }}{y} \left [  1-y^2 - \frac{  e^{\frac{y^2}{2}}
   \text{erfc}\left(\frac{y}{\sqrt{2}}\right)}{1-y^2 }  \right]  \ , \\
&{\cal G}(y|1) =  (y^2-2)[\gamma_{\rm E}+\log(2y^2)]+\mathcal{I}(y)  -2   \label{cond-G-1}
\ .\end{align}
Note that $P_H(y| 1) $ is also the  conditional  probability that a FBM path, starting at $x_0\ll 1$, and having survived up to time $T$ has final position $m= \sqrt 2 y T^H $. This reproduces Eqs.\ (9)-(10) of Ref.~\cite{WieseMajumdarRosso2010}.

The asymptotic behaviors for small $y$  are 
\be
P_{ H}(y| \vartheta ) \sim y^{\frac{1}{H}-1}\simeq y^{1-4\epsilon} +{\cal O}(\epsilon^2)
\ee
For large $y$, the situation is more complicated. For the two limiting cases the behavior is consistent with 
\bea
P_H(y| 0 ) &\sim&  y^{1+2 \epsilon} \rme^{-y^2/2 - \sqrt{8\pi} y \epsilon} +{\cal O}(\epsilon^2)\ , \\
P_H(y| 1 ) &\sim& y^{1-2 \epsilon} \rme^{-y^2/2}+{\cal O}(\epsilon^2)\ .
\eea
It would be interesting to understand this behaviour  from scaling arguments.

The conditional probability (\ref{70}) is plotted on figure \ref{f:cond-proba} for various value of $H$, supplemented by  results obtained via numerical integration of Eq.~\eqref{71} for    $\vartheta=0.1$, $0.5$, and  $0.9$. It varies smoothly as a function of $\vartheta$.

\section{Numerical Results} \label{s:Numerics}
To validate the perturbative approach used in this article, we tested our analytical results with direct numerical simulations of fBm paths. The discretized fBm paths are generated using the Davis and Harte procedure as described in  \cite{Dieker} (and references therein). The idea is to take advantage of the stationarity of the increments and use fast-Fourier transformations to compute efficiently the square root of its covariance function. This method is exact, \textit{i.e.}\ the samples generated have exactly the covariance function given in Eq.~\eqref{covariance}, and is  adapted to situations where the length of the path to generate is fixed.
Other simulation techniques exist,  reviewed in  Ref.\ \cite{Coeurjolly2000}.

\subsection{The third Arcsine Law}
\begin{figure*}[t]
        \fig{5.9cm}{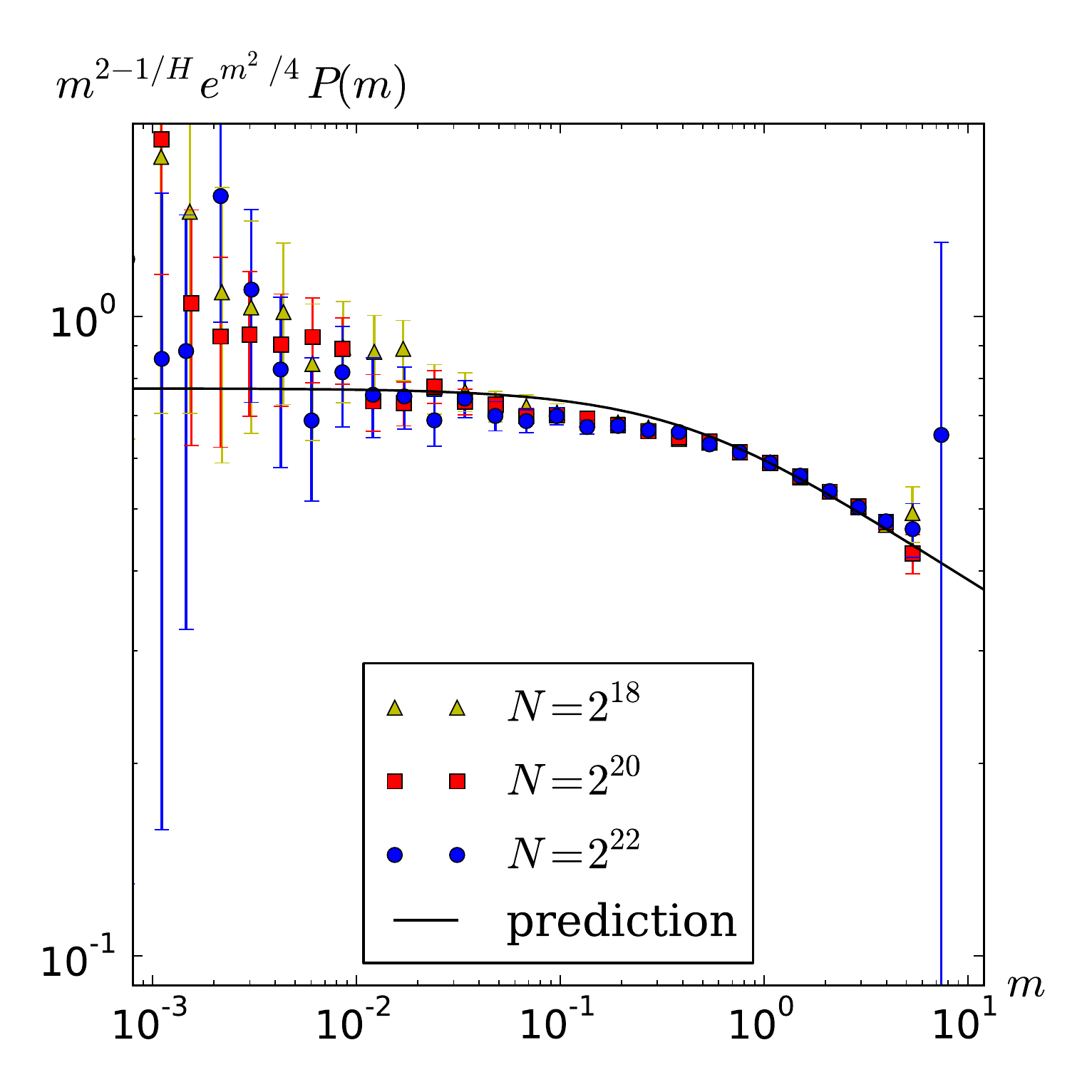}
        \fig{5.9cm}{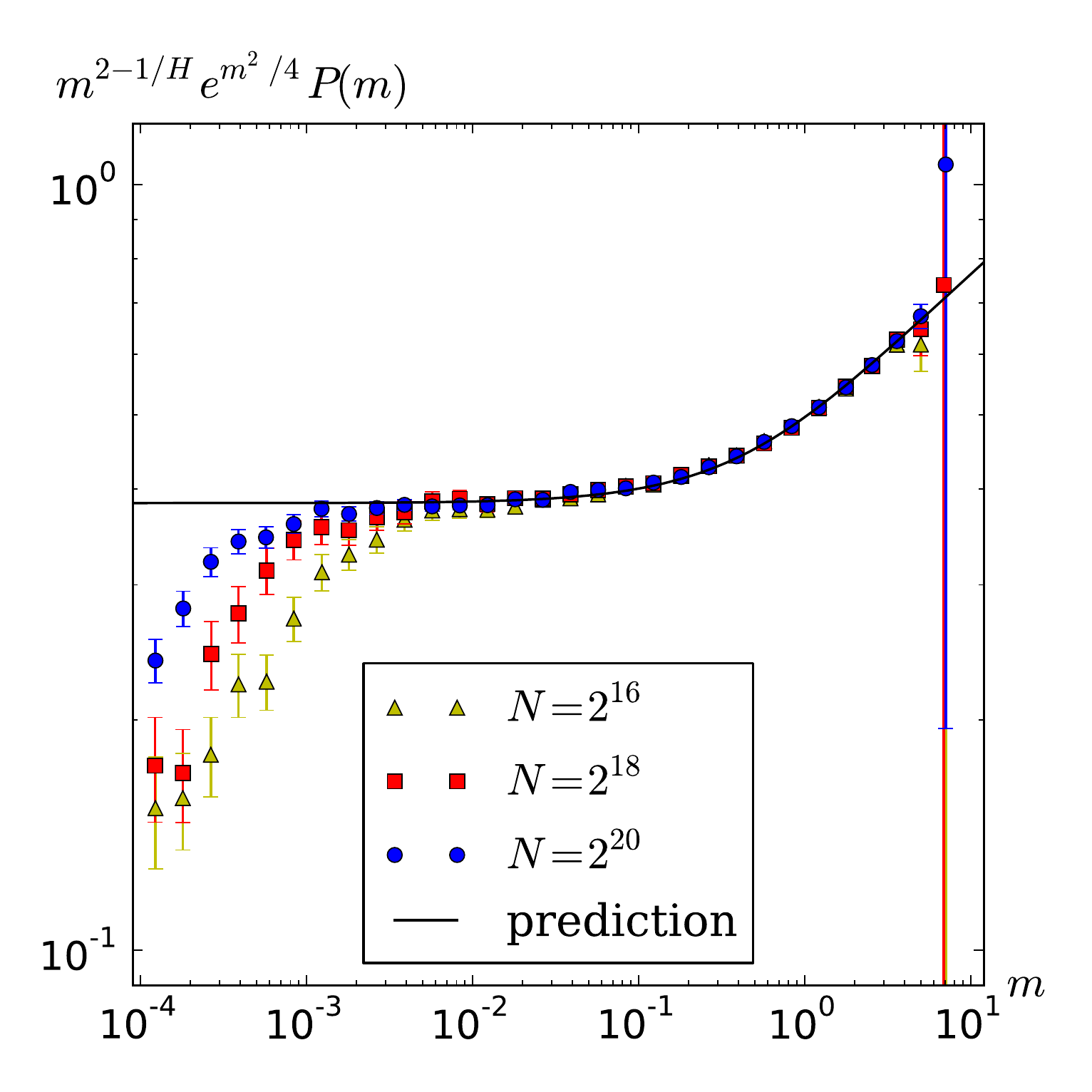}
        \fig{5.9cm}{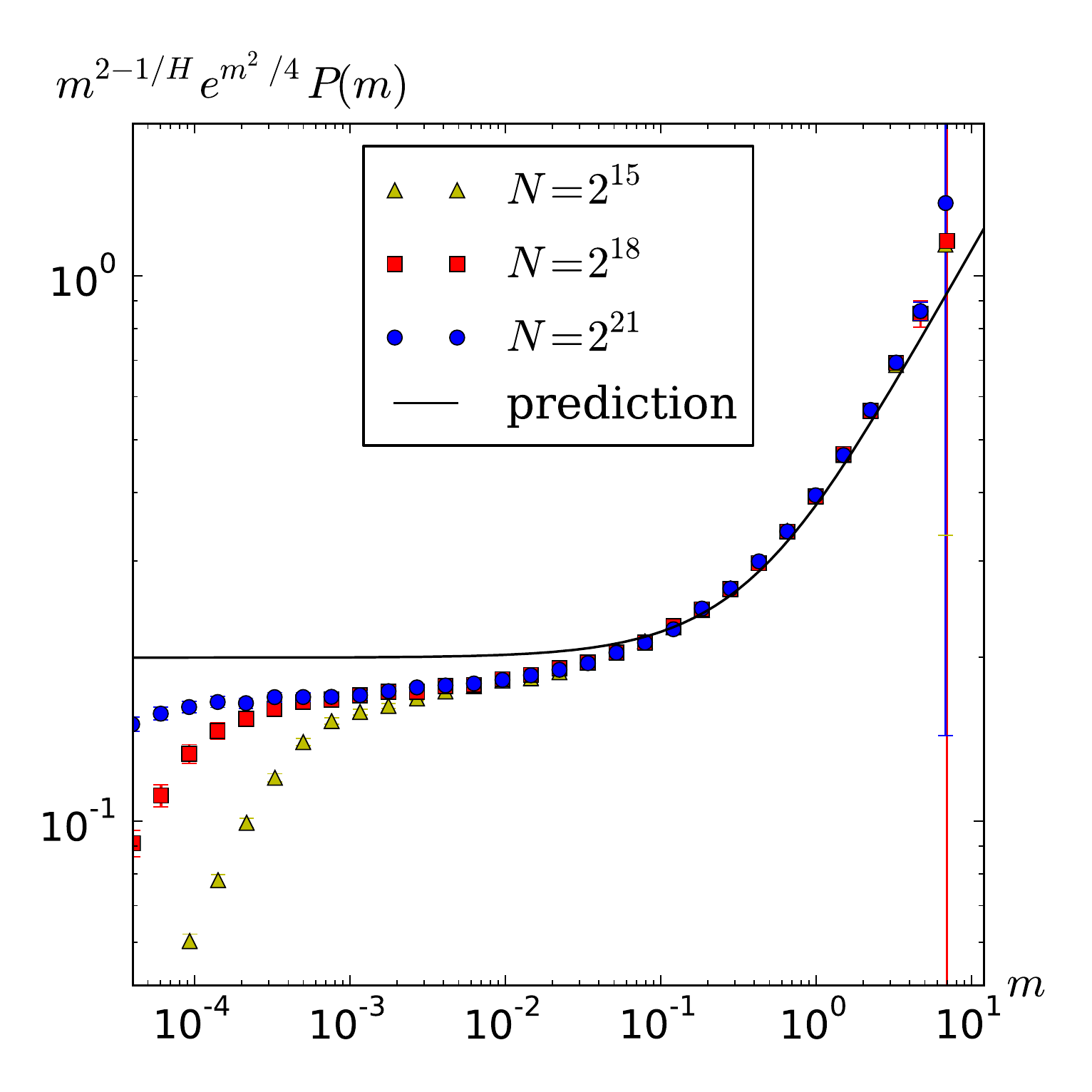}
        \caption{Middle: The combination \eqref{MaxDistribTest} for $H=0.6$. The plain line is the analytical prediction $\exp({\varepsilon [\mathcal{G}(m/\sqrt{2}) + 4 \log m ]+\mbox{cst})}$ of the  distribution of the maximum without its small-scale power law and large-scale Gaussian behavior. The symbols are numerical estimations for $T=1$ of the same quantity $m^{2-{1}/{H}}\exp({ {m^2}/{4}})P^{ T=1,H}_{\rm num}(m)$ for different sample sizes. At small scale discretization errors appear.  At large scales the statistics is poor due to the Gaussian prefactor. For the  four decades in between theory and  numerics are in very good agreement. 
                Left: {\em ibid} for $H=0.4$. Right {\em ibid} for $H=0.75$.~In all cases, the large scale-behavior on  both plots is consistent with  $m^{2\epsilon}$.}
        \label{MaxDistribNum}
\end{figure*}
For the distribution of $t_{\rm max}$, we want to test our analytical results given in Eqs.~\eqref{MaxPosDistrib}-\eqref{Fprediction}. Fig.~\ref{ArcSinLawfBm} shows  the good agreement between theory and numerics.
To perform a more precise comparison, we extract from the numerically computed distribution $P^{T,H}_{\rm num}(t)$ an estimation $\mathcal{F}_{{\rm num}}^{\varepsilon}$ of the function $\mathcal{F}$\,as
\begin{equation}
\mathcal{F}_{{\rm num}}^{\varepsilon}\!\left(\frac{t}{T-t}\right):= \frac{1}{\varepsilon} \log \Big(P^{T,H}_{\rm num}(t) \times [t(T-t)]^H \Big)\ .
\label{Fnum}
\end{equation}
This function should converge, as $\varepsilon \rightarrow 0$, to the theoretical prediction \eqref{Fprediction}. Obviously,  statistical errors become relevant in this limit due to  the factor of  $\varepsilon^{-1}$, while for larger $\varepsilon$ we expect to see deviation due to  order-$\epsilon^2$ (and larger) corrections, which are not taken into account in our analytical computations. As can be seen on Fig.~\ref{ArcSinLawfBmnum}, our numerical and analytical results are in  {\em remarkable} agreement for all values of $H$ studied, both for $\epsilon$ {\em positive} and {\em negative}. This means in particular that even for large values of $\varepsilon$ ($H=0.8$ or $H=0.2$ in the cases studied here), the order-$\varepsilon$ correction is large as compared to higher-order corrections.

The precision of our simulations allows us to numerically investigate these subleading $\mathcal{O}(\epsilon^2)$ corrections, extracted as follows, 
\begin{equation}\label{ArcsinLawOrdre2}
\begin{split}
\mathcal{F}^{\varepsilon}_2(u)&=\frac{1}{\varepsilon}\left(\mathcal{F}_{\rm num}^{\varepsilon}(u)- \mathcal{F}(u)\right)\\
&=\frac{1}{\varepsilon^2} \ln\!\left(\frac{P^{T,H}_{\rm num}(t) \times [t(T-t)]^H}{e^{\varepsilon \mathcal{F}(u)}}\right)\ .
\end{split}
\end{equation}
This is shown in Fig.~\ref{ArcSinLawfBmnum} (right). The collapse of the curves for different values of $\varepsilon$ (once rescaled by $\varepsilon^{-1}$), suggests that there exists a function $\mathcal{F}_2(u)$, which would be the limit as $\varepsilon \rightarrow 0$ of $\mathcal{F}_2^{\varepsilon}(u)$, such that the probability distribution can be written as
\begin{equation}\label{ArcsinLaw2}
P_H^T(t)=\frac{e^{\varepsilon \mathcal{F}(u)+\varepsilon^2 \mathcal{F}_2(u)}}{[t(T-t)]^H}+\mathcal{O}(\varepsilon^3)\ .
\end{equation}
Our estimation of $\mathcal{F}_2$ is plotted on figure \ref{ArcSinLawfBmnum} (right).
Our perturbative approach and its diagrammatic representation allows us   to write the integrals needed to compute $\mathcal{F}_2$ analytically;  this, however, is  left for future work \cite{DelormeWieseUnPublished}.

\subsection{The distribution of the maximum}
For the  distribution of the maximum we  rewrite  formula \eqref{MaxDistribPrediction} such that the small-$m$ behavior reproduces the exact scaling result \eqref{MaxScaling} without changing the result at $\varepsilon$-order,
\begin{equation}
f_H(y) = \sqrt{\frac{2}{\pi}} y^{\frac{1}{H}-2} e^{-\frac{y^2}{2}}   e^{\varepsilon \left[\mathcal{G}\left(y\right) + 4 \log y + {\rm  cst}\right]}  +\mathcal{O}(\varepsilon^2) \ .
\label{MaxDistribPredictionCorrected}
\end{equation}
To extract the non-trivial contribution from numerical simulations, we study  for $T=1$ (see Fig.~\ref{MaxDistribNum})
\begin{equation}
m^{2-\frac{1}{H}}e^{ \frac{m^2}{4}}P^{{1,H}}_{\rm num}(m)= e^{\varepsilon \left[\mathcal{G}\left(\frac{m}{\sqrt{2}}\right) + 4 \log m + {\rm  cst} \right]} +\mathcal{O}(\varepsilon^2)\ .
\label{MaxDistribTest}
\end{equation}
The left-hand side is evaluated from the normalized binned distribution of the maximum   for each fBm path, denoted $P^{1,H}_{\rm num}(m)$. The right-hand side is the analytical result; the constant term is evaluated by numerical integration such that $f_H(y)$, given in Eq.\ \eqref{MaxDistribPredictionCorrected}, is   normalized to $1$.

The sample size $N$ (\textit{i.e.}\ lattice spacing $ \rmd t=N^{-1}$) of the discretized fBm used for this numerical test is important, as the samples recover Brownian behavior for $m$ smaller than a cutoff of order $N^{-H}$. This can be  understood by assuming that the typical value of the first discretized point $X_{1/N}$ is of order $N^{-H}$; thus  for $m \ll N^{-H}$,
\begin{equation}
P^{1,H}_{\rm num}(m)\sim \text{prob}(X_{1/N}=m) \sim m^0
\end{equation}
Far small   $H$ the  system size necessary to obtain the asymptotic behavior at small scale is very large, so we focus our tests  on $H>0.4$. Figure \ref{MaxDistribNum} presents results for $H=0.4$, $H=0.6$ and $H=0.75$, without any fitting parameter. As   predicted,  convergence to the small-scale behavior is quite slow. For example, in the  $H=0.6$ plot   the convergence to the small-scale behavior is somewhere between $10^{-1}$ and $10^{-2}$ (in  dimensionless variables where we rescaled the total time to $T=1$). This might lead to a wrong numerical estimation of the persistence exponent or other related quantities, if the crossover to the large-scale behavior is not properly taken into account. At large scales, the  numerical data on Fig.~\ref{MaxDistribNum} grow as $m^{2\epsilon}$, consistent with the prediction~(\ref{asymp}).

As stated,  for $H<0.5$ the numerical simulations do not allow us  to    investigate the small-scale behavior of the distribution, as can be seen for  $H=0.4$     on figure \ref{MaxDistribNum}. Nevertheless,  the agreement  with the theoretical prediction is good in the crossover region and in the beginning of the tail. The numerical prefactor of the small-scale power law is also very sensitive to numerical errors (and probably to $\varepsilon^2$-corrections) due to a vanishing probability when $m \to 0$ for $H<0.5$, as can be seen in Fig.~\ref{MaxDistribPlot}.

\section{Conclusions}
\label{s:Conclusions}
To conclude, we developed a perturbative approach for the extreme-value statistics of  fractional Brownian motion. This allows to derive the, to our knowledge, first analytical results for generic values of $H$ in the range $0<H<1$, beyond scaling relations. The main, and most general result is the joint probability of the value of the maximum and the time when this maximum is reached, conditioned on the value of the end point, as given in Eq.~\eqref{mainresult}. From this, we extracted simpler result, as the unconditioned distribution of the value of the maximum, as well as distribution of the time when this maximum is reached. These two distributions have non-trivial features, which we compared to numerical simulations. The remarkable agreement of the simulations with our predictions is a valuable  check of  our  method. It also shows that the perturbative approach gives surprisingly good results, even far form the expansion point $H=\frac12$.

The method can be generalized to other cases of interest,       such as the other two Arcsine laws, linear and non-linear drift, and fractional Brownian bridges. Work in these directions is in progress.



\section{Acknowledgments}
We thank  Paul Krapivsky, Kirone Mallick, A.~Rosso and T.~Sadhu for stimulating discussions, and PSL for support through grant ANR-10-IDEX-0001-02-PSL.

\appendix
\begin{widetext}
\section{Details on the perturbative expansion}
\label{AppendixDetails}
We explicit here   details on the steps  transforming Eq.~\eqref{17} into Eq.~\eqref{ZgammaLaplace}. We have  to deal with   terms of the form
\begin{equation}
\begin{split}
\int_{X_{0}=x_1}^{X_{t}=x_2} \mathcal{D}[X] \Theta[X] \dot{X}_{0} e^{-S_0[X]}&= \lim\limits_{\delta \rightarrow 0}\int_{X_{0}=x_1}^{X_{t}=x_2} \mathcal{D}[X] \Theta[X] \frac{X_{\delta}-x_1}{\delta} e^{-S_0[X]}\\
&=\lim\limits_{\delta \rightarrow 0} \int_0^\infty \rmd x \frac{x-x_1}{\delta} P_0^+(x_1,x,\delta) P_0^+(x,x_2,t- \delta)\\
&=\lim\limits_{\delta \rightarrow 0} \int_0^\infty \rmd x\, 2 \partial_x P_0^+(x_1,x,\delta) P_0^+(x,x_2,t- \delta)\\
&= \int_0^\infty \rmd x\, \delta(x-x_1) 2 \partial_x P_0^+(x,x_2,t)\\
&=2\partial_{x_1} P_0^+(x_1,x_2,t)\ .
\end{split}
\end{equation}
We first introduced a discretized version of the derivative, then expressed the path integral in terms of propagators, did an integration by parts and finally took the limit of $\delta \rightarrow 0$.

With this result we can express every path integral in Eq.~\eqref{17} in terms of the bare propagator $P_0^+(x_1,x_2,t)$, 
\begin{equation}
\begin{split}
Z&_{\gamma}^+(m_1,t_1,x_0,t_2,m_2) =\\
&\frac{1}{2} \int_{t_1}^{T} d\tau_2 \int_0^{t_1} d\tau_1 \int_{x_1,x_2>0} \frac{P_0^+(m_1,x_1,\tau_1)\,2\partial_{x_1} P_0^+(x_1,x_0,t_1-\tau_1)\,P_0^+(x_0,x_2,\tau_2-t_1) \,2\partial_{x_2} P_0^+(x_2,m_2,T-\tau_2)}{\tau_2-\tau_1}\ .
\end{split}
\end{equation}
We now use the identity $\frac{1}{\tau_2-\tau_1} = \int_{y>0} e^{-y (\tau_2-\tau_1)}$, and perform two Laplace transformations ($t_1 \rightarrow s_1$ and $t_2 \rightarrow s_2$). It is important to note that the time integrals are in general  divergent at small times, thus we introduced a short-time cutoff $\tau$ in the action, \textit{c.f.} Eq.~\eqref{ActionExpansion}. The short-time cutoff $\tau$ corresponds to a large-$y$ cutoff $\Lambda = e^{-\gamma_{\rm E}} / \tau$. This value is imposed by the following equality, valid for all $T>0$, in the limit of $\Lambda \rightarrow \infty$ and $\tau \rightarrow 0$:
\begin{equation}
\int_0^T \rmd t \int_0^{\Lambda} e^{-y t} \rmd y =\log (T \Lambda)+\gamma_{\rm E}+ \mathcal{O}(e^{-T \Lambda }) \overset{!}{=} \log\!\left(\frac{T}{\tau}\right)= \int_\tau ^T  \frac{1}{t}\, \rmd t\ .
\end{equation}
To simplify the computations, we introduce new time variables,
\begin{equation}
T_1 = \tau_1, \, \,T_2 = t_1-\tau_1, \, \,T_3 = \tau_2-t_1, \, \,T_4 = t_1+t_2-\tau_2\ .
\end{equation}
This gives
\begin{align}
\tilde Z_{\gamma}^+(s_1,s_2) &= 2\!\! \int\limits_{t_1,t_2>0}\!\!\!e^{-s_1 t_1-s_2 t_2} \int\limits_{t_1}^{t_1+t_2}\!\!\!\rmd\tau_2 \int\limits_0^{t_1} \!\rmd\tau_1 \int\limits_0^{\Lambda}\!\rmd y\, e^{-y(\tau_2-\tau_1)} \,P_0^+(t_1)\,\partial P_0^+(\tau_1-t_1)\,P_0^+(\tau_2-t_1)\,\partial P_0^+(t_1+t_2-\tau_2)\nn\\
&=2 \int_0^{\Lambda}\!\!\rmd y\int_{T_i>0} e^{-(T_1 + T_2)s_1}e^{-(T_3+T_4)s_2}e^{-(T_2 + T_3)y}\,P_0^+(T_1)\,\partial P_0^+(T_2)\,P_0^+(T_3)\,\partial P_0^+(T_4)\ .
\end{align}
The space dependence (\textit{i.e.}\ $x_0$, $x_1$, $x_2$ dependence) is   omitted for notational clarity. The successive integrations over time variables transform this expression into a product of Laplace-transformed propagators with different Laplace variables,
\begin{equation}
\tilde Z_{\gamma}^+(m_1,s_1;x_0;m_2,s_2) = 2\int_0^{\Lambda} \!\!\! \rmd y \int_{x_1,x_2>0}\!\! \tilde P_0^+(m_1,x_1,s_1) \,\partial_{x_1} \tilde P_0^+(x_1,x_0,s_1+y) \tilde P_0^+(x_0,x_2,s_2+y) \,\partial_{x_2}  \tilde P_0^+(x_2,m_2,s_2)\ .
\end{equation}
This is the formula given in the main text in Eq.~\eqref{ZgammaLaplace}, apart that here we made explicit  the large-$y$ cutoff. As we will see, there is no large-$y$ divergence here, which render the cutoff irrelevant. The other   time orderings, corresponding to $Z_{\alpha}^+$ and $Z_{\beta}^+$, have a similar structure. For $Z_{\alpha}$, this gives
\begin{equation}
\begin{split}
Z&_{\alpha}^+(m_1,t_1,x_0,t_2,m_2) =\\
&\frac{1}{2} \int_{\tau_1}^{t_1} d\tau_2 \int_0^{t_1} d\tau_1 \int_{x_1,x_2>0} \frac{P_0^+(m_1,x_1,\tau_1)\,2\partial_{x_1} P_0^+(x_1,x_2,\tau_2-\tau_1)\,2\partial_{x_2}P_0^+(x_2,x_0,t_1-\tau_2) \, P_0^+(x_0,m_2,t_2)}{\tau_2-\tau_1}\ .
\end{split}
\end{equation}
This term is represented diagrammatically in Fig.~\ref{Diagrams} (right); computing the double Laplace transform gives
\begin{equation}\label{ZalphaLaplace}
\tilde Z_{\alpha}^+(m_1,s_1;x_0;m_2,s_2) =  \left [2\int_0^{\Lambda} \!\!\! \rmd y\int_{x_1,x_2>0}\!\! \tilde P_0^+(m_1,x_1,s_1) \,\partial_{x_1} \tilde P_0^+(x_1,x_2,s_1+y) \,\partial_{x_2}\tilde P_0^+(x_2,x_0,s_1)\right] \tilde P_0^+(x_0,m_2,s_2)\ .
\end{equation}
In this case, the integrations affect only the first three propagators. The term in square brackets is the correction to the constrained propagator from $m_1$ to $x_0$, with Laplace variable $s_1$. This object was at the center of Ref.\ \cite{WieseMajumdarRosso2010};  the results are recalled in the next appendix. Similarly for $Z_{\beta}$, after the Laplace transformations, the integrations affect only the last three propagators, giving
\begin{equation}\label{ZbetaLaplace}
\tilde Z_{\beta}^+(x_0,s_1;x_0;m_2,s_2) = \tilde P_0^+(m_1,x_0,s_1) \left [2\int_0^{\Lambda} \!\!\! \rmd y\int_{x_1,x_2>0}\!\! \tilde P_0^+(x_0,x_1,s_2) \,\partial_{x_1} \tilde P_0^+(x_1,x_2,s_2+y) \,\partial_{x_2}\tilde P_0^+(x_2,x_0,s_2)\right]\ .
\end{equation}

\section{Recall of the results for $Z_1^+(m,t)$}
\label{AppendixOldResult}

In Ref.\ \cite{WieseMajumdarRosso2010}, the propagator $Z^+(m,t)$ for fBm, conditioned to start at $x_0 \approx 0^+$,  to remain positive, and to finish in  $m$ at time $t$ was computed at order $\varepsilon$. For   standard Brownian motion, this conditioned propagator is
\begin{equation}
Z^{+}_0(m,t)= \lim\limits_{x_0 \rightarrow 0}\frac{1}{x_0} P_0^+(x_0,m,t) =\frac{m e^{- \frac{m^2}{4t}}}{2 \sqrt{\pi}t^{3/2}}\ .
\end{equation}
The term $x_0^{-1}$  is the normalisation (\textit{i.e.}\ one divides by the conditional probability). The order-$\epsilon$ correction of this propagator is given in Eq.~$(51)$ of \cite{WieseMajumdarRosso2010}, \
\begin{equation}\label{OldResult}
\begin{split}
Z^+_1(m,t)=& Z^+_0(m,t)  \left[ \left( \frac{m^2}{2t} -2 \right) \big(\ln(m^2) +\gamma_{\rm E} \big) + \mathcal{I}\left( \frac{m}{\sqrt{2t}}\right) + \log(t) -2\gamma_{\rm E} \right]\\
=&Z^+_0(m,t)  \left[\mathcal{I}(z)+z^2 \left(\log(2z^2)+\gamma_{\rm E} \right)  + (z^2-1)\log(t)-4 \log(z) -4\gamma_{\rm E} \right]\ .
\end{split}
\end{equation}
This result assumes a proper normalisation of $Z_1^+$ such that $x_0$ and $\log(x_0)$ terms cancel, \textit{i.e.} the limit $x_0 \rightarrow 0$ is well-defined, and the integral over $m$ is equal to unity. We introduced $z:=m/ \sqrt{2 t,}$ and $\mathcal{I}$ is the combination of special functions defined in Eq.\ (\ref{defI_maintext}), and  recalled in Eq.~\eqref{def_calI}. 

We can also use the diagrammatic rules introduced in this article to compute the Laplace-transformed correction to this propagator (without conditioning). This corresponds to the diagram represented in Fig.~\ref{Diagrams} (right) without the slice on the right,
\begin{equation}
\tilde P_1^+(x_0,m,s)= 2 \int_0^{\Lambda} \!\!\! \rmd y\int_{x_1,x_2>0}\!\!\tilde P_0^+(x_0,x_1,s)\partial_{x_1}\tilde P_0^+(x_1,x_2,s+y)\partial_{x_2}\tilde P_0^+(x_2,m,s)\ .
\end{equation}
This is   the term appearing in the square brackets in Eqs.~\eqref{ZalphaLaplace} and \eqref{ZbetaLaplace}.
The integrations over space can be done, giving the following integral, rescaling $y \rightarrow u s$, and setting $m=1$ for simplicity:
\begin{equation}
\begin{split}
\tilde P_1^+(&x_0,1,s)=\frac{1}{\sqrt{s}} \int_0^{\Lambda/s}\!\frac{\rmd u}{u^2} \Bigg\{\left[\left(\sqrt{s}-1\right) u-2\right] e^{-\sqrt{s}}\sinh (\sqrt{s} x_0)-x_0 u \sqrt{s}\, e^{-\sqrt{s}} \cosh (\sqrt{s} x_0)\\
&+\sqrt{u+1} \left[e^{-\sqrt{s} \sqrt{u+1} (1-x_0)}+e^{-\sqrt{s} \sqrt{u+1} (x_0+1)}-2 e^{-\sqrt{s} (\sqrt{u+1}+x_0)}-2 e^{-\sqrt{s} (x_0\sqrt{u+1} +1)}+2 e^{-\sqrt{s} (x_0+1)}\right]\Bigg\} .
\end{split}
\end{equation}
This is a logarithmically diverging integral at large $u$, which makes the UV cutoff necessary (\textit{cf.} Appendix \ref{AppendixDetails} where we explicit the link between the $y$ cutoff $\Lambda$ and the time cutoff $\tau$). Doing the integration over $u,$ and then taking the limit $x_0 \rightarrow 0$ as well as expressing the cutoff $\Lambda$ in term of $\tau$ gives  
\begin{equation}\label{PropagatorCorrectionLaplace}
\begin{split}
\frac{1}{x_0}\tilde P_1^+(x_0,m,s) \underset{x_0 \rightarrow 0}{\simeq}& e^{m \sqrt{s}} \left(m \sqrt{s}+1\right) \text{Ei}\left(-2 m \sqrt{s}\right)- e^{-m \sqrt{s}} \left(m \sqrt{s}+1\right) \log(m \sqrt{s})\\
&+ m \sqrt{s} e^{-m \sqrt{s}} \left[\log \left(\frac{m^2}{2 \tau }\right) -1\right]+ e^{-m \sqrt{s}} \left[\log \left(\frac{\tau ^2}{2 x_0^4}\right)-3 \gamma_{\rm E} +4\right]\ .
\end{split}
\end{equation}
This expression in Laplace variables for the correction to the propagator is a new result (in \cite{WieseMajumdarRosso2010}, a more complicated transformation was used to derive Eq.~\eqref{OldResult}). The inverse Laplace transform can be done, using Eqs.~\eqref{InvLaplace3}-\eqref{InvLaplace4} for the complicated terms, 
\begin{equation}
\frac{P^+_1(x_0,m,t)}{P^+_0(x_0,m,t)} \underset{x_0 \rightarrow 0}{\simeq}\mathcal{I}(z)+ z^2 \left[\log(2 z^2)+\gamma_{\rm E}\right] + (z^2-1) \left[\log \left(\frac{t}{\tau}\right)-1 \right]+\log \left(\frac{\tau ^2}{4 x_0^4 z^4}\right)-4 \gamma_{\rm E} +2\ .
\end{equation}
We still need to correct this with the rescaling of the diffusion constant, \textit{i.e.} taking into account the order-$\varepsilon$ correction in Eq.\ \eqref{ConstrainedPropagator} given the expression of the diffusive constant \eqref{Diffusive_constant}. This gives
\begin{equation}
2t \partial_t P_0^+(x_0,m,t)(1+\log(\tau))= P_0^+(x_0,m,t) (z^2-3)\big[1+\log(\tau)\big].
\end{equation}
A  check of consistency is that this cancels all dependence on $\tau$, and we find for the propagator at order $\varepsilon$,
\begin{equation}\label{PropagatorOrdre1}
P^+(x_0,m,t)\underset{x_0 \rightarrow 0}{\simeq}P^+_0(x_0,m,t)\Big\{1+\varepsilon\Big[\mathcal{I}(z)+ z^2 \left(\log(2 z^2)+\gamma_{\rm E}\right) + (z^2-1) \log (t)-\log \left(4 x_0^4 z^4\right)-4 \gamma_{\rm E}  \Big]\Big\}+\mathcal{O}(\varepsilon^2)\ .
\end{equation}
This propagator, integrated over $m,$   reads, both in time and Laplace variables\begin{equation}\label{IntegratedPropagator}
\begin{split}
\int_0^{\infty} \rmd m \, \tilde P_1^+(x_0,m,s) &\underset{x_0 \rightarrow 0}{\simeq} \frac{x_0}{\sqrt{s}}\left(3-3 \gamma_{\rm E} - \log(4s \tau)+\log \left(\frac{\tau^2}{x_0^4}\right)\right)\ ,\\
\int_0^{\infty} \rmd m \,  P_1^+(x_0,m,t) &\underset{x_0 \rightarrow 0}{\simeq} \frac{x_0}{\sqrt{\pi t}}\left(3-2 \gamma_{\rm E} + \log\left(\frac{t \tau}{x_0^4}\right)\right)\ .
\end{split}
\end{equation}

\section{Computation of $Z^+_{\gamma}$}
\label{AppendixZgamma}
\subsection{Outline of the Calculation}
We present here   details of the calculation of $Z_{\gamma}^+$, starting from its expression in Laplace variables \eqref{ZgammaLaplace}, graphically represented in Fig.~\ref{PathIntegralFigure}.
First, we introduce the notation\begin{equation}
\begin{split}
\mathcal{S}(m,x_0,s,y)&:= \frac{1}{x_0}\int_0^{\infty} \rmd x\, \tilde P_0^+(m,x,s) \,\partial_{x} \tilde P_0^+(x,x_0,s+y)\\
&= \frac{1}{x_0} \frac{e^{-(m-x_0) \sqrt{s+y}}-e^{-(m+x_0) \sqrt{s+y}}+2 e^{-x_0 \sqrt{s+y}-m \sqrt{s}}-e^{-(m-x_0)\sqrt{s} }-e^{- (m+x_0)\sqrt{s}}}{2 y}\ .
\end{split}
\end{equation}
The expression of $\tilde P^+_0$ is given in Eq.~\eqref{propagator}. We see from Eq.\ \eqref{ZgammaLaplace} that one can write $\tilde Z_{\gamma}^+(m_1,s_1;x_0;m_2,s_2)$ as
\begin{equation}
\tilde Z_{\gamma}^+(m_1,s_1;x_0;m_2,s_2)=-2 x_0^2 \int_{y>0}\mathcal{S}(m_1,x_0,s_1,y) \mathcal{S}(m_2,x_0,s_2,y)
\ .\end{equation}
The minus sign comes from an integration by parts. It is interesting to look at the asymptotics of $\mathcal{S}$ in the limit of $x_0 \rightarrow 0$,
\begin{equation}
\mathcal{S}(m,x_0,s,y)\underset{x_0 \rightarrow 0}{\simeq} \frac{1}{y} \left(  e^{-m \sqrt{s+y}}\sqrt{s+y}-e^{-m \sqrt{s}} \sqrt{s+y}\right) \underset{y\rightarrow \infty}{\sim} \frac{  e^{-m \sqrt{s}}}{ \sqrt{y}}
\ .\end{equation}
This implies that  the $x_0 \rightarrow 0$ limit {\em can not} be taken before integrating over $y$, as this induces  a new large-$y$, i.e.\ short-time divergence. Taking this limit before integration, and regularizing the new divergence with the large-$y$ cutoff $\Lambda$ would lead to a wrong result. This is expected as the scaling of the result in terms of $x_0$  depends on $H$, thus inducing a $\log(x_0)$ term at order $\varepsilon$.

In the following, we note $\mathcal{S}= \bar{\mathcal{S}} + \delta\mathcal{S}$ with
\begin{eqnarray}
\bar{\mathcal{S}}(m,x_0,s,y) &:=& \frac{1}{x_0}\frac{e^{-(m-x_0) \sqrt{s+y}}-e^{-(m+x_0) \sqrt{s+y}}+2 e^{-(x_0 +m)\sqrt{s}}-e^{-(m-x_0)\sqrt{s} }-e^{- (m+x_0)\sqrt{s}}}{2y}\ ,\\
\delta \mathcal{S}(m,x_0,s,y) &:=&  \frac{1}{x_0}\frac{e^{-x_0 \sqrt{s+y}-m \sqrt{s}}-e^{-x_0\sqrt{s}-m \sqrt{s}}}{y} 
\ .\end{eqnarray}
Denoting $\mathcal{S}_i := \mathcal{S}(m_i,x_0,s_i,y)$, the integration over $y$   is a sum of four terms (with the last two related by   exchanging  points $1$ and $2$),
\begin{equation}
\int_{y>0} \mathcal{S}_1 \mathcal{S}_2 =\int_{y>0} \bar{\mathcal{S}}_1\bar{\mathcal{S}}_2 + \int_{y>0} \delta \mathcal{S}_1 \delta \mathcal{S}_2+ \int_{y>0} \bar{\mathcal{S}}_1 \delta \mathcal{S}_2 +   \int_{y>0} \bar{\mathcal{S}}_2 \delta \mathcal{S}_1
\ .\end{equation}
This leads to the following decomposition of $Z_{\gamma}^+(m_1,t_1;x_0;m_2,t_2)$,
\begin{equation}\label{ZgammaDecomposition}
Z_{\gamma}^+=x_0^2\big[Z_A(m_1,t_1;m_2,t_2) +Z_B(m_1,t_1;x_0;m_2,t_2)+Z_C(m_1,t_1;m_2,t_2) +Z_C(m_2,t_2;m_1,t_1)\big]\ ,
\end{equation}
with
\begin{equation}\label{ZABC}
\begin{split}
Z_A(m_1,s_1;m_2,s_2)&=-2 \mathcal{L}_{s_2\rightarrow t_2}^{-1}\circ\mathcal{L}_{s_1\rightarrow t_1}^{-1}\!\!\left[\lim\limits_{x_0 \to 0} \int_{y>0}  \bar{\mathcal{S}}(m_1,x_0,s_1,y)  \bar{\mathcal{S}}(m_2,x_0,s_2,y)\right]\ ,\\
Z_B(m_1,s_1;x_0;m_2,s_2)&=-2 \mathcal{L}_{s_2\rightarrow t_2}^{-1}\circ\mathcal{L}_{s_1\rightarrow t_1}^{-1}\!\!\left[\lim\limits_{x_0 \to 0} \int_{y>0} \delta \mathcal{S}(m_1,x_0,s_1,y)  \delta \mathcal{S}(m_2,x_0,s_2,y)\right]\ ,\\
Z_C(m_1,s_1;m_2,s_2)&=-2 \mathcal{L}_{s_2\rightarrow t_2}^{-1}\circ\mathcal{L}_{s_1\rightarrow t_1}^{-1}\!\!\left[\lim\limits_{x_0 \to 0} \int_{y>0}  \bar{\mathcal{S}}(m_1,x_0,s_1,y) \delta \mathcal{S}(m_2,x_0,s_2,y)\right]\ .
\end{split}
\end{equation}
We anticipate here that $Z_A$ and $Z_C$ have a well-defined $x_0 \to 0$ limit, and only $Z_B$ has a divergence (as shown later). The next step consists in computing these three integrals over $y$, taking the limit of small $x_0$, and performing the inverse Laplace transforms \textit{w.r.t.}\ $s_1$ and $s_2$. The order of these manipulations can sometimes be inverted to simplify the calculations.

\subsection{The term $Z_A$}
In the first term of Eq.\ \eqref{ZABC} it is possible to take the $x_0 \rightarrow 0$ limit inside the integral, as this integrand  converges fast enough for large $y$, given the asymptotic of $\bar{\mathcal{S}}$, \begin{equation}\label{AsymptoticBarS}
\bar{\mathcal{S}} \underset{x_0 \rightarrow 0}{\simeq}  \frac{  e^{-m \sqrt{s+y}}\sqrt{s+y}- e^{-m \sqrt{s}}\sqrt{s}}{y}\ .
\end{equation}
This gives\begin{equation}
\int_{y>0} \bar{\mathcal{S}}_1\bar{\mathcal{S}}_2 \underset{x_0 \rightarrow 0}{\simeq} \int_{y>0}
\frac{\left( e^{-m_1 \sqrt{s_1+y}}\sqrt{s_1+y}-e^{-m_1 \sqrt{s_1}}\sqrt{s_1} \right) \left( e^{-m_2 \sqrt{s_2+y}}\sqrt{s_2+y}- e^{-m_2 \sqrt{s_2}}\sqrt{s_2}\right)}{y^2}
\ .\end{equation}
We can do the inverse Laplace transformations $s_1 \rightarrow t_1$ and $s_2 \rightarrow t_2$ before integrating over $y$, using
\begin{equation}
\mathcal{L}_{s \rightarrow t}^{-1}\left[- e^{-m \sqrt{s+y}}\sqrt{s+y}\right] = \frac{e^{-\frac{m^2}{4 t}} }{2 \sqrt{\pi } t^{3/2}}\left(1-\frac{m^2}{2t}\right)e^{- t y}\ .
\label{InverseLaplace1}
\end{equation}
One thus finds
\begin{equation}
\mathcal{L}_{s_2\rightarrow t_2}^{-1}\circ\mathcal{L}_{s_1\rightarrow t_1}^{-1}\int_{y>0} \bar{\mathcal{S}}_1\bar{\mathcal{S}}_2 \underset{x_0 \rightarrow 0}{\simeq}\frac{e^{-\frac{m_1^2}{4 t_1}-\frac{m_2^2}{4 t_2}} }{4 \pi t_1^{3/2}t_2^{3/2}}\left(1-\frac{m_1^2}{2t_1}\right)\left(1-\frac{m_2^2}{2t_2}\right) \int_{y>0}\frac{(1-e^{-t_1 y}) (1-e^{-t_2 y})}{y^2}\ .
\end{equation}
Integrating over $y$ and using the definition of $Z_A$, the final result for this term is
\begin{equation}\label{ResultZA}
Z_A(m_1,t_1;m_2,t_2)= \frac{e^{-\frac{m_1^2}{4 t_1}-\frac{m_2^2}{4 t_2}}}{ 2 \pi (t_1t_2)^{3/2}} \left(1 - \frac{m_1^2}{2 t_1} \right)\left(1- \frac{m_2^2}{2t_2}\right) \Big[t_1 \log (t_1)+t_2 \log (t_2)-(t_1+t_2) \log (t_1+t_2)\Big]\ .
\end{equation}

\subsection{The term $Z_B$}

For the second term of Eq.\ \eqref{ZABC}, the limit $x_0 \rightarrow 0$  cannot be  taken inside the integral, as
\begin{equation}\label{AsymptoticDeltaS}
\delta \mathcal{S}= \frac{1}{x_0}\frac{e^{-x_0 \sqrt{s+y}-m \sqrt{s}}-e^{-x_0\sqrt{s}-m \sqrt{s}}}{y} \underset{x_0 \rightarrow 0}{\simeq} \frac{ e^{-m \sqrt{s}}}{y} ( \sqrt{s}-\sqrt{s+y}) \underset{y \rightarrow \infty}{\sim} -\frac{e^{-m \sqrt{s}}}{\sqrt{y}}\ .
\end{equation}
However, we can extract the diverging part by writing
\begin{equation}\label{C15}
\int_{y>0} \delta \mathcal{S}_1  \delta \mathcal{S}_2=  e^{-m_1 \sqrt{s_1}-m_2 \sqrt{s_2}} \log(x_0^{-2}+1) + 
\int_{y>0} \left[ \delta \mathcal{S}_1 \delta \mathcal{S}_2- \frac{e^{-m_1 \sqrt{s_1}-m_2 \sqrt{s_2}}}{y+1} \, \Theta( y<x_0^{-2}) \right]
\ .\end{equation}
This expression is constructed such that for all $x_0>0$ the term added outside the integral and the term subtracted inside the integral cancel.  The diverging part when $x_0 \rightarrow 0$ is now the term outside the integral and the integral has a finite limit when $x_0 \rightarrow 0$. To proceed,  denote $\mathcal{K}:= e^{-m_1 \sqrt{s_1}-m_2 \sqrt{s_2}}$. We then decompose the integral as
a sum of three terms, 
\begin{equation}
\begin{split}
\int_{y>0} \bigg[ \delta \mathcal{S}_1 \delta \mathcal{S}_2 - \frac{\mathcal{K}}{y+1} \Theta (y<x_0^{-2}) \bigg] =&\int_0^{x_0^{-2}}\!\!\!\!\rmd y \left[\delta \mathcal{S}_1 \delta \mathcal{S}_2 - \mathcal{K} \frac{( \sqrt{s_1+y}-\sqrt{s_1})( \sqrt{s_2+y}-\sqrt{s_2})}{y^2}  \right]\\
+\mathcal{K}&\int_0^{x_0^{-2}}\!\!\!\!\rmd y \left[ \frac{( \sqrt{s_1+y}-\sqrt{s_1})( \sqrt{s_2+y}-\sqrt{s_2})}{y^2}- \frac{1}{y+1} \right] + \int_{x_0^2}^{\infty}\!\!\rmd y \,\delta \mathcal{S}_1 \delta \mathcal{S}_2\ .
\end{split}
\end{equation}
In the second term we can   take the limit  of $x_0 \rightarrow 0$ to obtain (without the $\mathcal{K}$ factor in front)
\begin{equation}
\begin{split}
\int_{y>0} &\left[\frac{( \sqrt{s_1+y}-\sqrt{s_1})( \sqrt{s_2+y}-\sqrt{s_2})}{y^2} - \frac{1}{y+1} \right]\\
&=-\left(2 + \sqrt{\frac{s_1}{s_2}} + \sqrt{\frac{s_2}{s_1}} \right)\log \left(\sqrt{s_1}+\sqrt{s_1}\right)
+\frac{1}{2}\sqrt{\frac{s_1}{s_2}} \log \left(s_1\right)+\frac{1}{2}\sqrt{\frac{s_2}{s_1}} \log \left(s_2\right)-1+\log (4)\ .
\end{split}
\end{equation}
For the first and third term, we first perform a rescaling of the integration variable ($y \rightarrow x_0^{-2} v$) and then take the limit of  $x_0 \rightarrow 0$,
\begin{eqnarray}
&&\int_0^{x_0^{-2}} \rmd y \left[\delta \mathcal{S}_1 \delta \mathcal{S}_2 - \mathcal{K} \frac{( \sqrt{s_1+y}-\sqrt{s_1})( \sqrt{s_2+y}-\sqrt{s_2})}{y^2}  \right] \underset{x_0 \rightarrow 0}{\simeq} \mathcal{K} \int_0^1 \rmd v \left[ \frac{\left( e^{-\sqrt{v}} -1 \right)^2}{v^2} -\frac{1}{v} \right]
\ ,\\
&& \int_{x_0^2}^{\infty} \rmd u \,\delta \mathcal{S}_1 \delta \mathcal{S}_2 \underset{x_0 \rightarrow 0}{\simeq} \mathcal{K} \int_1^{\infty} \rmd v  \frac{\left( e^{-\sqrt{v}} -1 \right)^2}{v^2} 
\ .
\end{eqnarray}
The sum of the  last two contributions in the limit of $x_0 \rightarrow 0$  is 
\begin{equation}
 \mathcal{K} \int_1^{\infty} \rmd v  \frac{\left( e^{-\sqrt{v}} -1 \right)^2}{v^2} + \mathcal{K} \int_0^1 \rmd v \left[ \frac{\left( e^{-\sqrt{v}} -1 \right)^2}{v^2} -\frac{1}{v} \right] = \mathcal{K}\left[ 3-2 \gamma_{\rm E}-2 \log (4)\right]\ .
\end{equation}
Summing all these contribution gives
\begin{equation}\label{ZB_Laplace}
\begin{split}
\int_{y>0} \delta \mathcal{S}_1&  \delta \mathcal{S}_2
\underset{x_0 \rightarrow 0}{\simeq} e^{-m_1 \sqrt{s_1}-m_2 \sqrt{s_2}} \,\times \\
&\left[  -\left(2 + \sqrt{\frac{s_1}{s_2}} + \sqrt{\frac{s_2}{s_1}} \right)\log \left(\sqrt{s_1}+\sqrt{s_2}\right)
+\sqrt{\frac{s_1}{s_2}} \log \left(\sqrt{s_1}\right)+\sqrt{\frac{s_2}{s_1}} \log \left(\sqrt{s_2}\right) -2\log(2 x_0) +2 - 2 \gamma_{\rm E}\right]\ .
\end{split}
\end{equation}
We now need a series of Inverse Laplace transforms obtained in appendix \ref{AppendixLaplace}.
To deal with the double Laplace inversion, we start with   formula \eqref{InvLaplace1} and use the special function $\mathcal{J}$ defined in Eq.\ \eqref{defCalJ}. Using commutativity of derivation and integration with the Laplace transform, we can use  the identity
\begin{equation}
\left(2 + \sqrt{\frac{s_1}{s_2}} + \sqrt{\frac{s_2}{s_1}} \right)  e^{-m_1 \sqrt{s_1}-m_2 \sqrt{s_2}}= \left( \partial_{m_1} + \partial_{m_2} \right) \left(\int_{m_1}  + \int_{m_2} \right) e^{-m_1 \sqrt{s_1}-m_2 \sqrt{s_2}} 
\end{equation}
to obtain
\begin{equation}
\begin{split}
\mathcal{L}_{s_2\rightarrow t_2}^{-1}\circ&\mathcal{L}_{s_1\rightarrow t_1}^{-1} \left[e^{-m_1 \sqrt{s_1}-m_2 \sqrt{s_2}}\left(2 + \sqrt{\frac{s_1}{s_2}} + \sqrt{\frac{s_2}{s_1}} \right) \log \left(\sqrt{s_1} + \sqrt{s_2} \right) \right]\\
&=\left(\partial_{m_1} + \partial_{m_2} \right)^2 \left\{ \frac{e^{-\frac{m_2^2}{4t_2}-\frac{m_1^2}{4t_1}}}{\pi \sqrt{t_1 t_2}} \left[\mathcal{J}\!\left( \frac{\left(m_2 t_1+m_1 t_2\right){}^2}{4 t_1 t_2 \left(t_1+t_2\right)} \right) + \frac{1}{2} \log\!\left( \frac{1}{4t_1} + \frac{1}{4t_2}\right) - \frac{\gamma_{\rm E}}{2}\right] \right\}\ .
\end{split}
\end{equation}
For the other terms, the inverse Laplace transforms are decoupled, and can be computed from Eq.\ \eqref{InvLaplace2}. We get
\begin{equation}
\mathcal{L}_{s_2\rightarrow t_2}^{-1}\circ\mathcal{L}_{s_1\rightarrow t_1}^{-1} \left[e^{-m_1 \sqrt{s_1}-m_2 \sqrt{s_2}} \sqrt{ \frac{s_1}{s_2}}  \log \left(\sqrt{s_1} \right) \right]= \partial_{m_1}^2 \left\{ \frac{e^{-\frac{m_2^2}{4t_2}-\frac{m_1^2}{4t_1}}}{\pi \sqrt{t_1 t_2}} \left[\mathcal{J}\!\left( \frac{m_1^2}{4 t_1} \right) + \frac{1}{2} \log \left( \frac{1}{4t_1} \right) - \frac{\gamma_{\rm E}}{2}\right] \right\}
\ .
\end{equation}
The sum of all terms, with a prefactor of $-2$  coming from the definition of $Z_B$, is
\begin{equation}
\begin{split}
Z_B(m_1,t_1;x_0;m_2,t_2) =&\; \frac{ m_1 m_2 e^{-\frac{m_2^2}{4t_2}-\frac{m_1^2}{4t_1}} }{2 \pi (t_1 t_2)^{3/2}}\big[ 2 \log(2 x_0) -2 + 2 \gamma_{\rm E} \big]\\
&+2 (\partial_{m_1} + \partial_{m_2})^2 \left\{ \frac{e^{-\frac{m_2^2}{4t_2}-\frac{m_1^2}{4t_1}}}{\pi \sqrt{t_1 t_2}} \left[\mathcal{J\!} \left( \frac{\left(m_2 t_1+m_1 t_2\right){}^2}{4 t_1 t_2 \left(t_1+t_2\right)} \right) + \frac{1}{2} \log\! \left( \frac{1}{4t_1} + \frac{1}{4t_2}\right) - \frac{\gamma_{\rm E}}{2}\right] \right\}\\
&-2\, \partial_{m_1}^2 \left\{ \frac{e^{-\frac{m_2^2}{4t_2}-\frac{m_1^2}{4t_1}}}{\pi \sqrt{t_1 t_2}} \left[\mathcal{J\!} \left( \frac{m_1^2}{4 t_1} \right) + \frac{1}{2} \log\! \left( \frac{1}{4t_1} \right) - \frac{\gamma_{\rm E}}{2}\right] \right\}  + (1 \leftrightarrow 2)\ .
\end{split}
\end{equation}
The derivatives can be computed explicitly, using the relation between  $\mathcal{I}$ and $\mathcal{J}$ given in Eq.~\eqref{I_J_relation},
\begin{equation}
\partial_{m_1}^2\!\left\{ \frac{e^{-\frac{m_2^2}{4t_2}-\frac{m_1^2}{4t_1}}}{\pi \sqrt{t_1 t_2}} \left[\mathcal{J}\!\left( \frac{m_1^2}{4 t_1} \right) + \frac{1}{2} \log \left( \frac{1}{4t_1} \right) - \frac{\gamma_{\rm E}}{2}\right] \right\}  =-  \frac{e^{-\frac{m_2^2}{4t_2}-\frac{m_1^2}{4t_1}}}{4 \pi (t_1 t_2)^{3/2}} t_2 \left[  \mathcal{I}\!\left(  \frac{m_1}{\sqrt{2t_1}}\right) + \left( \frac{m_1^2}{2 t_1} -1 \right) \left( \log \left(4 t_1 \right)  +\gamma_{\rm E} \right) \right]\ .
\end{equation}
The same result holds for the term involving $\partial^2_{m_2}$. For the term involving simultaneously $m_1$ and $m_2$, we  can  use almost the same trick,
\begin{equation}
\begin{split}
(\partial_{m_1} + \partial_{m_2})^2 &\left[ e^{-\frac{m_2^2}{4t_2}-\frac{m_1^2}{4t_1}} \mathcal{J} \left(\frac{\left(m_2 t_1+m_1 t_2\right){}^2}{4 t_1 t_2 \left(t_1+t_2\right)} \right)\right]\\
&= \frac{t_1 +t_2}{4 t_1 t_2}e^{-\frac{m_2^2}{4t_2}-\frac{m_1^2}{4t_1}} \left[ 2(z^2-1) \mathcal{J} \left( \frac{z^2}{2} \right) - 2(2 z^2-1) \mathcal{J}' \left( \frac{z^2}{2} \right) + 2 z^2 \mathcal{J}'' \left( \frac{z^2}{2} \right) \right]\\
&= - \frac{t_1 +t_2}{4 t_1 t_2}e^{-\frac{m_2^2}{4t_2}-\frac{m_1^2}{4t_1}}  \mathcal{I} \left( \frac{m_1 t_2 + m_2 t_1}{\sqrt{2 t_1 t_2 (t_1 +t_2)}}\right)\ .
\end{split}
\end{equation}
The second line is   the explicit derivative of the first line, expressed for simplicity in terms of the variable 
\begin{equation}
z = \frac{m_1 t_2 + m_2 t_1}{\sqrt{2 t_1 t_2 (t_1 +t_2)}}\ .
\end{equation}
The combination of $\mathcal{J}$ and its derivatives appearing in the second line is exactly the function $\mathcal{I}$, as can be checked from Eq.~\eqref{I_J_relation}.
After   these simplifications, \begin{align}\label{ResultZB}
Z_B\underset{x_0 \to 0}{\simeq} \frac{  e^{-\frac{m_2^2}{4t_2}-\frac{m_1^2}{4t_1}} }{2 \pi (t_1 t_2)^{3/2}}&\Bigg\{ 2m_1 m_2\big[ \log(2 x_0) + \gamma_{\rm E} -1\big]- (t_1+t_2) \left[   \mathcal{I}( z) + \left(z^2-1 \right) \left( \log\!\left( \frac{4 t_1 t_2}{t_1 +t_2} \right) + \gamma_{\rm E} \right)\right]\\
+t_2& \left[  \mathcal{I}\!\left(  \frac{m_1}{\sqrt{2t_1}}\right) + \left( \frac{m_1^2}{2 t_1} -1 \right) \big( \log (4 t_1 )  +\gamma_{\rm E}\big) \right] + t_1 \left[  \mathcal{I}\!\left(  \frac{m_2}{\sqrt{2t_2}}\right) + \left( \frac{m_2^2}{2 t_2} -1 \right) \big( \log (4 t_2 )  +\gamma_{\rm E}\big) \right]  \Bigg\}\ .\nn
\end{align}

\subsection{The term $Z_C$}

For this term,  we can take the limit $x_0 \rightarrow 0$ inside the integral, as it   converges for large $y$ using asymptotics \eqref{AsymptoticBarS} and \eqref{AsymptoticDeltaS}, giving 
\begin{equation}\label{ZC_Laplace}
\int_{y>0} \bar{\mathcal{S}}_1 \delta \mathcal{S}_2 \underset{x_0 \rightarrow 0}{\simeq} e^{- m_2 \sqrt{s_2}}\int_{y>0} \frac{e^{-m_1\sqrt{s_1+y}}\sqrt{s_1+y} -e^{-m_1\sqrt{s_1}}\sqrt{s_1} }{y} \frac{ \sqrt{s_2} - \sqrt{s_2+y}}{y}\ .
\end{equation}
To compute the Laplace inversion $s_1 \rightarrow t_1$, we use Eq. \eqref{InverseLaplace1}\,
\begin{equation}
\begin{split}
\mathcal{L}^{-1}_{s_1 \to t_1}\!\left[\int_{y>0} \mtilde{\mathcal{S}}_1 \delta \mathcal{S}_2 \right] &= \frac{e^{-\frac{m_1^2}{4 t_1} }}{2 \sqrt{\pi} t_1 ^{3/2}} \left(\frac{m_1^2}{2 t_1}-1\right)e^{- m_2 \sqrt{s_2}}\int_{y>0} \frac{(1-e^{- t_1 y})( \sqrt{s_2 +y} - \sqrt{s_2})}{y^2}\\
&=  \frac{e^{-\frac{m_1^2}{4 t_1} }}{2 \sqrt{\pi} t_1 ^{3/2}} \left(\frac{m_1^2}{2 t_1}-1\right) \frac{e^{- m_2 \sqrt{s_2}} } {\sqrt{s_2}} \int_{v>0} \frac{(1-e^{- t_1 s_2 v})( \sqrt{v +1} - 1)}{v^2}\ .
\end{split}
\end{equation}
We changed  variables  $y \rightarrow s_2 v$ between the two lines. To perform the inverse Laplace transform  w.r.t. $s_2$, we need\begin{equation}
\mathcal{L}^{-1}_{s_2 \rightarrow t_2} \left[  \frac{e^{- m_2 \sqrt{s_2}}}{ \sqrt{s_2}} e^{- t_1 s_2 v} \right ] = \theta (t_2 - v t_1) \frac{e^{- \frac{{m_2}^2}{4 (t_2 - v t_1)} }}{ \sqrt{ \pi(t_2 - v t_1 )}}
\end{equation}
Finally, to
compute $Z_C$, only the integration over $v$ remains to be done, 
\begin{align}
Z_C(m_1,t_1;t_2,m_2) &=  -\frac{e^{-\frac{m_1^2}{4 t_1} }}{ \sqrt{\pi} t_1 ^{3/2}} \left(\frac{m_1^2}{2 t_1}-1\right)  \int_{v>0} \left(\frac{e^{- \frac{{m_2}^2}{4 t_2}}}{\sqrt{\pi t_2}} -\Theta(t_2 - v t_1) \frac{e^{- \frac{{m_2}^2}{4 (t_2 - v t_1)}}}{\sqrt{\pi (t_2 - v t_1)}}\right) \frac{\sqrt{v+1} -1}{v^2}\nn \\
&=- \frac{ e^{-\frac{m_1^2}{4 t_1} }e^{-\frac{m_2^2}{4 t_2} }}{ 2\pi( t_1 t_2)^{3/2}} ( m_1^2 -2t_1)  \frac{t_2}{t_1} \int_{v>0} \left( 1 - \Theta\!\left(\frac{t_2}{t_1}-v\right) \frac{e^{-\frac{m_2^2}{4t_2} \left(\frac{1}{1-v t_1/t_2} -1 \right)}}{\sqrt{1- v \frac{t_1}{t_2}}}\right) \frac{\sqrt{v+1}-1}{v^2}\\
&=-\frac{ e^{-\frac{m_1^2}{4 t_1} }e^{-\frac{m_2^2}{4 t_2} }}{ 2\pi( t_1 t_2)^{3/2}} (m_1^2 - 2 t_1)\left[\nu \int_0^{\nu} \!\! \rmd v\,\left( 1 -  \frac{e^{-a \left(\frac{1}{1-v / \nu} -1 \right)}}{\sqrt{1- v/\nu}}\right) \frac{\sqrt{v+1}-1}{v^2} + \nu \int_{\nu}^{\infty}\!\!\! \rmd v \, \frac{\sqrt{v+1}-1}{v^2}\right]\ ;\nn
\end{align}
here we have introduced $\nu = t_2/t_1$ and $a=m_2^2/(4 t_2)$. Thus the following integrals   needs to be computed, 
\begin{equation}
I_1(a,\nu) = \nu \int_0^{\nu}  \rmd v\,\left( 1 -  \frac{e^{-a \left(\frac{1}{1-v / \nu} -1 \right)}}{\sqrt{1- v/\nu}}\right) \frac{\sqrt{v+1}-1}{v^2} \; \text{ \ and }\;\; I_2(\nu) = \nu \int_{\nu}^{\infty} \rmd v \, \frac{\sqrt{v+1}-1}{v^2}\ .
\end{equation}
The term $I_2$ is easy,  
\begin{equation}
I_2(\nu) = \nu \int_{\nu}^{\infty} \rmd v \, \frac{\sqrt{v+1}-1}{v^2} = \sqrt{\nu+1} -1+ \nu \,{\asinh}\! \left(\frac{1}{\sqrt{\nu}}\right)=\sqrt{ \frac{t_1+t_2}{t_1}} - 1 + \frac{t_2}{t_1} \,\asinh\! \left(\sqrt{\frac{t_1}{t_2}}\right)\ .
\end{equation}
The other integral is   more involved. To evaluate it, we perform a change of variables
\begin{equation}
I_1  (a ,\nu  ) = \nu \int_0^{\nu}  \rmd v\,\left( 1 -  \frac{e^{-a \left(\frac{1}{1-v /\nu} -1 \right)}}{\sqrt{1- \frac{v}{\nu}}}\right) \frac{\sqrt{v+1}-1}{v^2}=\int_0^{\infty} \rmd x\,\left( \frac{1}{\sqrt{x+1}}- e^{-a x } \right) \frac{\sqrt{(\nu+1)x+1}-\sqrt{x+1}}{x^2}\ .
\end{equation}
To simplify the   integrand, we then  take its second derivative w.r.t. $a$,
\begin{equation}
\partial^2_a I_1(a,\nu) =- \int_0^{\infty}\rmd x\,e^{-ax} \left( \sqrt{(\nu+1)x+1}-\sqrt{x+1}\right)=-\frac{\sqrt{\pi } \left(\sqrt{\nu +1} e^{\frac{a}{\nu +1}} \text{erfc}\left(\sqrt{\frac{a}{\nu +1}}\right)-e^a \text{erfc}\left(\sqrt{a}\right)\right)}{2 a^{3/2}}\ .
\end{equation}
The function
\begin{equation}
\begin{split}
f(a) 
=\,& \frac{1}{2} \,\mathcal{I}\!\left(\sqrt{2a} \right) + 3a -1 + a \log(a)
\end{split}
\end{equation}
where $\mathcal{I}$ is defined in \eqref{def_calI}, satisfies
\begin{equation}
f''(a)= -\frac{\sqrt{\pi}}{2} \frac{e^a}{a^{3/2}}\text{erfc}(\sqrt{a})\ .
\end{equation}
We can then express the second derivative of $I_1$ in terms of  $f$,  
\begin{equation}
\partial^2_a I_1(a,\nu) =\frac{1}{1+\nu} f''\left(\frac{a}{1+\nu}\right)-f''(a)\ .
\end{equation}
After   two integrations over $a$ we obtain, with yet unknown functions $A(\nu)$ and $B(\nu)$, 
\begin{equation}
I_1(a,\nu) =  (\nu +1) \,f\! \left( \frac{a}{\nu +1} \right)-f(a) + B(\nu) a + A(\nu)\ .
\end{equation}
The small-$a$ behavior of $f$ can be obtained as 
\begin{equation}
f(a)=2 \sqrt{\pi } \sqrt{a}+a \log (a)-\frac{2 \sqrt{\pi}}{3} a^{3/2}+\frac{a^2}{3}+ {\cal O}(a^{5/2})\ .
\end{equation}
We can compare this to the limit when $a$ goes to $0$ of the initial integral to determinate the integration constants $A$ and $B$. The limit is computed by taking the limit inside the integral, with result
\begin{equation}
\lim_{a \rightarrow 0} I_1(a,\nu) =1- \sqrt{\nu +1}+ \frac{1}{2}(\nu +1) \log (\nu +1)- \vartheta  \log \left(\sqrt{\nu +1}+1\right)\ .
\end{equation}
Finally, we get
\begin{equation}
\begin{split}
I_1\! \left( \frac{m_2^2}{4t_2}, \frac{t_2}{t_1}\right) =&\left(1 + \frac{t_2}{t_1} \right) f\! \left( \frac{{m_2}^2}{4 t_2} \frac{ t_1}{t_2+t_1}\right) - f\! \left( \frac{{m_2}^2}{4 t_2}\right)\\
&+ 1 -\sqrt{\frac{t_2+t_1}{t_1}}+\frac{t_2+t_1}{t_1} \log \left( \sqrt{\frac{t_2+t_1}{t_1}}\right) - \frac{t_2}{t_1}\log \left( \sqrt{\frac{t_2+t_1}{t_1}}+1 \right)\ .
\end{split}
\end{equation}
This has been checked numerically with excellent precision. 

There are a few terms that cancel between $I_1$ and $I_2$, and expressing  $\asinh$ in terms of $\ln$, and $f$ in terms of $\mathcal{I}$ finally gives 
\begin{equation}\label{ResultZC}
\begin{split}
Z_C(m_1,&t_1;m_2,t_2) =  \frac{ e^{-\frac{m_1^2}{4 t_1} }e^{-\frac{m_2^2}{4 t_2} }}{ 2\pi( t_1 t_2)^{3/2}} \left(1- \frac{m_1^2}{2 t_1} \right)\\
\times& \bigg[ (t_1+t_2)\,\mathcal{I}\!\left( \frac{m_2}{\sqrt{2t_2}} \sqrt{\frac{ t_1}{t_2+t_1}}\right) - t_1 \mathcal{I}\!\left( \frac{m_2}{\sqrt{2t_2}}\right) -2 t_2 + t_1 \left( \frac{{m_2}^2}{2 t_2} -1\right)\log\! \left( \frac{t_1}{t_2 + t_1}\right) + t_2 \log\! \left( \frac{t_1+t_2}{t_2}\right) \bigg]\ .
\end{split}
\end{equation}\\
We   computed numerically the double Laplace transform of \eqref{ResultZC}, and checked   with high precision agreement with \eqref{ZC_Laplace}, where the integral over $y$ is   evaluated numerically.

\section{Correction to the third Arcsine Law}
\label{AppendixArcsine_Law}
As stated in the main text, the distribution of $t_{\rm max}$ can be extracted from our path integral \eqref{PathIntegral} as follows:
\begin{equation}
P^T_H(t) = \lim\limits_{x_0\rightarrow 0}\frac{1}{Z^N(T,x_0)} \int_{m_1,m_2>0} Z^{+}(m_1,t;x_0;m_2,T-t)\ .
\end{equation}
The order-$0$ contribution \eqref{Zorder0} gives for the normalisation
\begin{equation}
Z^N= \int_0^T \rmd t \int_{m_1,m_2>0} Z_0^+(m_1,t;x_0;m_2,T-t) +\mathcal{O}(\varepsilon) = x_0^2 +\mathcal{O}(\varepsilon)
\ .\end{equation}
We recover the well-known {\em\ Arcsine Law} distribution for standard Brownian motion,
\begin{equation}\label{ArcsinLawBrownien}
P_{\half}^T(t)= \lim\limits_{x_0 \rightarrow 0} \frac{\int_{m_1,m_2>0} Z_0^+(m_1,t;x_0;m_2,T-t)}{x_0^2} =  \int_{m_1,m_2>0}\frac{m_1 m_2 e^{-\frac{m_1^2}{4 t_1}-\frac{m_2^2}{4 t_2}}}{4 \pi t_1^{3/2}t_2^{3/2}}= \frac{1}{\pi \sqrt{t(T-t)}}\ .
\end{equation}\\
Let us now write every term in the $\varepsilon$-expansion: $Z^N = Z_{(0)}^N+ \varepsilon Z_{(1)}^N +\mathcal{O}(\varepsilon^2)$ and  $Z^+ = Z_{(0)}^++ \varepsilon Z_{(1)}^+ +\mathcal{O}(\varepsilon^2)$. It is important to note that these terms slightly differ from  those in Eq.~\eqref{PathIntegralExpansion}, where the expansion was done \textit{w.r.t.} the non-local perturbation in the action. As computed in Eq.~\eqref{Zorder0Bis}, the term $Z_0^+$ contains some order-$\varepsilon$ correction, contrary to $Z^+_{(0)}$ which is defined as the constant part of $Z^+$ in its $\varepsilon$ expansion.

Using these new notations, we have
\begin{equation}\label{DistribExpension}
P^T_H(t) = \lim\limits_{x_0\rightarrow 0}\frac{\int Z_{(0)}^+}{Z^N_{(0)}} \left[ 1 + \varepsilon\left(\frac{\int Z_{(1)}^+}{\int Z_{(0)}^+}-\frac{Z_{(1)}^N}{Z_{(0)}^N}\right)\right]+\mathcal{O}(\varepsilon^2) = P_{\half}^T(t) \lim\limits_{x_0\rightarrow 0}\left[ 1 + \varepsilon\left(\frac{\int Z_{(1)}^+}{\int Z_{(0)}^+}-\frac{Z_{(1)}^N}{Z_{(0)}^N}\right)\right]+\mathcal{O}(\varepsilon^2)\ ,
\end{equation}
where $\int$ symbol implicitly denotes integration over $m_1$ and $m_2$. The normalisation  ensures that the correction to the probability
\begin{equation}
\delta P^T(t) = P_{\half}^T(t)\lim\limits_{x_0 \to 0}\left(\frac{\int Z_{(1)}^+}{\int Z_{(0)}^+}-\frac{Z_{(1)}^N}{Z_{(0)}^N}\right)
\end{equation}
does not change the normalisation, \textit{i.e.} its integral over $t$ vanishes.

To compute the order-$\varepsilon$ correction to the distribution \eqref{ArcsinLawBrownien}, we have to compute the integral over $m_1$ and $m_2$ of $Z_{\alpha}^+$, as well as $Z_{\beta}^+$ and $Z_{\gamma}^+(m_1,t_1;x_0;m_2,t_2)$. The last term,   computed in Appendix \ref{AppendixZgamma}, was decomposed in four terms, see Eq.~\eqref{ZgammaDecomposition}. The expressions for these terms are given in Eqs.~\eqref{ResultZA},  \eqref{ResultZB} and \eqref{ResultZC}.
Using the identity $\int_{z>0} e^{-\frac{z^2}{2}}(z^2-1) =0$, we find the  simplifications
\begin{equation}
\int_{m_1,m_2>0} Z_A = \int_{m_1,m_2>0} Z_C = 0\ .
\end{equation}
Thus,  the only contribution of $Z^+_{\gamma}$ comes   from $Z_B$, defined in \eqref{ZABC},
\begin{equation}
\begin{split}
\frac{1}{x_0^2}\int_{m_1,m_2>0}\!\!\!\!Z_{\gamma}^+=\int_{m_1,m_2>0}\!\!\!\! Z_B(m_1,t_1;x_0;m_2,t_2) =&  -\frac{2}{\pi \sqrt{t_1 t_2}} \left(1 +\log\!\left(\frac{4 t_1 t_2}{t_1 + t_2} \right) - 2\log(2 x_0)  + 2\gamma_{\rm E} \right) + \frac{1}{t_1} +\frac{1}{t_2}\\
&-\frac{t_1+t_2}{2 \pi (t_1 t_2)^{3/2}} \int_{m_1,m_2>0} e^{-\frac{m_1^2}{4 t_1}-\frac{m_2^2}{4 t_2}} \mathcal{\,I\!}\left(z= \frac{m_1 t_2 + m_2 t_1}{\sqrt{2 t_1 t_2 (t_1 +t_2)}}\right)
\end{split}
\end{equation}
We have used the identity $\int_0^{\infty} \rmd z e^{-z^2/2} \mathcal{I}(z) = \sqrt{2 \pi}$. To compute
the last integral,  we use   relation \eqref{I_J_relation}, which in this case gives
\begin{equation}
\frac{t_1+t_2}{2 \pi (t_1 t_2)^{3/2}} \int_{m_1,m_2>0} e^{-\frac{m_1^2}{4 t_1}-\frac{m_2^2}{4 t_2}} \mathcal{I}( z) = -\frac{2}{\pi \sqrt{t_1 t_2}} \int_{m_1,m_2>0} (\partial_{m_1} + \partial_{m_2})^2 \left[e^{-\frac{m_1^2}{4 t_1}-\frac{m_2^2}{4 t_2}} \mathcal{J}\left( \frac{(m_1 t_2 + m_2 t_1)^2}{4 t_1 t_2 (t_1 +t_2)}\right) \right]
\ .\end{equation}
Only the cross term of the derivatives (\textit{i.e.} the term with $2 \partial_{m_1} \partial_{m_2}$) is not a total derivative and gives a non-zero contribution,
\begin{equation}
\frac{2}{\pi \sqrt{t_1 t_2}} \int_{m_2>0} e^{-\frac{m_2^2}{4 t_2}}\,\partial_{m_1}\!\left.\mathcal{J}\left(\frac{\left(m_2 t_1+m_1 t_2\right){}^2}{4 t_1 t_2 \left(t_1+t_2\right)} \right)\right|_{m_1=0} =  \frac{2}{\pi t_1} \arctan\! \left( \sqrt{\frac{t_2}{t_1}}\right)
\ .\end{equation}
The final result for this correction is
\begin{equation}
\begin{split}
\frac{1}{x_0^2}\int_{m_1,m_2>0}\!\!\!\!Z_{\gamma}^+ =&  \frac{-2}{\pi \sqrt{t_1 t_2}} \left[\log\left(\frac{4 t_1 t_2}{t_1 + t_2} \right)- 2 \log(2 x_0)+1  + 2\gamma_{\rm E} \right] + \frac{1}{t_1} +\frac{1}{t_2}- \frac{2}{\pi t_1} \arctan\! \left( \sqrt{\frac{t_2}{t_1}}\right) - \frac{2}{\pi t_2} \arctan\! \left( \sqrt{\frac{t_1}{t_2}}\right)\ .
\end{split}
\end{equation}
The contributions to the correction from $Z_{\alpha}^+$ and $Z_{\beta}^+$ are easily computed from their expressions in terms of propagators given in the main text, \textit{c.f.} Eqs.~\eqref{Zalpha} and \eqref{Zbeta}, and then using formula \eqref{IntegratedPropagator},
\begin{equation}
\frac{1}{x_0^2}\int_{m_1,m_2>0}P_0^+(x_0,m_1,t_1)P_1^+(x_0,m_2,t_2)+(1 \leftrightarrow 2) \underset{x_0 \to 0}{\simeq} \frac{1}{\pi \sqrt{t_1 t_2}} \left[6- 4\gamma_{\rm E}+ \log(t_1 t_2)+\log\left(\frac{\tau^2}{x_0^8}\right)  \right] \ .
\end{equation}
The last term of order $\varepsilon$ comes from the rescaling of the diffusive constant, which was made explicit in Eq.~\eqref{Zorder0Bis},
\begin{equation}
2[1+ \log(\tau)\\ ] (t_1 \partial_{t_1}+t_2 \partial_{t_2}) \frac{1}{x_0^2}\int_{m_1,m_2>0} Z_0^+ = -2 \frac{ [1+ \log(\tau)]}{\pi\sqrt{t_1 t_2}}\ .
\end{equation}
Summing all these contributions at order $\varepsilon,$ and taking into account the correction from normalisation gives the final result for the order-$\varepsilon$ term of the probability, 
\begin{equation}
\begin{split}
\delta P^T(t)= \frac{1}{\pi \sqrt{t_1 t_2}} \Bigg\{&-\log(t_1 t_2)+ \sqrt{\frac{t_1}{t_2}}  \left[\pi -2 \arctan\left(\sqrt{\frac{t_1}{t_2}}\right)\right]+\sqrt{\frac{t_2}{t_1}} \left[\pi -2 \arctan\left(\sqrt{\frac{t_2}{t_1}}\right)\right]\\
 &+2 \log(T) + 4 - 6 \gamma_{\rm E} + \log \left(\frac{\tau^2}{x_0^4}\right) -\frac{Z_{(1)}^N(T,x_0)}{x_0^2}-2 \big[1+\log(\tau)\big] \Bigg\}\ ,
\label{ArcineLawExpension2}
\end{split}
\end{equation}
with $t_1=t$ and $t_2=T-t$. As expected, the dependence in $\tau$ vanishes at the end of the computation, and the order  $\varepsilon$ of the normalisation factor $Z_{(1)}^N$ is fixed by the condition $\int_0^T \rmd t\, \delta P^T(t)=0$, which gives
\begin{equation}
Z_{(1)}^N = x_0^2 \big[ 8 \log(2) +2 - 6 \gamma_{\rm E} - 4 \log(x_0)\big]
\ .\end{equation}
Equivalently, the constant term, i.e.\ the second line of Eq.~\eqref{ArcineLawExpension2}, becomes  $-8 \log(2)$. The interpretation of this result as well as a comparison to numerical simulations is presented in the main text.

\section{Distribution of the maximum of the fractional BM}
\label{AppendixMaxDistrib}
Similarly to the distribution of $t_{\rm max}$, the distribution of $m$ can be computed from the path integral $Z^+(m_1,t_1,x_0,m_2,t_2)$. This is done by taking the limit of small $x_0$, the integral over $m_2$ and the integral over $t_1$ at $t_1+t_2=T$ fixed,
\begin{equation}
P^T_H(m)=\lim_{x_0 \rightarrow 0}\frac{1}{Z^N(T,x_0)} \int_0^{m_2} \rmd m_2 \int_0^T \rmd t \, Z^+(m,t,x_0,m_2,T-t)
\ .\end{equation}
It is useful to note that the integration over $t=t_1$ at fixed $T=t_1+t_2$ can be replaced by taking the Laplace transform of $Z^+$ at equal arguments ($s_1=s_2=s$) and then performing the inverse Laplace transform  $s \rightarrow T$. The normalisation $Z^N(T,x_0)$ is the same as the one for the distribution of $P_H^T(t)$;  expanding in $\varepsilon$ thus gives the same structure as \eqref{DistribExpension}, with the $\int$ symbol now being the integrals over $m_2>0$ and $t_1 \in[0,T]$.

We start with the contribution of $Z_{\gamma}$. As before, the integral over $m_2$ of $Z_A$ vanishes, so this term does not contribute. The correction from $Z_B$ can be computed starting with Eq.~\eqref{ZB_Laplace},  taken at equal Laplace variables (\textit{i.e.} $s_1=s_2=s$),
\begin{equation}
\int_{m_2} \int_{t} Z_B = 4 \frac{e^{-m \sqrt{s}}}{\sqrt{s}}\left[ \log(x_0)-1+ \gamma_{\rm E} + 2 \log(2) + \log(\sqrt{s})\right]\ .
\end{equation}
To take the inverse Laplace transform, we use Eq.\ \eqref{InvLaplace3}. This gives
\begin{equation}
\int_{m_2} \int_{t} Z_B = 4 \frac{e^{-\frac{m^2}{4T}}}{\sqrt{\pi T}}\left[\mathcal{J\!}\left(\frac{m^2}{4T}\right)+ \log\! \left( \frac{4 x_0}{\sqrt{T}}\right)+ \frac{\gamma_{\rm E}}{2}-1  \right]
\ .\end{equation}For the contribution of $Z_C$, it is easier to compute the inverse Laplace transform of Eq.\ \eqref{ZC_Laplace} ($s_1=s_2=s \rightarrow T$) before integrating over $y$. This gives
\begin{equation}
\int_{m_2} \int_{t} Z_C = -2\frac{e^{-\frac{m^2}{4T}}}{\sqrt{\pi T}} \int_{0}^{ \infty} \frac{\rmd y}{y^2}\left [e^{-\frac{m^2}{4T}y}\left(\sqrt{1+y}-1-y\right) + \sqrt{1+y}-1\right]
\ .\end{equation}
Let us define
\begin{equation}
I_C(a):= \int_{0}^{ \infty} \frac{\rmd u}{u^2}\left(e^{-a u}\left(\sqrt{1+u}-1-u\right) + \sqrt{1+u}-1\right)
\ .\end{equation}
After deriving twice \textit{w.r.t.} $a$, then integrating twice, and fixing the integration constants, we get\begin{equation}
\begin{split}
I_C(a)= &\,\gamma_{\rm E} +1+\log (4)+ a[3 -\gamma_{\rm E} -\log (4)]\\
&-\frac{a^2}{3}  \, _2F_2\!\left(1,1;\frac{5}{2},3;a\right)+\frac{ \pi}{2}  (2 a-1) \text{erfi}(\sqrt{a})- e^a \sqrt{\pi a} +(1-a) \log (a)\ .
\end{split}
\end{equation}
We can express this in terms of the special function $\mathcal{I}$\,,
\begin{equation}
I_C\!\left(\frac{z^2}{2}\right)= \gamma_{\rm E}+2+\log(4)- \frac{z^2}{2}\left[\gamma_{\rm E}+\log(4)\right]- \frac{1}{2} \mathcal{I}(z)+\left(1- \frac{z^2}{2}\right) \log \left(\frac{z^2}{2}\right)
\ ,\end{equation}\\
This has been checked numerically. The final result for this correction is (with $z:=m/\sqrt{2T}$),
\begin{equation}
\int_{m_2} \int_{t} Z_C = \frac{e^{-\frac{z^2}{2}}}{\sqrt{\pi T}} \left[\mathcal{I}(z)+ (z^2-2)\left(\gamma_{\rm E}+\log \left(2 z^2\right)\right)-4\right] 
\ .\end{equation}The last corrections are: $x_0^{-2}\int_{m_2} \int_{t} Z^+_{\alpha}$ and $x_0^{-2}\int_{m_2} \int_{t} Z^+_{\beta}$. The first one is   easy to compute using the results for the correction to the propagator recalled in Eq.\ \eqref{IntegratedPropagator}, and the inverse Laplace transform \eqref{InvLaplace3},
\begin{equation}
\begin{split}
\frac{1}{x_0^2}\int_0^T \rmd t \int_0^{\infty} \rmd m_2\, P_0^+(x_0,m,t) P_1^+(x_0,m_2,T-t) \underset{x_0 \to 0}{\simeq}\,&\mathcal{L}^{-1}_{s \rightarrow T} \left[\frac{e^{-m \sqrt{s}}}{\sqrt{s}}\left(3 - \log(4s \tau)-3 \gamma_{\rm E} + \log\left(\frac{\tau^2}{x_0^4}\right)\right)\right]\\
\underset{x_0 \to 0}{\simeq}\,& \frac{e^{- \frac{m^2}{4T}}}{\sqrt{\pi T}} \left[-2 \,\mathcal{J}\! \left(\frac{m^2}{4T}\right)+\log\left(\frac{T}{\tau}\right)+2 - 2 \gamma_{\rm E} + \log\left(\frac{\tau^2}{x_0^4}\right) \right]\ .
\end{split}
\end{equation}
For the correction from $Z_{\beta}^+$, we start with the Laplace expression of the correction to the propagator \eqref{PropagatorCorrectionLaplace}, where the integration over $m_2$ simplifies the last slice to $\frac{x_0}{\sqrt{s}}$. Then, the needed inverse Laplace  transform  is 
\begin{equation}
\begin{split}
\frac{1}{x_0^2}\int_0^T \rmd t \int_0^{\infty} \rmd m_2\, P_1^+(x_0,m,t) P_0^+(x_0,m_2,T-t) 
\underset{x_0 \to 0}{\simeq}\,& \frac{1}{x_0} \mathcal{L}^{-1}_{s \rightarrow T}\left[\frac{P_1^+(x_0,m,s)}{\sqrt{s}}\right]\\
\underset{x_0 \to 0}{\simeq}\,& \frac{e^{-\frac{m^2}{4T}}}{\sqrt{\pi T}} \left[ -2 \, \mathcal{J}\!\left(\frac{m^2}{4 T}\right) + \frac{m^2}{2T}\log\left( \frac{T}{\tau}\right) +2 -2 \gamma_{\rm E}+ \log\left(\frac{\tau^2}{x_0^4}\right)\right]\ .\\
\end{split}
\end{equation}
The final result for this is obtained using Eqs.~\eqref{InvLaplace3}-\eqref{InvLaplace4}.\\

We now give a summary of all corrections, in the limit of $x_0 \rightarrow 0$:\begin{equation}
\begin{split}
\frac{1}{x_0^2}\int_t \int_{m_2}\, P_1^+(x_0,m,t) P_0^+(x_0m_2,T-t) &\underset{x_0 \to 0}{\simeq} \frac{e^{-\frac{m^2}{4T}}}{\sqrt{\pi T}} \left[ -2 \mathcal{J}\left(\frac{m^2}{4 T}\right) + \frac{m^2}{2T}\log\left( \frac{T}{\tau}\right) +2 -2 \gamma_{\rm E}+ \log\left(\frac{\tau^2}{x_0^4}\right)\right]\ ,\\
\frac{1}{x_0^2}\int_t \int_{m_2}\, P_0^+(x_0,m,t) P_1^+(x_0,m_2,T-t)
&\underset{x_0 \to 0}{\simeq} \frac{e^{- \frac{m^2}{4T}}}{\sqrt{\pi T}} \left[-2 \mathcal{J} \left(\frac{m^2}{4T}\right)+\log\left( \frac{T}{\tau}\right)+2 - 2 \gamma_{\rm E} + \log\left(\frac{\tau^2}{x_0^4}\right)\right]\ ,\\
\int_{t} \int_{m_2} Z_C(m,t;m_2,T-t) &\underset{x_0 \to 0}{\simeq} \frac{e^{-\frac{m^2}{4T}}}{\sqrt{\pi T}} \left[\mathcal{I}\left(\frac{m}{\sqrt{2T}}\right)+ \left( \frac{m^2}{2T}-2\right)\left(\gamma_{\rm E}+\log \left(\frac{m^2}{T}\right)\right)-4\right]\ ,\\
\int_{t} \int_{m_2} Z_B(m,t;m_2,T-t) &\underset{x_0 \to 0}{\simeq}  \frac{e^{-\frac{m^2}{4T}}}{\sqrt{\pi T}}\left[4\mathcal{J}\left(\frac{m^2}{4T}\right)+ 4\log \left(\frac{4 x_0}{\sqrt{T}}\right)+ 2 \gamma_{\rm E}-4  \right]\ ,\\
\frac{4(1+\ln(\tau))}{x_0^2}\,T \partial_T \int_t \int_{m_2} Z_0^+&\underset{x_0 \to 0}{\simeq}\frac{e^{-\frac{m^2}{4T}}}{\sqrt{\pi T}}\left[1+\ln(\tau)\left(\frac{m^2}{2 T}-1\right)\right]\ .
\end{split} 
\end{equation}
The last line is the correction to the diffusion constant, \textit{i.e.} the order-$\epsilon$ term appearing in Eq.\ \eqref{Zorder0Bis}. The final result at order $\varepsilon$ is 
\begin{equation}\label{F12}
\int_{m_2} \int_t \, Z^+
 = \frac{e^{-\frac{m^2}{4T}}}{\sqrt{\pi T}} \left\{1+ \varepsilon \left[\mathcal{I}\!\left(\frac{m}{\sqrt{2T}}\right)+ \left( \frac{m^2}{2T}-2\right)\left(\gamma_{\rm E}+\log\! \left(\frac{m^2}{T}\right)\right) + \left(\frac{m^2}{2T}-1\right)\log(T) + \mbox{cst}\right]\right\}+\mathcal{O}(\varepsilon^2)\ .
\end{equation}
 To better interpret   the different terms, we   recast the corrections, and especially those as $\frac{m^2}{2 T} \log(T)$ and $\log(T), $ 
into an exponential form,
\begin{equation}
\frac{e^{-\frac{m^2}{4T}}}{\sqrt{\pi T}}\left[1+\varepsilon\left(\frac{m^2}{2T}-1\right)\log(T)\right]+\mathcal{O}(\varepsilon^2)=\frac{e^{-\frac{m^2}{4T}}}{\sqrt{\pi T}}e^{\varepsilon\frac{m^2}{2T}\log(T)}T^{-\varepsilon}+O(\varepsilon^2)=\frac{e^{-\frac{m^2}{4T^{1+2\varepsilon}}}}{\sqrt{\pi }T^{1/2+\varepsilon}}+\mathcal{O}(\varepsilon^2)\ .
\end{equation}
\end{widetext}
This part of the correction  gives the correct dimension to the variables in the  order-0 result,
\begin{equation}
z=\frac{m}{\sqrt{2t}} \rightarrow y=\frac{m}{\sqrt{2} t^H} = \frac{m}{\sqrt{\langle x_t^2 \rangle}}\ .
\end{equation}
The other parts of the correction,
which are a function of $z=\frac{m}{\sqrt{2t}}$ and which we   call $\mathcal{G}(z)$, give a non-trivial change to the scaling function of the distribution,
\begin{eqnarray}\label{F15}
P_H^T(m) &=&\frac{e^{-\frac{m^2}{4T^{2H}}}}{\sqrt{\pi} T^H} e^{\varepsilon \left[\mathcal{G}\left(z = \frac{m}{\sqrt{2t}}\right) + {\rm cst} \right]} +\mathcal{O}(\varepsilon^2)\nn\\ &=& \frac{e^{- \frac{y^2}{2}}}{\sqrt{\pi}T^H} e^{\varepsilon \left[ \mathcal{G}\left(y\right) + {\rm cst}\right]}+\mathcal{O}(\varepsilon^2)\ .
\end{eqnarray}
We changed the variable in $\mathcal{G}$ from $z$ to $y$ as it does not change the result at order $\varepsilon$ and since it is more consistent in terms of dimensions. The function $\mathcal{G}$ is given by
\begin{equation}
\mathcal{G}(y) = \mathcal{I}(y) + (y^2-2)\left[\log(2y^2)+\gamma_{\rm E}\right]
\ ;\end{equation} 
The function $\mathcal{I}$  is regular at $y=0$, and its asymptotic
behavior is given in Eq.~\eqref{Iasymptotic}; this gives the asymptotics for $\cal G$ as
\begin{equation}
\mathcal{G}(y) \sim
\begin{cases} 
        \hfill -2 \ln(y)  \hfill & \text{ for } y \rightarrow \infty \\
        \hfill -4\ln(y) \hfill & \text{ for } y \rightarrow 0\ .\\
\end{cases}
\end{equation}
  Since these  asymptotics are logarithmic  new power laws are obtained  for the density distribution, both at $m \rightarrow 0$ and $m \rightarrow \infty$, which multiply the Gaussian term, with
\begin{equation}
P_{\half + \varepsilon}^T(m)\times e^{ \frac{m^2}{4T^{1+2\varepsilon}}} \sim
\begin{cases}
m^{-4 \varepsilon} \text{ for } m \to 0\\
m^{-2 \varepsilon} \text{ for } m \to \infty\ .
\end{cases}
\end{equation}
The constant term in Eq.~\eqref{F12} is fixed by normalisation. Instead of computing it at order $\varepsilon$, we can also evaluate it numerically such that \eqref{F15} is exactly normalized, and not only at order $\varepsilon$.
This is appropriate for numerical checks and the procedure we adopted for the latter.

\section{Survival distribution}
\label{AppendixSurvival}
To compute the survival probability up to  time $T$ of a fBm starting in $m$, we need to take the primitive function \textit{w.r.t.} $m$ of \eqref{F12}. We can deal with the terms involving $\mathcal{I}$ using \eqref{I_J_relation}; the difficult part comes from
\begin{equation}
\int_0^y \!\!\rmd m\, e^{-\frac{m^2}{2}}(2-m^2)\ln(m)\ .
\end{equation}
To deal with this integration, we  consider $e^{-\frac{m^2}{2}}m^a$, compute the primitive function \textit{w.r.t.} $m$, and then take the derivative \textit{w.r.t.} $a$, at $a=0$ and $a=2$.

The final result can  be written as
\begin{equation}
S(y)=\textrm{erf}\!\left(\frac{y}{\sqrt{2}}\right)+\varepsilon \mathcal{M}(y)+\mathcal{O}(\varepsilon^2)
\end{equation}
This is at leading order in $\epsilon$ equivalent to the exponentiated form given in the main text \eqref{survivalscaling}, with the function $\mathcal{M}$   given by
Eq.~(\ref{Mexpression}).

\section{Special functions and some inverse Laplace transforms}
\label{AppendixLaplace}

In our computations  there are two combinations of special functions which appear frequently, and which we denote $\mathcal{I}$ and $\mathcal{J}$. Their expressions in terms of hypergeometric functions and error functions are\begin{eqnarray}\label{def_calI}
\mathcal{I}(z) &=& \frac{z^4}{6}  \,_2F_2\!\left(1,1; \frac{5}{2},3; \frac{z^2}{2} \right) + \pi (1-z^2) \mathrm{erfi}\!\left( \frac{z}{\sqrt{2}} \right) \nn\\&&- 3z^2 + \sqrt{2 \pi} e^{\frac{z^2}{2}}z +2
\\
\label{defCalJ}
\mathcal{J}(x)&=&\frac{\pi}{2}  \text{erfi}\left(\sqrt{x}\right)-x \, _2F_2\!\left(1,1;\frac{3}{2},2;x\right)
\end{eqnarray}
These functions are linked by
\begin{equation}\label{I_J_relation}
\partial_z^2 \left[e^{- \frac{z^2}{2}} \mathcal{J}\!\left( \frac{z^2}{2}\right) \right] = - \frac{1}{2}e^{- \frac{z^2}{2}} \mathcal{I}(z)\ .
\end{equation}
It is   useful to give their asymptotics, as their natural definition in terms of a series does not allow for an efficient  evaluation at large arguments, 
\begin{eqnarray}\label{Jasymptotic}
\mathcal{J}(x) & {\underset{x \rightarrow \infty}{ \simeq}} &\frac{1}{2} \Big[\log (4 x)+\gamma_{\rm E}\Big] +\frac{1}{4 x}-\frac{3}{16 x^2}+\frac{5}{16 x^3}\nn\\&&-\frac{105}{128 x^4}    + {\cal O}\! \left(\frac{1}{x^5}\right) \\
\label{Iasymptotic}
\mathcal{I}(z) &{\underset{z \rightarrow \infty}{ \simeq}}& -z^2 \left[\log \left(2 z^2\right)+\gamma_{\rm E} \right]+ \log (2z^2)+\gamma_{\rm E} +3\nn\\&&+\frac{1}{2 z^2}-\frac{1}{2 z^4}+\mathcal{O}\!\left(\frac{1}{z^5}\right)
\ .\end{eqnarray}
These functions appear in the inverse Laplace transforms involving $\log(x)$ or $\text{Ei}(x)$ functions. We give here the main non-trivial formulas used to deal with   Laplace inversions:
\begin{widetext}
\begin{equation}\label{InvLaplace1}
\begin{split}
\mathcal{L}_{s_2\rightarrow t_2}^{-1}\circ\mathcal{L}_{s_1\rightarrow t_1}^{-1}& \left[e^{-m_1 \sqrt{s_1}-m_2 \sqrt{s_2}} \log \left(\sqrt{s_1} + \sqrt{s_2} \right) \right]\\
&=\partial_{m_1} \partial_{m_2} \left\{ \frac{e^{-\frac{m_2^2}{4t_2}-\frac{m_1^2}{4t_1}}}{2\pi \sqrt{t_1 t_2}} \left[2\mathcal{J}\!\left( \frac{\left(m_2 t_1+m_1 t_2\right)^2}{4 t_1 t_2 \left(t_1+t_2\right)} \right) +  \log \! \left( \frac{1}{4t_1} + \frac{1}{4t_2}\right) - \gamma_{\rm E}\right] \right\} \ ,\\
\end{split}
\end{equation}
\begin{equation}\label{InvLaplace2}
\mathcal{L}_{s_2\rightarrow t_2}^{-1}\circ\mathcal{L}_{s_1\rightarrow t_1}^{-1} \left[e^{-m_1 \sqrt{s_1}-m_2 \sqrt{s_2}} \log \left(\sqrt{s_1} \right) \right]= \partial_{m_1} \partial_{m_2}\left\{ \frac{e^{-\frac{m_2^2}{4t_2}-\frac{m_1^2}{4t_1}}}{2\pi \sqrt{t_1 t_2}} \left[2\mathcal{J}\!\left( \frac{m_1^2}{4 t_1} \right) -  \log ( 4t_1) - \gamma_{\rm E}\right] \right\}\ ,
\end{equation}
\begin{equation}\label{InvLaplace3}
\mathcal{L}_{s \rightarrow t}^{-1} \left[ \frac{e^{-m \sqrt{s}}}{m \sqrt{s}}\log(m^2 s)\right] = \frac{e^{- \frac{m^2}{4t}}}{m\sqrt{\pi t}}\left[2\mathcal{J}\left(\frac{m^2}{4t}\right)+\log\left(\frac{m^2}{4t}\right) - \gamma_{\rm E}\right]\ ,
\end{equation}
\begin{equation}
\mathcal{L}^{-1}_{s\rightarrow t}\left[m\sqrt{s} e^{-m \sqrt{s}} \log(m^2s)\right]= \frac{m e^{-\frac{m^2}{4t}}}{2 \sqrt{\pi} t^{3/2}} \left\{-\mathcal{I}\! \left(\frac{m}{\sqrt{2t}}\right)+\left(\frac{m^2}{2t}-1\right)\left[\log\left(\frac{m^2}{4t}\right)-\gamma_{\rm E}\right]\right\}\ ,
\end{equation}
\begin{equation}\label{InvLaplace5}
\mathcal{L}_{s \rightarrow t}^{-1} \left[ \frac{e^{m\sqrt{s}}}{m\sqrt{s}} \text{Ei}\left(-2m \sqrt{s}\right)\right] =\frac{e^{-\frac{m^2}{4 t}}}{2m \sqrt{\pi t }} \left[ -2\mathcal{J}\!\left(\frac{m^2}{4 t}\right) + \log\!\left(\frac{m^2}{t}\right)+\gamma_{\rm E}\right]\ ,
\end{equation}
\begin{equation}\label{InvLaplace4}
\mathcal{L}_{s \rightarrow t}^{-1} \left[ e^{m\sqrt{s}} \text{Ei}\left(-2m \sqrt{s}\right)\right] =\frac{m e^{-\frac{m^2}{4 t}}}{4 \sqrt{\pi }t^{3/2}} \left[ 2\mathcal{J}\!\left(\frac{m^2}{4 t}\right) - \log\left(\frac{m^2}{t}\right)-\gamma_{\rm E}-\frac{2\sqrt{\pi t}}{m}e^{\frac{m^2}{4t}}\text{erfc}\left(\frac{m}{2 \sqrt{t}}\right)\right]\ .
\end{equation}
To derive Eq.~\eqref{InvLaplace1}, we start with an integral representation of the logarithm,
\begin{equation}
\log \left( \sqrt{s_1}+\sqrt{s_2} \right) = \int_0^{\infty} \frac{\text{d} \alpha}{\alpha} \left( e^{-\alpha} - e^{-\alpha \left( \sqrt{s_1}+\sqrt{s_2} \right) } \right)
\ .\end{equation}
We compute now the inverse Laplace transform of this integral representation, with the exponential prefactor
\begin{equation}
\begin{split}
\mathcal{L}_{s_2\rightarrow t_2}^{-1}&\circ\mathcal{L}_{s_1\rightarrow t_1}^{-1} \left[e^{-m_1 \sqrt{s_1}-m_2 \sqrt{s_2}} \left(e^{-\alpha}- e^{-\alpha\left(\sqrt{s_1}+\sqrt{s_2}\right)} \right) \right] \\
&= \frac{m_1 m_2 e^{- \frac{m_1^2}{4t_1} - \frac{m_2^2}{4t_2}}}{4 \pi (t_1 t_2)^{3/2}} \left[ e^{-\alpha} - \left( 1 + \frac{\alpha}{m_2} \right) \left( 1 + \frac{ \alpha}{m_2}\right)e^{ - \alpha^2 \left( \frac{1}{4 t_1} + \frac{1}{4 t_2} \right) -  \alpha \left( \frac{m_1}{2 t_1}+ \frac{m_2}{2t_2}\right)} \right]\ .
\end{split}
\end{equation}
To simplify this expression, it is useful to take the primitive \textit{w.r.t.} $m_1$ and $m_2$,
\begin{equation}
\begin{split}
\int_{m_1,m_2} \mathcal{L}_{s_2\rightarrow t_2,s_1\rightarrow t_1}^{-1}\left[e^{-m_1 \sqrt{s_1}-m_2 \sqrt{s_2}} \left(e^{-\alpha}- e^{-\alpha\left(\sqrt{s_1}+\sqrt{s_2}\right)} \right) \right] = \frac{e^{-\frac{m_2^2}{4t_2}-\frac{m_1^2}{4t_1}}}{\pi \sqrt{t_1 t_2}} \frac{ e^{-\alpha} - e^{- \alpha^2 \left( \frac{1}{4 t_1}+ \frac{1}{4 t_2} \right) - \alpha \left( \frac{m_1}{2 t_1} + \frac{m_2}{2 t_2}\right)} }{ \alpha}
\end{split}
\ .\end{equation}
We still have to deal with the integration over $\alpha$ which is now an integral of the form 
\begin{equation}
\int_{\alpha >0} \frac{e^{-\alpha}- e^{-\alpha^2 A -\alpha B }}{\alpha}
\ .\end{equation}
We can compute this integral by deriving w.r.t $A$, integrating over $\alpha$, and then integrating over $A$; alternatively, we can use the same strategy with $B$. The two results are
\begin{equation}
\int_{\alpha >0} \frac{e^{-\alpha}- e^{-\alpha^2 A -\alpha B }}{\alpha}=\frac{1}{2}\left(\pi \, \text{erfi}\!\left(\frac{B}{2 \sqrt{A}}\right)+\log (A)-2 \log (B)-\gamma_{\rm E} \right)-\frac{B^2 \, _2F_2\!\left(1,1;\frac{3}{2},2;\frac{B^2}{4 A}\right)}{4A}+C_B\ ,
\end{equation}
\begin{equation}
\int_{\alpha >0} \frac{e^{-\alpha}- e^{-\alpha^2 A -\alpha B }}{\alpha}=\frac{\pi}{2} \text{erfi}\!\left(\frac{B}{2 \sqrt{A}}\right)-\frac{B^2 \, _2F_2\!\left(1,1;\frac{3}{2},2;\frac{B^2}{4 A}\right)}{4 A} +  C_A \ .
\end{equation}
Thus 
\begin{equation}
C_A - C_B = \frac{1}{2} \Big[\log (A)-2 \log (B)-\gamma_{\rm E} \Big]
\ ,\end{equation}
and the  case $A=0$, $B=1$, allows us to conclude on $ C_A= \frac{1}{2} \log(A) -\frac{\gamma_{\rm E}}{2}$ and $ C_B= \log(B)$. The final result for the integral is
\begin{equation}
\begin{split}
\int_{\alpha >0} \frac{e^{-\alpha}- e^{-\alpha^2 A -\alpha B }}{\alpha}&=\frac{\pi}{2}  \text{erfi}\!\left(\frac{B}{2 \sqrt{A}}\right)-\frac{B^2 \, _2F_2\!\left(1,1;\frac{3}{2},2;\frac{B^2}{4 A}\right)}{4 A} + \frac{1}{2} \log(A) - \frac{\gamma_{  \rm E}}{2}\\
&= \mathcal{J\!} \left(\frac{B^2}{4A} \right) + \frac{1}{2} \log(A) - \frac{\gamma_{\rm E}}{2}\ .
\end{split}
\end{equation}
We checked this result  numerically with   very good precision.

Applying this formula to the integral over $\alpha$ and specifying $A= \frac{1}{4t_1} + \frac{1}{4t_2}$ and $B=\frac{m_1}{2t_1} + \frac{m_2}{2t_2}$, we obtain Eq.~\eqref{InvLaplace1}. The same computation,  with $A= \frac{1}{4 t_1}$, and $B = \frac{m_1}{2 t_1}$ gives Eq.~\eqref{InvLaplace2}.\

To derive Eq.\ \eqref{InvLaplace5} (with $m=1$ for simplicity), we start with the integral representation of the exponential integral function,\begin{equation}
e^{\sqrt{s}} \text{Ei}\left(-2 \sqrt{s}\right)=-\int_0^{\infty } \frac{e^{-\sqrt{s}-x}}{\sqrt{s} \left(2 \sqrt{s}+x\right)} \, \rmd x = - \int_0^{\infty}  \frac{e^{-\sqrt{s} (2 y+1)}}{y+1}\rmd y\ .
\end{equation}
Doing the inverse Laplace transform  inside the integral leads to
\begin{equation}
\begin{split}
\mathcal{L}^{-1}_{s \rightarrow t}\left[e^{\sqrt{s}} \text{Ei}\left(-2 \sqrt{s}\right)\right]&=-\int_0^{\infty } \frac{(2 y+1) e^{-\frac{(2 y+1)^2}{4 t}}}{2 \sqrt{\pi } t^{3/2} (y+1)} \rmd y= -\frac{e^{-\frac{1}{4 t}}}{\sqrt{\pi } t^{3/2}} \int_0^{\infty}\frac{t e^{-u}}{\sqrt{4 t u+1}+1} \rmd u\ .\\
&=\frac{e^{-\frac{1}{4 t}} \left[6 t \left(\pi  \text{erfi}\left(\frac{1}{2 \sqrt{t}}\right)+\log (t)-\gamma +2\right)-\, _2F_2\left(1,1;2,\frac{5}{2};\frac{1}{4 t}\right)\right]}{24 \sqrt{\pi } t^{5/2}}-\frac{1}{2 t}\ .
\end{split}
\end{equation}
To express this result in terms of our special function $\mathcal{J}$, we can use the following relation between Hypergeometric functions, 
\begin{equation}
 _2F_2\left(1,1;2,\frac{5}{2};a\right)= 3 \, _2F_2\left(1,1;\frac{3}{2},2;a\right)-\frac{3 \left [e^a \sqrt{\frac{\pi }{4 a}} \text{erf}(\sqrt{a})-1\right]}{a}\ .
\end{equation}
This can be checked by Taylor expansion. With that, and the definition of $\mathcal{J}$ in Eq.\ \eqref{defCalJ}, we obtain  the announced result \eqref{InvLaplace4}. Equation~\eqref{InvLaplace5} is obtained from there by taking one derivative.
        
\section{Check of the covariance function}
\label{AppendixCheck}
As a check of the   action, we   computed  the two-point correlation function (\textit{i.e.}\ the covariance function). The needed path integral is
\begin{equation}
\langle X_{t_1} X_{t_2} \rangle =\int_{x} \int_{X_0=0}^{X_T=x} \mathcal{D}\left[X \right]X_{t_1} X_{t_2} e^{-S\left[X \right]}
\ .\end{equation}At first order in $\varepsilon$, we can expand this path integral using Eq.~\eqref{ActionExpansion}\,, 
\begin{equation}
\langle X_{t_1} X_{t_2} \rangle = \langle X_{t_1} X_{t_2} \rangle_0 + \frac{\varepsilon}{2} \int_0^{t-\tau} \rmd \tau_1 \int_{\tau_1+\tau}^t \rmd \tau_2 \frac{\left\langle X_{t_1} X_{t_2}\dot{X}_{\tau_1} \dot{X}_{\tau_2} \right\rangle_0}{\tau_2 -\tau_1}  + O(\varepsilon^2)
\ .\end{equation}
Here, averages $\langle \bullet \rangle_0$ are performed with the action $S_0[X]$ given in Eq.~\eqref{Action0}, \textit{i.e.\ }the action of standard Brownian motion   with  diffusive constant $D_{\varepsilon,\tau} = 1 + 2 \varepsilon [1 + \ln(\tau)] + O(\varepsilon^2)$. This action is quadratic, and using Wick contractions allows us to write
\begin{equation}
\langle X_{t_1} X_{t_2}\dot{X}_{\tau_1} \dot{X}_{\tau_2} \rangle_0 = 4 \big(\min (t_1,t_2) \delta (\tau_1 -\tau_2) +  \theta(t_1 - \tau_1) \theta(t_2 - \tau_2) +\theta(t_1 - \tau_2)\theta(t_2 - \tau_1) \big) + O(\varepsilon)\ .
\end{equation}
In this equation, we used only the zeroth order for the diffusive constant ($D_{\varepsilon,\tau} = 1 + O(\varepsilon)$); the first term does not contribute since  $\tau_1$ and $\tau_2$ do not  coincide due to the time regularization.

The last two terms require to compute the integrals\begin{equation}
\begin{split}
\int_0^{\min(t_1,t_2-\tau)}\rmd\tau_1   \int_{\tau_1+\tau}^{t_2}\rmd\tau_2  \frac{ 1}{\tau_2 -\tau_1}+&\int_0^{\min(t_2,t_1-\tau)} \rmd\tau_1  \int_{\tau_1+\tau}^{t_1} \rmd\tau_2 \frac{ 1}{\tau_2 -\tau_1}\\
&= t_1 \ln(t_1) + t_2 \ln(t_2) - |t_1 -t_2| \ln|t_1 -t_2| - 2 \min(t_1,t_2)( \ln(\tau) +1)\ .
\end{split}
\end{equation}
We   now sum all contributions   to order $\varepsilon$, the Brownian result with the rescaled diffusive constant being $\langle X_{t_1} X_{t_2}\rangle_0= 2 D_{\varepsilon,\tau}\min(t_1,t_2)$. This gives
\begin{equation}
\begin{split}
\langle X_{t_1} X_{t_2} \rangle &= 2 D_{\varepsilon,\tau}\min(t_1,t_2) + 2 \varepsilon\, (t_1 \ln(t_1) + t_2 \ln(t_2) - |t_1 -t_2| \ln|t_1 -t_2|) - 4 \varepsilon \min(t_1,t_2) (\ln(\tau) +1) + O(\varepsilon^2) \\
&= 2 \min(t_1,t_2) + 2 \varepsilon\, (t_1 \ln(t_1) + t_2 \ln(t_2) - |t_1 -t_2| \ln|t_1 -t_2|) + O(\varepsilon^2)\\
&= t_1^{1 + 2 \varepsilon} + t_2^{1 + 2 \varepsilon} - |t_1-t_2|^{1 + 2 \varepsilon} +O(\varepsilon^2)\ .
\end{split}
\end{equation}
The $\tau$ dependence in the diffusive constant and in the first correction to the action cancel,  and we recover the fBm correlation function at first order in $\varepsilon$. We also see   that the correction to the diffusive constant is equivalent to setting  $\log(\tau)=-1$.

\end{widetext}


\tableofcontents

\end{document}